\documentclass[11pt, DIV10,a4paper]{article}
\usepackage{color}
\usepackage{float}
\usepackage[bf,small]{caption2}
\usepackage{comment}
\usepackage{amsmath}
\usepackage{bigints}
\usepackage{gensymb}
\usepackage{rotating, booktabs}
\usepackage{amsopn}
\usepackage{amsfonts}
\usepackage{amsthm}
\usepackage{amssymb}
\usepackage{amsbsy}
\usepackage{multirow}
\usepackage{natbib}
\usepackage{tocbibind}
\usepackage[a4paper,colorlinks,breaklinks,bookmarksopen,bookmarksnumbered]{hyperref}
\usepackage[refpage]{nomencl}
\usepackage{makeidx}
\usepackage{pst-all}
\usepackage{epsfig,psfrag}
\usepackage{graphicx}
\usepackage{amsmath}
\usepackage{epstopdf}
\usepackage{tabularx}
\usepackage{subfigure}
\usepackage{setspace}
\usepackage[mathscr]{euscript}
\usepackage[margin=1in]{geometry}
\usepackage{xr-hyper}
\usepackage{amsfonts} 
\usepackage{calc}
\usepackage{enumitem}

\newcommand{\ueq}[1][]{%
  \if\relax\detokenize{#1}\relax
    \sbox0{$\underbrace{=}_{}$}%
    \mathrel{\mathmakebox[\wd0]{=}}
  \else
    \mathrel{\underbrace{=}_{\mathclap{#1}}}
  \fi}

\newcommand{\bzero}{\boldsymbol{0}}
\newcommand{\bone}{\boldsymbol{1}}

\newcommand {\ctn}{\citet}       

\usepackage{url}
\newcommand{\e}{\ensuremath{\epsilon}}

\newcommand {\btheta}{\mbox{\boldmath $\theta$}}

\newcommand {\bmu}{\mbox{\boldmath $\mu$}}
\newcommand {\bbeta}{\mbox{\boldmath $\beta$}}
\newcommand {\bSigma}{\mbox{\boldmath $\Sigma$}}

\newcommand {\bepsilon}{\mbox{\boldmath $\epsilon$}}

\newcommand{\bI}{\mathbf I}

\newcommand{\bA}{\mathbf A}
\newcommand{\bB}{\mathbf B}
\newcommand{\bC}{\mathbf C}

\newcommand{\bv}{\mathbf v}

\newcommand{\bx}{\mathbf x}

\newcommand{\bz}{\mathbf z}
\newcommand{\bd}{\mathbf d}
\newcommand{\bh}{\mathbf h}
\newcommand{\bR}{\mathbf R}
\newcommand{\bD}{\mathbf D}
\newcommand{\bH}{\mathbf H}
\newcommand{\bG}{\mathbf G}
\newcommand{\bs}{\mathbf s}

\newcommand{\bof}{\mathbf f}

\numberwithin{equation}{section}
\numberwithin{algo}{section}
\numberwithin{table}{section}
\numberwithin{figure}{section}

\normalsize

\begin{document}

\title{\textbf{How Ominous is the Premonition of Future Global Warming?}}
\author{Debashis Chatterjee$^{\dag}$ and Sourabh Bhattacharya$^{\ddag, +}$ }
\date{}
\maketitle
\begin{center}
$^{\dag}$  Visva Bharati University \\
$^{\ddag}$ Indian Statistical Institute\\
$+$ Corresponding author:  \href{mailto: bhsourabh@gmail.com}{bhsourabh@gmail.com}
\footnote{This article is intended for Dr. C. R. Rao Special Issue.}
\end{center}

\begin{abstract}

Global warming -- the rise in global average temperatures observed in recent decades -- has drawn significant attention due to its profound and far-reaching 
impacts on the climate system. A critical question is whether this warming trend will continue into the future. General circulation models (GCMs) are the primary 
tools for projecting such future climate scenarios, and nearly all of them forecast an alarming increase in global temperatures.

While the reality of current global warming is undeniable, the reliability of GCM forecasts remains open to scrutiny. In this study, we undertake a systematic 
evaluation of these forecasts using our recently developed Bayesian multiple testing framework for model selection in inverse regression problems. Our central question is: 
{\it How likely is the current global warming pattern, assuming the future projections of the GCMs are correct?} This reframes the typical forecasting paradigm 
into an inverse regression setting, where the present is treated as unknown and inferred from future outcomes. Our framework coherently combines this inverse formulation 
with conventional forward modeling to identify the best-fitting models.

To model the temporal dynamics of global temperature, we adopt a nonparametric compositional Gaussian process (GP) emulator that treats the climate system as an unknown 
black-box process. Using data from the Intergovernmental Panel on Climate Change (IPCC), we find that GCMs which perform best under various future scenarios still fail to 
convincingly account for the observed warming trend -- if one assumes only their projected future data are accurate.

We further analyze ensemble forecasts from all GCMs under each scenario as multivariate time series governed by multidimensional GPs. The inverse-fit results from 
this multivariate framework strongly reinforce the conclusions drawn from the univariate analysis: assuming the GCMs’ future predictions are valid, the observed global 
warming pattern appears highly improbable. This casts serious doubt on the representativeness of IPCC-endorsed GCM projections.

Lastly, we offer our own forecasts of future global temperatures based solely on the historical data, using our GP-based model. These forecasts do not support the drastic 
warming trends predicted by most GCMs. In fact, only the projections under the ``Commitment" scenario fall within the high-density regions of our Bayesian forecast distributions.
\\[2mm]
\end{abstract}

\section{Introduction}
\label{sec:intro}

The gradual warming of the earth's average surface temperature,
known as global warming, is perhaps the gravest concern for environmental scientists. 
Overwhelming evidence from multiple and independent sources of data has led the U.S. Global Change Research Program, the National Academy of Sciences, and 
the Intergovernmental Panel on Climate Change (IPCC) to independently conclude that global warming, particularly, in the recent decades, is undeniable.
As per the records (see \ctn{IPCC18}, for example), compared to the pre-industrial baseline $1850-1900$, the $2009-2015$ time period was warmer by about $0.87\degree$C, 
and that each decade is getting warmer by about $0.2\degree$C. 
Such an alarming rate of increase is unprecedented, and even the prehistorical rates of global warming, such as the Paleocene-Eocene Thermal Maximum, fails to match
the current rate of global warming (see, for example, \ctn{Masson13}). However, see \ctn{Idso13a} and the references therein who argue, 
providing details on past temperature records, that this global warming is not unprecedented.

Global warming is considered responsible for increasing droughts, heat waves, increase in extremely wet or dry events within the monsoon period in India and East Asia,
increase in frequencies of hurricanes and typhoons, increase in global sea level as a result of melting glaciers, expansion of deserts and much more.
According to the IPCC, ``{\it human influence on climate has been the dominant cause of observed warming since the mid-20th century}", and this conclusion has been upheld by
all scientific bodies. In fact, human activities are estimated to have caused approximately 1.0$\degree$C of global warming above pre-industrial levels.
Scientific investigations reveal that (see \ctn{Olivier19}) the emission of greenhouse gases, with over 90\% of the impact 
of carbon dioxide and methane, has been a major contributing factor to global warming by human activities such as fossil fuel burning, 
agricultural emissions and deforestation. But also see \ctn{Idso13a} who write  
``{\it The empirical observations cited above reveal a relationship opposite of what is expected if carbon dioxide and methane were the powerful greenhouse
gases the IPCC claims them to be. Clearly, if there is anything at all that is unusual, unnatural, or unprecedented about Earth's current surface air
temperature, it is that it is so {\it cold}}." and \ctn{Lange13} who mention in their key findings section
``{\it There appears to be nothing unusual about the extremes of wetness and dryness experienced during the twentieth century, or about recent changes in
ocean circulation, sea level, or heat content, that would require atmospheric carbon dioxide forcing to
be invoked as a causative factor. Natural variability in the frequency or intensity of precipitation extremes and sea-level change occurs largely on decadal and
multidecadal time scales, and this variability cannot be discounted as a major cause of recent changes where they have occurred.}"

The IPCC has warned that if the warming increases by $1.5\degree$C compared to the pre-industrial era $1850-1900$, 
then human and natural systems would be at grave risk. The concerning news is that under the current conditions
global warming is projected to surpass $2.8\degree$C by the year $2100$ (see \ctn{CAT19}).

The climate projections are performed by the general circulation models (GCMs) that attempt to model the major climate system components, namely, atmosphere, land surface, ocean 
and sea ice, and the interactions among them. Expressing great confidence in such models, the IPCC has claimed that (see \ctn{Lupo13})
``{\it development of climate models has resulted in more realism in the representation of many quantities and aspects of the climate system}," adding,
``{\it it is extremely likely that human activities have caused more than half of the observed increase in global average surface temperature since the 1950s}".
However, \ctn{Lupo13} writes ``{\it Confidence in a model is further based on the careful evaluation of its performance, in which model output is compared against 
actual observations. A large portion of this chapter, therefore, is devoted to the evaluation of climate models against real-world climate and other biospheric data. 
That evaluation, summarized in the findings of numerous peer-reviewed scientific papers described in the different subsections of this chapter, reveals the 
IPCC is overestimating the ability of current state-of-the-art GCMs to accurately simulate both past and future climate. The IPCC's stated confidence in the models,
as presented at the beginning of this chapter, is likely exaggerated. The many and varied model deficiencies discussed in this chapter indicate much work remains
to be done before model simulations can be treated with the level of confidence ascribed to them by the IPCC.}"
This was written quite a few years ago, and by now we expect the GCMs to have reduced their deficiencies and to yield more reliable climate projections.

The current GCM predictions by different GCMs available from the IPCC website \url{http://www.ipcc-data.org/sim/gcm_global/index.html}, 
under the assumptions of several future climate scenarios associated with greenhouse gas emissions, pertaining to the  
Special Report on Emissions Scenarios (SRES), a report by the IPCC published in 2000. According to the IPCC Fourth Assessment Report (AR4), published in 2007,
there are three SRES, namely, A1B, A2 and B1. Brief descriptions of the assumptions, obtained from the IPCC website, are reproduced below for the reader's convenience.

The key assumption for A1B is a future world of very rapid economic growth, low population growth and rapid introduction of new and more efficient technology. 
Major underlying themes are economic and cultural convergence and capacity building, with a substantial reduction in regional differences in per capita income. 
In this world, people pursue personal wealth rather than environmental quality.

SRES A2 corresponds to a very heterogeneous world. The underlying theme is that of strengthening regional cultural identities, with an emphasis on 
family values and local traditions, high population growth, and less concern for rapid economic development.

In SRES B1, a convergent world with the same global population as in the A1B is assumed 
but with rapid changes in economic structures toward a service and information economy, with reductions in materials intensity, 
and the introduction of clean and resource-efficient technologies. 

Commitment is a non-SRES idealised scenario in which the atmospheric burdens of long-lived greenhouse gases are held fixed at AD2000 levels.

The scenarios A1B, A2, B1 and Commitment consist of $21$, $17$, $21$ and $16$ GCMs, respectively, each yielding a simulated 
global mean temperature time series in the duration $1900-2099$.
The HadCRUT4 observed near surface average global temperature dataset during 
the years $1850-2020$ is also available
from the IPCC website; see \url{https://www.metoffice.gov.uk/hadobs/hadcrut4/data/current/download.html}. At the time of writing this paper, 
the year $2020$ was ongoing, and so we found reasons
to doubt the reliability of the last few data points, and as such, here we shall consider the dataset ranging from $1850-2016$. 
This dataset pertains to temperature anomalies in degree celsius relative to the years $1961-1990$. 
Now, the most widely quoted value for the global average temperature for the $1961-1990$ period is $14\degree$C, which has been developed by \ctn{Jones99}.
Hence, we convert the HadCRUT4 temperature anomalies data to (approximate) 
actual temperatures by adding $14\degree$C to the anomalies.
We also convert the GCM-simulated actual temperatures, originally available in Kelvin, to degree celsius.  

\begin{figure}
	\centering
	\subfigure [$21$ GCMs.]{ \label{fig:a1b}
	\includegraphics[width=7.5cm,height=6.7cm]{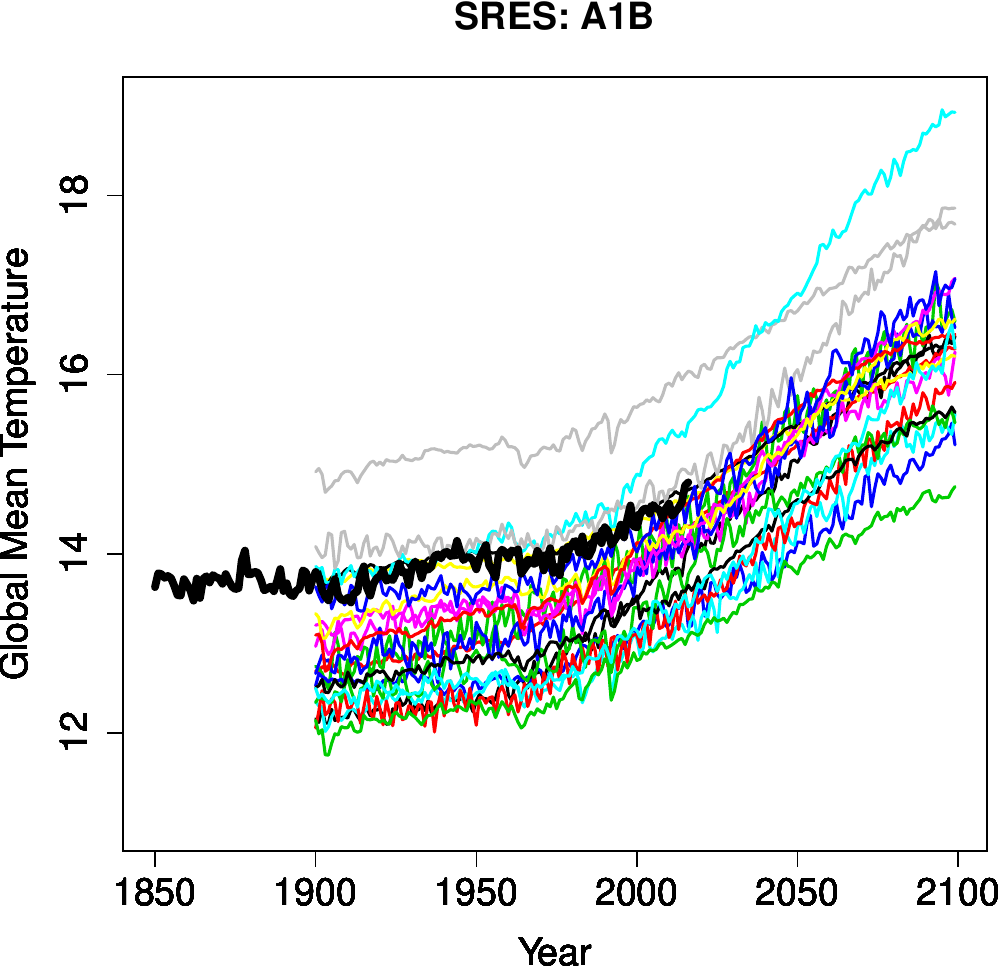}}
	\hspace{2mm}
	\subfigure [$17$ GCMs.]{ \label{fig:a2}
	\includegraphics[width=7.5cm,height=6.7cm]{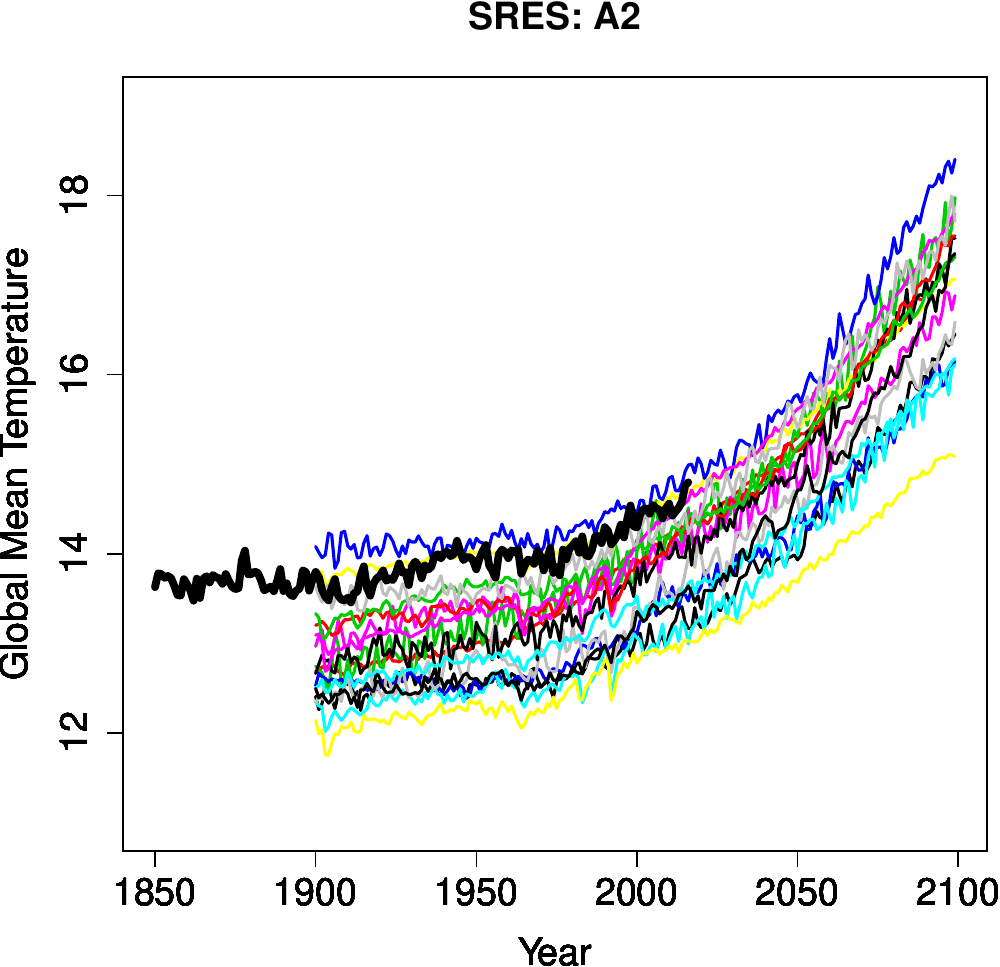}}\\
	\vspace{2mm}
	\subfigure [$21$ GCMs.]{ \label{fig:b1}
	\includegraphics[width=7.5cm,height=6.7cm]{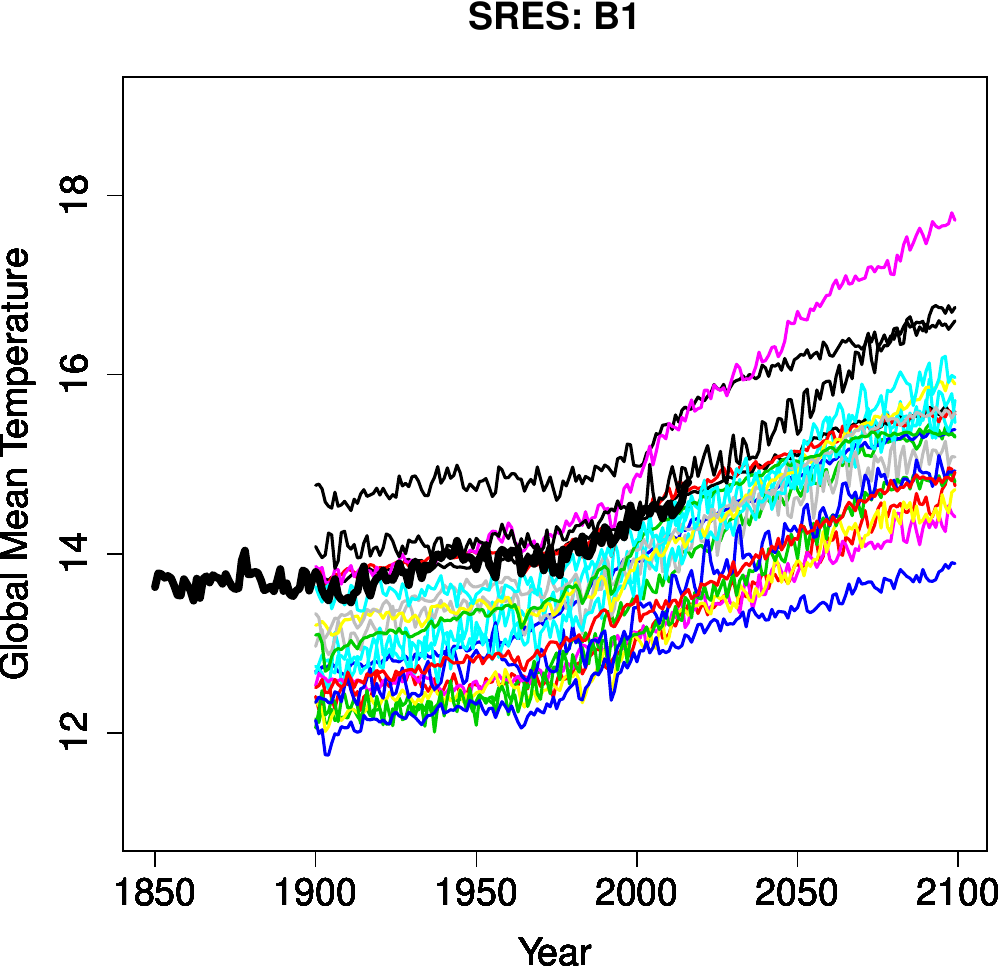}}
	\hspace{2mm}
	\subfigure [$16$ GCMs.]{ \label{fig:commit}
	\includegraphics[width=7.5cm,height=6.7cm]{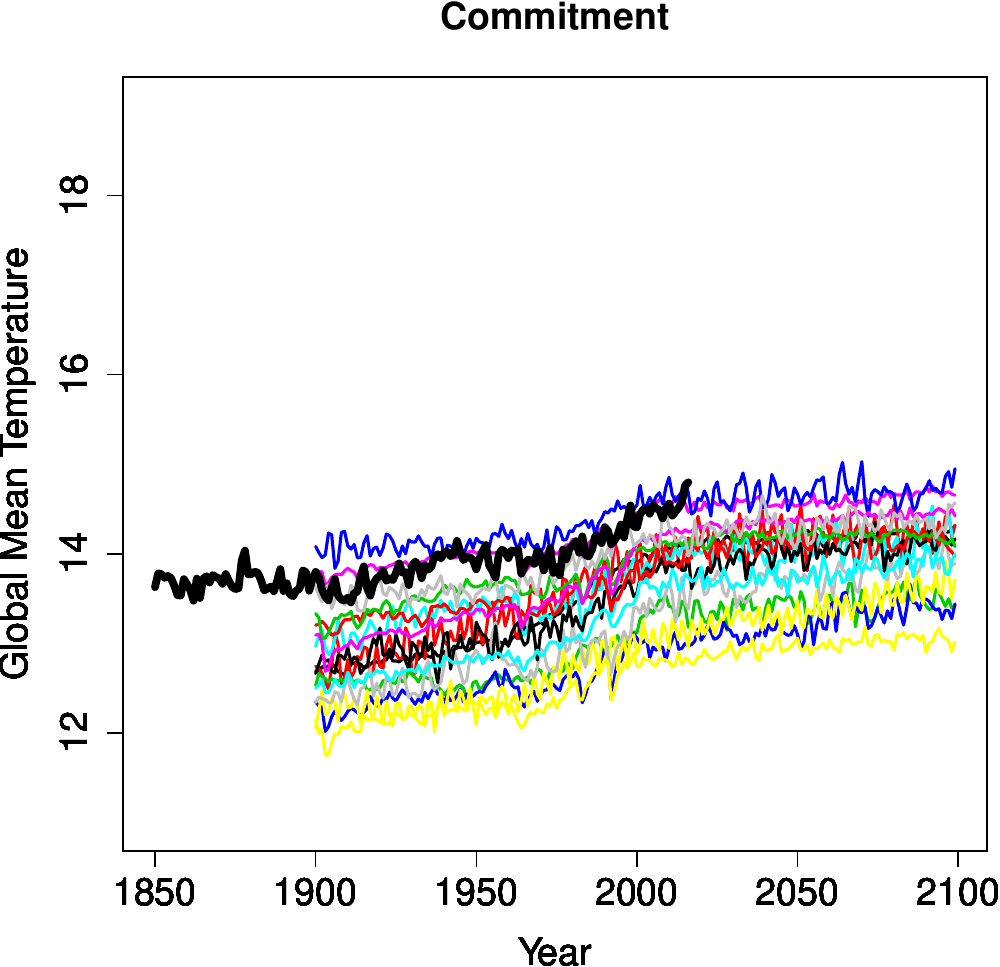}}
	\caption{Visualization of the HadCRUT4 data (thick, black line) and the GCM based time series. The temperature is in $\degree$C.}
	\label{fig:gcm_models}
\end{figure}

Figure \ref{fig:gcm_models} presents the
diagrams of the HadCRUT4 dataset (thick, black line) and the GCM predictions. Observe that the GCM based global temperatures seem to significantly underestimate the observed 
global temperatures during the years $1900-2016$. Moreover, their rates of increase seem to be much faster than that of the observed dataset. 
Hence, the sharp increase of most of the GCM based future temperatures till the end of this century, is potentially unreliable.
Observe that the future predictions of the Commitment models are more stable compared to the others.  

Perhaps the most important ingredient in any statistical learning is quantification of uncertainty. The GCM results displayed in Figure \ref{fig:gcm_models}
are devoid of any uncertainty quantification; at least we are unable to find any in the IPCC website. 
In the observed HadCRUT4 data context, an ensemble of $100$ time series 
are available, which has been recommended by climatologists to  quantify uncertainty in the observations to some extent.
It seems that ensembles can be obtained even for GCM models, provided they are run with different initial conditions. But the models are deterministically dynamic,
and non-probabilistic, so that rigorous statistical ways of uncertainty quantification need not apply. It is thus not clear how believable the future global warming
forecasts presented in Figure \ref{fig:gcm_models} are. In fact, as detailed in \ctn{Lupo13}, the leading scientific experts have placed no faith in 
the GCMs. For instance, Freeman Dyson has written (see \ctn{Dyson07}),
``{\it I have studied the climate models and I know what they can do. The models solve the equations of fluid dynamics, and they do a very good
job of describing the fluid motions of the atmosphere and the oceans. They do a very poor job of describing the clouds, the dust, the chemistry, and the biology of
fields and farms and forests. They do not begin to describe the real world that we live in}". \ctn{Green09} tested
whether the warming-trend forecasts used by the IPCC are more accurate than the standard benchmark
forecast that there will be no change, using the historical HadCRUT3 observed dataset, which exhibited clear global warming till the present years.
To their surprise, they found that the errors from the IPCC warming trend forecasts
were nearly eight times greater than the errors from the no-change forecasts. Consequently, \ctn{Green09} recommend that the best policy is to do nothing
about global warming.

The evaluation method of \ctn{Green09} was not based upon model based statistical or probabilistic methods and thus calls for more sophisticated analyses.
In this article, we evaluate the global warming forecasts shown in Figure \ref{fig:gcm_models} in a rigorous footing using our recently-developed Bayesian methods. 
An important question in this regard is if the observed HadCRUT4 time series is plausible, given the GCM forecasts. This gives rise to an inverse regression problem
in the following sense. The future temperature depends upon the present; our goal is to learn about the present, pretending it to be unknown, while the future
is assumed to be known. Given each climate scenario, we then select the best GCM using our 
Bayesian multiple testing paradigm for model selection in inverse regression problems (\ctn{Chatterjee20b}). The multiple testing procedure, it must be mentioned,
not only considers the inverse aspect; it combines the inverse aspect with the forward in a coherent Bayesian compound decision theoretic sense, to compare the
models under consideration. Once the best models are selected, we then show that even for such best GCMs, 
the Bayesian posterior time series for the current years ($1850-2016$) do not convincingly support the observed HadCRUT4 data, given the future forecasts
for the years $2017-2099$. 

It is important to discern that the actual model for climate dynamics must be infeasibly complex and, in fact, unknown. 
Even the GCMs, which are complex computer models, are nothing but black boxes to us. The purpose of this discussion is to make it clear
that standard time series models are inappropriate for climate dynamics. As such, we consider modeling the logarithm of the global temperature at any year as a  
function of that at the previous year, plus some random error, where the function is assumed to be unknown and modeled appropriately by Gaussian process (GP).
The key idea has parallels with \ctn{Bhattacharya07} and \ctn{Ghosh14}. 
It is important to appreciate that although our time series model seems to be a Marovian model at the first glance, it is actually made up of 
compositions of GPs, and as we shall clarify, has highly structured non-Markovian dependence, with non-Gaussian, intractable distribution. 

Apart from the Bayesian model selection framework, we also treat the different GCM time series in any given climate scenario as an ensemble, and extend
our univariate climate dynamics modeling to the multivariate situation, with multidimensional GPs replacing the previous
one-dimensional GPs. The posterior distribution of the mean of the logarithm of the time series during $1850-2016$, averaged over the dimensions (ensembles) 
in the corresponding climate scenario, is of interest in these cases. Our results in the multidimensional context very emphatically bear out
that the HadCRUT4 data with its global warming trend must be highly implausible if the GCM forecasts are believed to be true.

Furthermore, given the observed HadCRUT4 data and our GP emulation model, we also provide Bayesian forecasts for the years $2017-2099$, 
which show no evidence of drastic future global warming. Interestingly, as can be anticipated from panel (d) of Figure \ref{fig:gcm_models}, 
only the forecasted time series by the best GCM model in the Commitment scenario fall in the high density regions of our Bayesian forecasted time series.

The general reader is likely to anticipate from the above discussions that computations associated with a study of such a proportion must be infeasibly complex.
We assure this is not so. We wrote all our codes in the C language as efficiently as possible, parallelizing them using the Message Passing Interface (MPI) protocol
whenever relevant, for example, in the case of the Bayesian multiple testing procedure. In such a case, we implemented the GP models associated with the large
number of GCM forecasts in the parallel computing architecture (VMWare) available at our institution. Very efficient and time-saving computations are the results
of our parallel processing. Details will be presented in due course.

The rest of our article is structured as follows. 
We begin with an overview of our methodological contributions in Section \ref{sec:overview}.
In Section \ref{sec:gp_time_series} we introduce our GP emulation model for climate dynamics,
and discuss relevant prior choices in Section \ref{sec:prior_univariate}. The methods for Bayesian posterior 
inference regarding the current temperature time series 
given the future GCM simulations, and regarding future forecasts given the current temperature time series, are detailed in Section \ref{sec:current_future}.
In Section \ref{sec:mult_inv} we introduce our Bayesian multiple testing procedure in the context of best GCM selection in different climate scenarios, and provide
details on our method of implementation in Section \ref{sec:test_implementation}. The results of our best GCM selections and their detailed analyses are provided in  
Section \ref{sec:gcm_results}. In Section \ref{sec:multivariate}, we model the ensemble of GCM-based future temperature time series in each climatic scenario 
as nonparametric multidimensional time series, driven by multidimensional GPs, and present the relevant theory and methods.
The results and detailed analyses of our Bayesian multivariate GP emulation of climate dynamics are presented in Section \ref{sec:results_mult}.
In Section \ref{sec:forecast} we forecast the future global temperature with our Bayesian GP approach, conditional on the HadCRUT4 data, and compare
our results with the GCM forecasts as well as with the analysis of \ctn{Green09}. Finally, in Section \ref{sec:conclusion},
we summarize our contributions, along with relevant discussions. 

\section{An overview of our methodological contributions}
\label{sec:overview}
Amid this complex intersection of methodology and application, it is important to delineate our new methodological contributions. 

First,
while GPs are widely used in spatial statistics (see, for example, \ctn{Cressie93} and \ctn{Banerjee14}), Bayesian machine
learning (\ctn{Ras06}) and complex computer experiments (see, for example, \ctn{Santner03}), their compositions for direct nonparametric time series modeling have not been
explored in the literature. This is true not only in climate science but in general statistics, for both fequentist and Bayesian paradigma.
The closest related work involves modeling latent states (\ctn{Ghosh14}). In fact, compositional GPs for dynamic black-box computer experiments 
were first proposed by \ctn{Bhattacharya07}, and our approach generalizes this idea to a broader time-series setting.

Second, the inverse regression perspective we adopt is itself a methodological innovation in time-series analysis. It allows a rigorous Bayesian framework to evaluate future 
forecasts -- regardless of how those forecasts were generate -- by assessing how probable the observed past would be if the projected future were true. While inverse problems 
have been addressed in other contexts, \ctn{Chatterjee22} notes that such formulations are rare in time-series modeling. Our recognition of the inverse paradigm's relevance 
to global warming studies is a key conceptual leap in this paper.

This new perspective introduces challenges in both inference and model selection: given only forecasts from multiple GCMs, how do we determine 
which ones best support the observed past? Traditional model selection techniques compare different models based on a single dataset. Here, however, we are faced with 
a new challenge in the inverse regression setup -- given only multiple forecast data  generated by as many competing models, the best model needs to be selected. 
Our solution is to construct an averaged forecast time series from each GCM and fit our compositional GP models 
tuned to individual GCM forecasts, 
to this ensemble. Each GCM thus yields a distinct black-box model, which we then compare via Bayesian inverse model selection.

However, model selection in the inverse Bayesian regression setup remains unexplored in the literature. In his regard, we adopt and extend  
our Bayesian multiple testing framework for inverse regression (\ctn{Chatterjee20b}), enabling principled comparison across GCMs. 
This framework also allows us to incorporate forward modeling in tandem with the inverse approach.

Third, implementing this analysis demanded efficient computational tools. We designed and employed a parallel algorithm to manage the computational burden of 
multiple model comparisons under complex compositions of  GP priors. To further validate our findings, we analyzed the ensemble forecasts from each GCM scenario 
as a multivariate time series, modeled using compositions of multidimensional GPs. This multivariate framework does not require model selection. 
Instead, we applied a 
Bayesian decision-theoretic model adequacy test (\ctn{Bhattacharya13}) to evaluate the inverse fit. Our multivariate analysis strongly supports the conclusions drawn 
from the univariate models.

Lastly, our contributions are not limited to critical assessment of th GCM projections. We provide independent Bayesian forecasts of future global mean temperatures 
based on historical observations (1850--2016). Our forecasts for 2017–2099, constructed using nonparametric GP composition models, show that the drastic 
warming predicted by most GCMs lies outside the high-probability regions of our posterior distributions. This offers additional evidence of possible misalignment 
between GCM projections and historical trends.

For further details regarding various novel aspects of Bayesian inverse regression, including those discussed above, see \ctn{Chatterjee22}, which of course includes 
a detailed chapter on inverse Bayesian treatment of this global warming problem.

\section{Compositional GP emulator for nonparametric climate dynamics}
\label{sec:gp_time_series}
Let $\{x_t:~t=0,1,2,\ldots\}$ denote the time series representing the logarithm of the global temperature over time. For time $t\geq 1$, we model $x_t$ as
\begin{equation}
x_t=f_t(x_{t-1})+\e_t, \label{eq:evo2}
\end{equation}
where $\e_t\sim N(0,\sigma^2_{\e})$ independently, for $t\geq 1$. In this article, we assume that $x_0$ is known.
Crucially, we assume that $f_t$ is an unknown function dependent on time $t$. For any real $z$, we write $f_t(z)=f(t,z)$, 
where $f(\cdot)$ is considered an unknown function on $\mathbb R^+\times\mathbb R$, which we shall model as a GP.
Here $\mathbb R^+=[0,\infty)$ and $\mathbb R=(-\infty,\infty)$.

To simplify notation, define the input vector $x^*_{t,u}=(t,x_u)$, so that the model becomes
\begin{equation}
	x_t=f(x^*_{t,t-1})+\e_t,\hspace{2mm}\e_t\sim N(0,\sigma^2_{\e})~\mbox{independently}. \label{eq:evo3}
\end{equation}

%
We model $f(\cdot)$ as a GP with mean function $\mu_f(\cdot)=\bh(\cdot)'\bbeta_f$ and 
with $\bh(x^*)=(1,x^*)'$ for any $x^*\in\mathbb R^+\times\mathbb R$,
and covariance
function of the form $\sigma^2_f c_f(\cdot,\cdot)$. Here $\sigma^2_f$ is the process variance and $c_f$ is the correlations function.
Typically, for any $z^*_1,z^*_2\in\mathbb R^+\times\mathbb R$, 
$c_f(z^*_1,z^*_2)=\exp\{-(z^*_1-z^*_2)'\bR_f(z^*_1-z^*_2)\}$, where $\bR_f$ is a $2\times 2$-dimensional diagonal matrix 
consisting of respective smoothness parameters $\{r_{1,f},r_{2,f}\}$ that control the rate at which the correlation decays with distance. 

Although the exponential kernel implies infinite smoothness, this assumption is not problematic in our context, since the temperature data are global 
averages, inherently smoothed over space and time (year). However, for applications with more granular or volatile data, this assumption may be less tenable.

A well-known challenge with GP models is the computational cost associated with inverting large covariance matrices. This issue becomes more 
pronounced with compositional GPs, where function evaluations are nested or time-evolving. To mitigate this, we introduce auxiliary variables 
that facilitate efficient sampling and improve numerical stability. This strategy has been successfully used in prior work on dynamic Gaussian proceeses
(\ctn{Bhattacharya07}; \ctn{Ghosh14}).

This emulation model allows us to fit a fully nonparametric representation of the global temperature series, capturing complex nonlinear dependencies 
over time without committing to rigid functional forms. It also forms the core of both our forward forecasting and inverse model assessment frameworks discussed 
in the next sections.

Our model is thus associated with the parameter set $\btheta=(\btheta_f,\sigma^2_{\e})$, where $\btheta_f=(\bbeta_f,\sigma^2_f,r_{1f},r_{2f})$. 
The choice of the priors on the parameters will be discussed subsequently, but we shall assume that all the components of $\btheta$ are {\it a priori} independent.
Henceforth, abusing notation, we shall denote densities and distributions using the notation $[\cdot]$ and conditional densities and distributions by
$[\cdot|\cdot]$.

\subsection{Hierarchical structure induced by our GP approach}
\label{subsec:hierarchy}
Thus, for any $T>1$, our modeling strategy can be described in the following hierarchical form:
\begin{align}
[x_t|f,\btheta_f,x_{t-1}]&\sim N\left(f(x^*_{t,t-1}),\sigma^2_{\e}\right);~t=1,\ldots,T;\label{eq:x_t_dist}\\
[f(\cdot)|\btheta_f]&\sim GP\left(\bh(\cdot)'\bbeta_f,\sigma^2_fc_f(\cdot,\cdot)\right);\label{eq:gp_f}\\
[\bbeta_f,\sigma^2_f,\bR_f,\sigma^2_{\e}]
	&=[\bbeta_f][\sigma^2_f][r_{1f}][r_{2f}][\sigma^2_{\e}],\label{eq:theta_prior}
\end{align}
where the components of $\bbeta_f$ will also be considered independent {\it a priori}.
Forms of the prior distributions in (\ref{eq:theta_prior}) are provided in Section \ref{sec:prior_univariate}.

\subsection{Joint distribution of $\{x_t:t=1,\ldots,T\}$}
\label{subsec:data_dist}

Note that
$[x_1\mid x_0]\sim N(\bh(x_0)'\bbeta_f,\sigma^2_f+\sigma^2_{\e})$, 
but $[x_2\mid x_1,x_0]$$=$$[f(2,x_1)+\e_2\mid x_1,x_0]$$=$$[f(2,f(1,x_0)+\e_1)+\e_2\mid f(1,x_0)+\e_1,x_0]$. Hence, the conditional distribution of 
$[x_t|x_{t-1},x_0]$, for $t\geq 2$, need not be straightforward	to get hold of.
In this regard, we adopt the procedure introduced by \ctn{Bhattacharya07} which has also been successfully exploited in the nonparametric state-space modeling approach
of \ctn{Ghosh14}, to deal with this problem. 
The key idea is to conceptually simulate the entire function $f$ modeled by the GP, and use the simulated process as a look-up table to
obtain the conditional distributions of $\{x_t:~t\geq 2\}$. 


\subsubsection{The key concept}
\label{subsubsec:key}
For simplicity of illustration, let $x_t=f(x^*_{t,t-1})$. 
Now consider a table with the first column $z\in\mathbb R^+\times\mathbb R$ and the second column $f(z)$.
Existence of this table hinges on the implicit assumption that the entire process $f(\cdot)$ is available. 
Given this table, conditional on $x^*_{t,t-1}$ (equivalently, conditional on $x_{t-1}$), $x_t=f(x^*_{t,t-1})$ can be obtained 
by looking-up the input $x^*_{t,t-1}$ from the first column of the table and 
getting hold of the corresponding output value $f(x^*_{t,t-1})$, located in the second column of the table. 
Thus, we refer to such a hypothetical table as a ``look-up table". 
In practice, we can construct a look-up table by simulating a realization of the GP $f$ 
on a fine enough grid of inputs. Given this look-up table realization, simulation from the conditional distribution of $f(x^*_{t,t-1})$, fixing $x^*_{t,t-1}$ as known, 
will approximate $x_t$ as accurately as we desire by making the grid as fine as required, thanks to the well-known interpolation property
of GPs. Formalization of this key concept leads to the following detailed steps. 



\subsubsection{Auxiliary variables for emulating the look-up table}
\label{subsec:auxiliary}

Note that given $x_0$ we can simulate $x_1=f(x^*_{1,0})\sim N(\bh(x^*_{1,0})'\bbeta_f,\sigma^2_f)$, the marginal
distribution of the GP prior. 
To simulate the rest of the dynamic sequence, we first need to generate the
rest of the process $\{f(x^*):~x^*\neq x^*_{1,0}\}$ for the look-up table approach.

In practice, it is not possible to have a simulation of this entire set $\{f(x^*):~x^*\neq x^*_{1,0}\}$. 
We only have available a set of grid points $\bG_n=\{z^*_1,\ldots,z^*_n\}$ where $z^*_i\in\mathbb R^+\times\mathbb R$ for $i=1,\ldots,n$, obtained, perhaps, 
by Latin hypercube sampling 
(see, for example, \ctn{Santner03})
and a corresponding
simulation of $f$, given by $\bD^*_n=\{f(z^*_1),\ldots,f(z^*_n)\}$, the latter having a joint
multivariate normal distribution with mean 
\begin{equation}
E\left[\bD^*_n\mid\btheta_f\right]=\bH_{D^*_n}\bbeta_f\label{eq:mean1}
\end{equation}
and covariance matrix 
\begin{equation}
V\left[\bD^*_n\mid\btheta_f\right]=\sigma^2_f\bA_{f,D^*_n},
\label{eq:var1}
\end{equation}
where
$\bH^{\prime}_{D^*_n}$=$[\bh(z_1),\ldots,\bh(z_n)]$ and $\bA_{f,D^*_n}$ is a 
correlation matrix with the $(i,j)$-th element $c_f(z^*_i,z^*_j)$.

Given $(x_0,f(x^*_{1,0}))$, we simulate $\bD^*_n$ from $[\bD^*_n\mid\btheta_f, f(x^*_{1,0}),x_{0}]$.
Note that the conditional $[\bD^*_n\mid f(x^*_{1,0}),x^*_{1,0}]$
has an $n$-variate normal distribution with mean vector 
\begin{equation}
E[\bD^*_n\mid \btheta_f, f(x^*_{1,0}),x_{0}]=\bmu_{g,D^*_n}=\bH_{D^*_n}\bbeta_f+\bs_{f,D^*_n}(x^*_{1,0})(f(x^*_{1,0})-\bh(x^*_{1,0})'\bbeta_f) \label{eq:mean2}
\end{equation}
and covariance matrix
\begin{equation}
V[\bD^*_n\mid \btheta_f, f(x^*_{1,0}),x_{0}]=\sigma^2_f\bSigma_{f,D^*_n},
\label{eq:var2}
\end{equation}
where
$\bs_{f,D^*_n}(\cdot)=\left(c_f(\cdot,z^*_{1}),\ldots,c_f(\cdot,z^*_{n})\right)'$ and 
\begin{equation}
\bSigma_{f,D^*_n}=\bA_{f,D^*_n}-\bs_{f,D^*_n}(x^*_{1,0})\bs_{f,D^*_n}(x^*_{1,0})'.
\label{eq:Sigma_g_Dz}
\end{equation}


\subsubsection{Distribution of $x_t$ given $\bD^*_n$}
\label{subsubsec:dist_x_t}

Let us now deal with the conditional distribution $[x_t=f(x^*_{t,t-1})\mid \bD^*_n,x_{t-1},x_{t-2},\ldots,x_1]$. 
Since the look-up table idea supports conditional independence, that is, given a simulation of the entire
random function $f$, $x_t$ depends only upon $x_{t-1}$ via $x_t=f(x^*_{t,t-1})$, 
it is sufficient to obtain the conditional distribution of 
$[f(x^*_{t,t-1})\mid \bD^*_n,x_{t-1}]$; see \ctn{Bhattacharya07} and \ctn{Ghosh14} for detailed arguments. 
This distribution is of course normal with mean
\begin{equation}
\mu_t=\bh(x^*_{t,t-1})^\prime \bbeta_f+\bs_{g,D^*_n}(x^*_{t,t-1})^\prime \bA_{f,D^*_n}^{-1}(\bD^*_n-\bH_{D^*_n}\bbeta_f)
\label{eq:cond_mean_xt}
\end{equation}
and variance 
\begin{equation} \sigma_t^2=\sigma^2_f\left\{1-\bs_{f,D^*_n}(x^*_{t,t-1})^\prime \bA_{f,D^*_n}^{-1}\bs_{f,D^*_n}(x^*_{t,t-1})\right\}.
\label{eq:cond_var_xt}
\end{equation}
%
%
For mathematical theory on the accuracy of the Markov approximation of the distributions of $x_t$ given $\bD^*_n$, see \ctn{Ghosh14}.

\subsubsection{Summary of the look-up table procedure}
\label{subsubsec:lookup_summary}

The look-up table idea involves the following steps, given that $x_0$ is known:
\begin{itemize}
\item[(1)] Draw $x_1=f(x^*_{1,0})\sim N(\bh(x^*_{1,0})'\bbeta_f,\sigma^2_f)$.
\item[(2)] Given $x_0$, and $x_1=f(x^*_{1,0})$, draw $\bD^*_n\sim [\bD^*_n\mid\btheta_f, f(x^*_{1,0}),x_{0}]$.
\item[(3)] For $t=2,3,\ldots$, draw $x_t\sim [x_t=f(x^*_{t,t-1})\mid\btheta_f,\bD^*_n,x_{t-1}]$.
\end{itemize}

\subsubsection{Joint distribution of $\{x_1,\ldots,x_{T},\bD^*_n\}$} 
\label{subsubsec:lookup_prior_conditionals}
So far we have discussed the situations where $\e_t=0$, but our actual model (\ref{eq:evo3}) consists of non-zero $\e_t$ which are normally distributed with mean zero
and variance $\sigma^2_{\e}$. In such case, once $\bG_n$ and $\bD^*_n$ are available, we write down the joint distribution
of $\{x_1,\ldots,x_T,\bD^*_n\}$ conditional on the other parameters as
\begin{align}
[x_1,\ldots,x_{T},\bD^*_n\mid\btheta_f,\sigma^2_{\e}]
&=[x_1=f(x^*_{1,0})+\e_1\mid x_0,\sigma^2_{\e}]
[\bD^*_n\mid\btheta_f]\notag\\
	&\qquad\times\prod_{t=1}^{T-1} [x_{t+1}=f(x^*_{t+1,t})+\e_{t+1}\mid \bD^*_n,x_{t},\btheta_f,\sigma^2_{\e}].
\label{eq:x_joint}
\end{align}
In (\ref{eq:x_joint}), 
$[x_1=f(x^*_{1,0})+\e_1\mid x^*_{1,0},\sigma^2_{\e}]$$\sim N(\bh(x^*_{1,0})'\bbeta_f,\sigma^2_f+\sigma^2_{\e})$
and the distribution of $\bD^*_n$ is multivariate normal with mean and variance given by (\ref{eq:mean1}) 
and (\ref{eq:var1}).
The conditional distribution $[x_{t+1}=f(x^*_{t+1,t})+\e_{t+1}\mid \bD^*_n,x_t,\btheta_f,\sigma^2_{\e}]$
is normal with mean
\begin{equation}
\mu_{x_t}=\bh(x^*_{t+1,t})^\prime \bbeta_f+\bs_{f,D^*_n}(x^*_{t+1,t})^\prime \bA_{f,D^*_n}^{-1}(\bD^*_n-\bH_{D^*_n}\bbeta_f)
\label{eq:cond_mean_xt1}
\end{equation}
and variance 
\begin{equation} 
\sigma^2_{x_t}=\sigma^2_{\e}+\sigma^2_f\left\{1-\bs_{f,D^*_n}(x^*_{t+1,t})^\prime \bA_{f,D^*_n}^{-1}\bs_{f,D^*_n}(x^*_{t+1,t})\right\}.
\label{eq:cond_var_xt1}
\end{equation}
Observe that in this case even if $x^*_{t+1,t}\in\bG_n$, due to the presence of the additive error term 
$\e_{t+1}$, 
the conditional variance of $x_{t+1}$ is non-zero, equalling $\sigma^2_{x_t}=\sigma^2_{\e}$,
the error variance.

\subsubsection{Non-Markovian dependence structure of $\{x_1,\ldots,x_{T}\}$}
\label{subsubsec:non_Markovian}

Note that although conditionally on $\bD^*_n$ the variables $x_t$ have a Markovian structure, if $\bD^*_n$ is integrated out from (\ref{eq:x_joint}), then 
the marginalized distribution of $\{x_1,\ldots,x_{T}\}$ is non-Markovian.
In fact, the marginalized conditional distribution of $x_{t+1}$ depends upon $\{x_k:~k<{t+1}\}$; (see also \ctn{Bhattacharya07} and \ctn{Ghosh14}).
An important issue discussed in this context by \ctn{Bhattacharya07} and \ctn{Ghosh14} is that this strong marginalized dependence structure is the root of
all numerical instabilities associated with the model implementation. 
Essentially, by sample path continuity of the underlying GP, $x_0,x_1,\ldots,x_{t}$ will be often close to each other with high probability,
particularly if $\sigma^2_f$ and $\sigma^2_{\e}$ are small. This would render the relevant correlation matrix almost singular, which would be difficult to
invert. Since such inversions are required for every $t\in\{2,\ldots,T\}$ and at every iteration of any Monte Carlo simulation method, 
progress would be almost impossible when $T$ is relatively large, with increasing computational cost for each $t$, further aggravating the situation.

In contrast, if $\bD^*_n$ is retained, it is required to deal with $[x_{t+1}|\bD^*_n,x_t,\btheta_f,\sigma^2_f]$, which requires computation of 
$\bA^{-1}_{f,D^*_n}$ only once, for all $t\geq 2$, for any MCMC iteration. 
Moreover, invertibility of $\bA_{f,D^*_n}$, given $r_{1,f}$ and $r_{2.f}$ is largely controlled by the user, since
the $(i,j)$-th element of $\bA_{f,D^*_n}$ is of the form $c_f(z^*_i,z^*_j)$, where $z^*_1,\ldots,z^*_n$ are fixed constants, which can be
judiciously chosen by the user. 
Thus, retaining $\bD^*_n$ significantly mitigates the issues of numerical instability and computational burden
inherent in the marginalized distribution of $\{x_1,\ldots,x_{T}\}$. It is hence no wonder that retaining $\bD^*_n$ in the model is the only sensible decision.

\section{Prior distributions for $\btheta_f$ and $\sigma^2_\e$}
\label{sec:prior_univariate}
We assume the following forms of the prior distributions:
\begin{align}
[\bbeta_f]&\sim N_3\left(\bbeta_{f,0},\bSigma_{\beta_{f,0}}\right);
\label{eq:betaf_prior}\\
[\sigma^2_f]&\propto\left(\sigma^2_f\right)^{-\left(\frac{\alpha_f+2}{2}\right)}\exp\left\{-\frac{\gamma_f}{2\sigma^2_f}\right\};
~\alpha_f,\gamma_f>0;
\label{eq:sigma_f_prior}\\
[\sigma^2_{\epsilon}]&\propto\left(\sigma^2_{\epsilon}\right)^{-\left(\frac{\alpha_{\epsilon}+2}{2}\right)}\exp\left\{-\frac{\gamma_{\epsilon}}{2\sigma^2_{\epsilon}}\right\};
~\alpha_{\epsilon},\gamma_{\epsilon}>0;
\label{eq:sigma_epsilon_prior}\\
[\log(r_{i,f})]&\sim N\left(\mu_{r_{i,f}},\sigma^2_{r_{i,f}}\right);\ \ \mbox{for}\ \ i=1,2.
\label{eq:rf_prior}
\end{align}
%
All the prior parameters are assumed to be known. Now we discuss our approach to
selecting the prior parameters for the applications of our Bayesian model.

As per (\ref{eq:betaf_prior}), we set the prior of $\bbeta_f$ to be trivariate normal with the identity matrix as the variance, 
that is, we set $\bSigma_{\beta_{f,0}}=\bI_3$, where
$\bI_3$ is the $3$-dimensional identity matrix. This choice turned out to be appropriate as larger variances in the diagonal caused the posterior time series to
explode with increasing time.
For the mean $\bbeta_{f,0}$, except the first component associated with the intercept, we set the rest of the components to zero. We set the first component
of $\bbeta_{f,0}$ to be the mean of the underlying logarithm of the time series data to be modeled, after thinning by $5$ observations. 
This ensures that the intercept corresponds to the overall mean of the log time series. 

For the choice of the parameters of the priors of $\sigma^2_f$ and $\sigma^2_{\epsilon}$
we first note that the mean is of the form $\gamma/(\alpha-2)$ and the variance is of the form
$2\gamma^2/\{(\alpha-2)^2(\alpha-4)\}$. Thus, if we set $\gamma/(\alpha-2)=a$, then the variance becomes
$2a^2/(\alpha-4)$. Here we set $a=\hat\sigma^2/2$ for both $\sigma^2_f$ and $\sigma^2_{\epsilon}$, 
where $\hat\sigma^2$ is the variance of the underlying log time series obtained after thinning by $5$ observations. 
Again, this strategy is to ensure that the expected variability matches the data variability.
For each of these priors we set $\alpha=4.01$, so that the 
variance is of the form $200a^2$. 

In order to choose the parameters of the log-normal priors of the smoothness parameters $r_{1f}$ and $r_{2f}$, we
set the mean of the log-normal prior with parameters $\mu$ and $\sigma^2$,
given by $\exp(\mu+\sigma^2/2)$, to 1. This yields $\mu= -\sigma^2/2$. Since the variance
of this log-normal prior is given by $(\exp(\sigma^2)-1)\exp(2\mu+\sigma^2)$, the relation
$\mu=-\sigma^2/2$ implies that the variance is $\exp(\sigma^2)-1=\exp(-2\mu)-1$. We set
$\sigma^2=1$, so that $\mu=-0.5$. This implies that the mean is 1 and the variance is approximately 2,
for the priors of each smoothness parameter $r_{i,f}$; $i=1,2$.
This prevents the smoothness parameters from being too large or too small. Indeed, if the smoothness parameters are too large then 
$c_f(z^*_1,z^*_2)\approx 0$ for $z_1\neq z_2$, so that the correlation matrix is rendered almost the identity matrix. On the other hand, if the smoothness
parameters are close to zero, then $c_f(z^*_1,z^*_2)\approx 1$ for $z^*_1,z^*_2$, making the correlation matrix almost singular. Both these undesirable situations
are ruled out by our prior choice.

\section{Posterior distributions of current and future time series in our dynamic GP approach}
\label{sec:current_future}

\subsection{Posterior of current given the future}
\label{subsec:current_given_future}
Let us assume that for any given GCM, the logarithms of the future mean global temperatures $\{x_t:~t=T_0+1,\ldots,T\}$ are given, where $1\leq T_0\leq T-1$. 
In our case, the times $\{0,\ldots,T_0\}$ correspond to the current years $\{1850,\ldots,2016\}$ and the times $\{T_0+1,\ldots,T\}$
correspond to the future years $\{2017,\ldots,2099\}$.
Then assuming that $x_0$ is known, we can obtain the posterior distribution
of the logarithms of the current mean global temperatures $\{x_t:~t=1,\ldots,T_0\}$ as follows:
\begin{align}
	&[x_1,\ldots,x_{T_0}|x_{T_0+1},\ldots,x_{T}]\notag\\
	&=\int[x_1,\ldots,x_{T_0}|\bD^*_n,x_{T_0+1},\ldots,x_{T},\btheta_f,\sigma^2_{\e}]d[\bD^*_n,\btheta_f,\sigma^2_{\e}|x_{T_0+1},\ldots,x_{T}]\notag\\
	&\approx\int [x_1,\ldots,x_{T_0}|\bD^*_n,\btheta_f,\sigma^2_{\e}]d[\bD^*_n,\btheta_f,\sigma^2_{\e}|x_{T_0+1},\ldots,x_{T}].
	\label{eq:curr_post}
\end{align}
The second approximate equality follows from the first equality since given $\bD^*_n$, $\{x_1,\ldots,x_{T_0}\}$ are conditionally approximately 
independent of $\{x_{T_0+1},\ldots,x_{T}\}$; ``approximate" because $x_{T_0}$ and $x_{T_0+1}$ are not independent, even when $\bD^*_n$ is conditioned upon. 
This approximate conditional independence ensures
$[x_1,\ldots,x_{T_0}|\bD^*_n,x_{T_0+1},\ldots,x_{T},\btheta_f,\sigma^2_{\e}]\approx[x_1,\ldots,x_{T_0}|\bD^*_n,\btheta_f,\sigma^2_{\e}]$.
In our practical applications, however, we shall replace this approximate equality with equality. For well-chosen fine enough grid $\bG_n$ this is not at all a serious issue. 

Hence, if we can have simulations from the posterior $[\bD^*_n,\btheta_f,\sigma^2_{\e}|x_{T_0+1},\ldots,x_{T}]$, then we can easily simulate from
(\ref{eq:curr_post}) using
\begin{equation*}
[x_1,\ldots,x_{T_0}|\bD^*_n,\btheta_f,\sigma^2_{\e}]=\prod_{t=0}^{T_0-1} [x_{t+1}=f(x^*_{t+1,t})+\e_{t+1}\mid \bD^*_n,x_{t},\btheta_f,\sigma^2_{\e}],
\end{equation*}
where $[x_{t+1}=f(x^*_{t+1,t})+\e_{t+1}\mid \bD^*_n,x_{t},\btheta_f,\sigma^2_{\e}]$ is normally distributed with mean and variance given by 
(\ref{eq:cond_mean_xt1}) and (\ref{eq:cond_var_xt1}), respectively, for $t=0,1,\ldots,T_0-1$.

To obtain samples from the posterior 
\begin{align}
	&[\bD^*_n,\btheta_f,\sigma^2_{\e}|x_{T_0+1},\ldots,x_{T}]\notag\\
	&\propto[\bD^*_n|\btheta_f][\btheta_f][\sigma^2_{\e}][x_{T_0+1},\ldots,x_{T}|\bD^*_n,\btheta_f,\sigma^2_{\e}]\notag\\
	&=[\bD^*_n|\btheta_f][\btheta_f][\sigma^2_{\e}]\prod_{t=T_0}^{T-1} [x_{t+1}=f(x^*_{t+1,t})+\e_{t+1}\mid \bD^*_n,x_{t},\btheta_f,\sigma^2_{\e}],\notag
\end{align}
we resort to Markov Chain Monte Carlo (MCMC) where
we sample $\bbeta_f$ and $\bD^*_n$ from their respective full conditional distributions and the remaining parameters $\{r_{1f},r_{2f},\sigma^2_f,\sigma^2_\e\}$ using
Transformation based Markov Chain Monte Carlo (TMCMC) introduced by \ctn{Dutta13}. In particular, we use the additive transformation, with judicious choice
of the tuning constants.

\subsection{Posterior of future given the current}
\label{subsec:future_given_current}
Now, given $\{x_1,\ldots,x_{T_0}\}$, which may be interpreted as the current observed log global mean temperatures, we can obtain the posterior distribution 
of the future log global mean temperatures $\{x_{T_0+1},\ldots,x_{T}\}$ in a similar manner. That is,
\begin{align}
	&[x_{T_0+1},\ldots,x_{T}|x_1,\ldots,x_{T_0}]\notag\\
	&=\int[x_{T_0+1},\ldots,x_{T}|\bD^*_n,x_1,\ldots,x_{T_0},\btheta_f,\sigma^2_{\e}]d[\bD^*_n,\btheta_f,\sigma^2_{\e}|x_1,\ldots,x_{T_0}]\notag\\
	&=\int [x_{T_0+1},\ldots,x_{T}|\bD^*_n,x_{T_0},\btheta_f,\sigma^2_{\e}]d[\bD^*_n,\btheta_f,\sigma^2_{\e}|x_1,\ldots,x_{T_0}].
	\label{eq:future_post}
\end{align}
Thus, after obtaining MCMC samples from 
\begin{align}
	&[\bD^*_n,\btheta_f,\sigma^2_{\e}|x_1,\ldots,x_{T_0}]\notag\\
	& \propto[\bD^*_n|\btheta_f][\btheta_f][\sigma^2_{\e}][x_{1},\ldots,x_{T_0}|\bD^*_n,\btheta_f,\sigma^2_{\e}]\notag\\
	&=[\bD^*_n|\btheta_f][\btheta_f][\sigma^2_{\e}]\prod_{t=0}^{T_0-1} [x_{t+1}=f(x^*_{t+1,t})+\e_{t+1}\mid \bD^*_n,x_{t},\btheta_f,\sigma^2_{\e}]\notag
\end{align}
using the same techniques as for 
$[\bD^*_n,\btheta_f,\sigma^2_{\e}|x_{T_0+1},\ldots,x_{T}]$, we simulate from 
\begin{equation*}
	[x_{T_0+1},\ldots,x_{T}|\bD^*_n,x_{T_0},\btheta_f,\sigma^2_{\e}]=\prod_{t=T_0}^{T-1} [x_{t+1}=f(x^*_{t+1,t})+\e_{t+1}\mid \bD^*_n,x_{t},\btheta_f,\sigma^2_{\e}],
\end{equation*}
where $[x_{t+1}=f(x^*_{t+1,t})+\e_{t+1}\mid \bD^*_n,x_{t},\btheta_f,\sigma^2_{\e}]$ is normally distributed with mean and variance given by 
(\ref{eq:cond_mean_xt1}) and (\ref{eq:cond_var_xt1}), respectively, for $t=T_0,1,\ldots,T-1$.
This yields simulations from (\ref{eq:future_post}).

\section{A Bayesian multiple testing framework for GCM selection in any given climate scenario}
\label{sec:mult_inv}
Given any climate scenario, let us consider GCMs $\mathcal M_k$; $k=1,\ldots,K$, from among which the best model 
needs to be selected. For our purpose, we adopt and extend the novel Bayesian multiple testing procedure for model selection introduced by \ctn{Chatterjee20b} 
that respects the inverse regression perspective of the models, in coherence with the forward aspect.  

It is important to mention that in statistics, model selection pertains to choosing the best model from among a set of models that attempt to fit a single dataset.
However, in our present GCM case, there are $K$ datasets generated by $K$ GCMs in a given climate scenario. Our strategy will be to combine the $K$ datasets
into a single dataset by taking averages over the $K$ GCMs for each time point, and then to invoke our GP
based dynamics for the averaged time series, where the hyperparameters of the model are fixed using the mean and variance of the
original GCM-specific simulated time series. This yields $K$ different GP based models for the averaged time series, inheriting the main characteristics
of the GCM-specific time series. The design of our Bayesian multiple testing procedure ensures that 
the GP models will be compared with respect to their abilities to fit the averaged simulated future global
temperature data in the forward sense, as well as their abilities to capture the HadCRUT4 data given 
the averaged GCM-simulated future global temperature data, in the inverse sense. Details follow.

%

Let us denote the logarithms of the observed current global mean temperatures (the HadCRUT4 data) by $\left\{x^{(0)}_t:~t=1,\ldots T_0\right\}$.
For GCM $\mathcal M_k$, let $\left\{x^{(k)}_t:~t=0,1,\ldots\right\}$ denote the logarithms of its simulated global mean temperature time series, for $k=1,\ldots,K$. 
For $t=0,1,\ldots$, let $\bar x_t=K^{-1}\sum_{k=1}^Kx^{(k)}_t$, and let this averaged time series  $\{\bar x_t:~t=0,1,\ldots\}$ be also modeled by the 
GP emulation procedure given by (\ref{eq:x_t_dist}), (\ref{eq:gp_f}) and (\ref{eq:theta_prior}), with parameters denoted by 
$\btheta^{(k)}=(\btheta^{(k)}_f,{\sigma^{(k)}_\e}^2)$. The rationale behind this modeling strategy is simple: if the functional forms $f(\cdot)$ associated with
the individual time series $\left\{x^{(k)}_t:~t=0,1,\ldots\right\}$ are unknown, then the functional form driving the dynamics of their average must also be unknown,
which is again best modeled by a GP.
In this regard, let $[\bar x_{T_0+1},\ldots,\bar x_T|\btheta^{(k)},\mathcal M_k]$ denote the density of
the logarithms of the future global mean temperatures, averaged over all the models in the climate scenario 
under GP emulation model $\mathcal M_k$, with its associated parameters $\btheta^{(k)}$.  


We combine the competing models in the following mixture form:
\begin{equation}
	[\bar x_{T_0+1},\ldots,\bar x_T|\btheta]
	=\sum_{k=1}^Kp_k[\bar x_{T_0+1},\ldots,\bar x_T|\btheta^{(k)},\mathcal M_k],
	\label{eq:mix1}
\end{equation}
where  $\btheta=(\btheta^{(1)},\ldots,\btheta^{(K)})$, $0\leq p_k\leq 1$, for $k=1,\ldots,K$ and $\sum_{k=1}^Kp_k=1$.
Letting $\zeta$ denote the allocation variable (model index), with $P(\zeta=k)=p_k$, note that 
$[\bar x_{T_0+1},\ldots,\bar x_T|x^{(0)}_1,\ldots,x^{(0)}_{T_0},\btheta,\zeta=k]=[\bar x_{T_0+1},\ldots,\bar x_T|x^{(0)}_1,\ldots,x^{(0)}_{T_0},\btheta^{(k)},\mathcal M_k]$.
We consider the Dirichlet prior for $(p_1,\ldots,p_K)$ with parameters $(\alpha_1,\ldots,\alpha_K)$, where $\alpha_k>0$, for $k=1,\ldots,K$. 
In our problem, we shall set $\alpha_k=1$, for all $k=1,\ldots,K$, for all the climate scenarios. Thus, the prior is uniform over the simplex, indicating no preference
for any specific GCM {\it a priori}.
The priors for the parameters $\btheta^{(k)}$ remain the same as described in Section \ref{sec:prior_univariate}. Since for different $k$ the prior
depends upon the mean and variance of the underlying entire $k$-th GCM-simulated time series, the priors are all very distinct from one another. In fact,
the distinctions among the priors induces distinctions among the competing Bayesian models, since otherwise all of them have the same dynamic structure driven
by GPs, started at the same known initial value $x_0$.

We let $\left\{\bar x_t:~t=1,\ldots T_0\right\}$ stand for the random quantities corresponding to $\left\{x^{(0)}_t:~t=1,\ldots T_0\right\}$, whose posterior
distribution will be of interest to us. In particular, it is of interest in evaluating how well this posterior captures the observed current log global mean
temperatures, which we shall formalize in our multiple testing procedure. Towards this goal, 
for any $T_0$-dimensional vector $\bv_{T_0}=(v_1,\ldots,v_{T_0})$, and for some $c>0$, let us define the following discrepancy measures in the spirit of 
\ctn{Chatterjee20b}:
\begin{align}
	S^{(k)}_1(\bv_{T_0}) &=\frac{1}{T_0} \sum_{t=1}^{T_0}\frac{\left|v_t - M(\bar x_t|\bar x_{T_0+1},\ldots,\bar x_T,\mathcal M_k)\right|}
	{\sqrt{Var(\bar x_t|\bar x_{T_0+1},\ldots,\bar x_T,\mathcal M_k)+c}},\label{eq:T1}
\end{align}
where $M(\bar x_t|\bar x_{T_0+1},\ldots,\bar x_T,\mathcal M_k)$ stands for the posterior mode of $[\bar x_t|\bar x_{T_0+1},\ldots,\bar x_T,\mathcal M_k]$.
Similarly, let
\begin{align}
	S^{(k)}_2(\bv_{T_0}) &=\frac{1}{T_0} \sum_{t=1}^{T_0}\frac{\left(v_t - M(\bar x_t|\bar x_{T_0+1},\ldots,\bar x_T,\mathcal M_k)\right)^2}
	{Var(\bar x_t|\bar x_{T_0+1},\ldots,\bar x_T,\mathcal M_k)+c}.\label{eq:T2}
\end{align}
In our examples, we set $c=0.01$.
Various other measures of discrepancy can be defined (see \ctn{Bhattacharya13} for a discussion on such discrepancy measures; see also \ctn{Sabya13}), 
but for brevity we focus on these two measures in this paper. 

Importantly, using discrepancy measures \ctn{Bhattacharya13} introduced a novel Bayesian decision-theoretic methodology for  
Bayesian model assessment in inverse regression problems, which we shall adopt to assess goodness-of-fit of the best GCMs with respect to fitting
the HadCRUT4 data, conditioned on the future GCM projections and our Bayesian dynamic GP emulation strategy.

With $\bar\bx_{T_0}=(\bar x_1,\ldots,\bar x_{T_0})$ and $\bx^{(0)}_{T_0}=\left(x^{(0)}_1,\ldots,x^{(0)}_{T_0}\right)$, for a given discrepancy measure $S^{(k)}$, 
let $[\bar\ell_{k},\bar u_{k}]$ denote the $100(1-\alpha)\%$ credible interval for the posterior distribution of 
$S^{(k)}(\bar\bx_{T_0})$ for any desired $\alpha\in (0,1)$; in our application, we set $\alpha=0.05$. 
Following the recommendation of \ctn{Chatterjee20b} for practical purposes (see (8.1) and (8.2) of
Section 8 of their article) we now define the appropriate multiple hypotheses that we shall
test for our Bayesian model selection purpose. 
%
For $k=1,\ldots,K$, 
\begin{equation}
	H_{0k}:\zeta=k,
	S^{(k)}\left(\bar\bx_{T_0}\right)-S^{(k)}\left(\bx^{(0)}_{T_0}\right)\in [\bar\ell_{k},\bar u_{k}]
	\label{eq:H0_1}
\end{equation}
versus
\begin{align}
	&H_{1k}:\left\{\zeta\neq k\right\}
	\bigcup\left\{\zeta=k,
	S^{(k)}\left(\bar\bx_{T_0}\right)-S^{(k)}\left(\bx^{(0)}_{T_0}\right)\in [\bar\ell_{k},\bar u_{k}]^c\right\},
	\label{eq:H1_1}
\end{align}
where, for any set $A$, $A^c$ stands for its complement.

The hypotheses are so designed that the best model is chosen on the basis of both forward and inverse perspectives.
To elucidate, note that to select the best model we first need to choose a model 
$[\bar x_{T_0+1},\ldots,\bar x_T|\btheta^{(k)},\mathcal M_k]$ 
indexed by $\zeta= k$ 
which has high marginal posterior probability. This reflects the forward perspective of the model selection problem. 
Indeed, the posterior probability of $\{\zeta= k\}$ is proportional to its corresponding marginal density 
$[\bar x_{T_0+1},\ldots,\bar x_T|\mathcal M_k]=\int [\bar x_{T_0+1},\ldots,\bar x_T|\btheta^{(k)},\mathcal M_k]d[\btheta^{(k)}]$
(see (\ref{eq:marginal_density}) for details). 
This marginal density has interpretation in the forward sense only since it is not associated with the posterior distribution 
$[\bar x_1,\ldots, \bar x_{T_0}|\bar x_{T_0+1},\ldots,\bar x_T,\mathcal M_k]$,
the latter to be interpreted as the inverse aspect of the problem. 

The inverse sense in our multiple testing formalization is made explicit in the following way.
In addition to selecting $\zeta= k$ with high marginal posterior probability, we demand that for such model 
\begin{equation}
	S^{(k)}\left(\bar\bx_{T_0}\right)-S^{(k)}\left(\bx^{(0)}_{T_0}\right)\in [\bar\ell_{k},\bar u_{k}]
	\label{eq:inv_test1}
\end{equation}
is also satisfied. Roughly, this condition demands that for $\mathcal M_k$ to qualify as a good inverse regression model, 
the observed discrepancy measure $S^{(k)}\left(\bx^{(0)}_{T_0}\right)$ must be included in the
desired credible interval of the reference discrepancy measure $S^{(k)}\left(\bar\bx_{T_0}\right)$.
This reflects the inverse perspective since the reference discrepancy measure explicitly deals with the posterior 
$[\bar x_1,\ldots, \bar x_{T_0}|\bar x_{T_0+1},\ldots,\bar x_T,\mathcal M_k]$
associated with the inverse regression problem. The key idea of the Bayesian goodness-of-fit test devised by \ctn{Bhattacharya13} is based on the same principle. 

Note that our Bayesian multiple hypotheses formulation (\ref{eq:H0_1}) and (\ref{eq:H1_1}) does not involve cross-validation, even though \ctn{Chatterjee20b}
formulated the general Bayesian multiple testing framework for model and variable selection in problems involving covariates using inverse leave-one-out cross-validation
with respect to posteriors associated with the covariates (see also \ctn{Bhattacharya13}). Indeed, as must be evident from the very beginning, 
our current global climate change problem is not the traditional model selection problem. However, our Bayesian multiple testing procedure is based on similar principles
introduced in \ctn{Chatterjee20b}.

\subsection{The Bayesian multiple testing procedure} 
\label{subsec:muller}



Let
\begin{align*}
d_k=&\begin{cases}
1&\text{if the $k$-th null hypothesis is rejected;}\\
0&\text{otherwise;}
\end{cases}\\
r_k=&\begin{cases}
1&\text{if $H_{1k}$ is true;}\\
0&\text{if $H_{0k}$ is true.} 
\end{cases}
\end{align*}
Following \ctn{Chatterjee20b} (see also \ctn{muller04}) and (\ctn{Guindani09}), let us define the true positives as 
\begin{equation}
TP=\sum_{k=1}^Kd_kr_k,
\label{eq:tp}
\end{equation}
the posterior expectation of which is to be maximized subject to controlling the posterior expectation of the error term
\begin{equation}
E=\sum_{k=1}^Kd_k(1-r_k).
\label{eq:e}
\end{equation}
From the above notions it is clear that the optimal decision configuration can be obtained by minimizing the function
\begin{align}
	\xi(\bd)&=-\sum_{k=1}^Kd_kE(r_k|\bar x_{T_0+1},\ldots,\bar x_T)
	+\lambda\sum_{k=1}^Kd_k E\left[ (1-r_k)|\bar x_{T_0+1},\ldots,\bar x_T\right]\notag\\
	&= -(1+\lambda)\sum_{k=1}^Kd_k\left(v_k-\frac{\lambda}{1+\lambda}\right),\notag
\end{align}
with respect to all possible decision configurations of the form $\bd=\{d_1,\ldots,d_K\}$, where
$\lambda>0$,
and
\begin{equation*}
	v_{k}=E(r_k|\bar x_{T_0+1},\ldots,\bar x_T)
	= \left [H_{1k}\big | \bar x_{T_0+1},\ldots,\bar x_T\right ]
\end{equation*}
is the posterior probability of the $k$-th alternative hypothesis. 
Letting $\beta=\lambda/(1+\lambda)$ denote the penalizing constant, one can equivalently maximize
\begin{equation}
	f_{\beta}(\bd)=\sum_{k=1}^K d_k\left(v_k-\beta\right)\label{eq:beta1}
\end{equation}
with respect to $\bd$ and obtain the optimal decision configuration.
In this case, the optimal decision configuration $\widehat\bd=\{\widehat d_1,\ldots,\widehat d_K\}$ is given by the following: for $k=1,\ldots,K$,
\begin{equation}
\widehat d_k=\begin{cases}
	1&\text{if $v_{k}>\beta$;}\\
0&\text{otherwise.} 
\end{cases}
	\label{eq:optimal_decision}
\end{equation}

In our model selection setup, the least value of the penalty $\beta\in (0,1)$ for which the decision configuration $\hat d_{\tilde k}=0$ and
$\hat d_k=1$ for all $k\in\{1,\ldots,K\}\backslash\{\tilde k\}$ is obtained, for some $\tilde k\in\{1,\ldots,K\}$, yields the best model $\mathcal M_{\tilde k}$.
This is because in such a case, $v_{\tilde k}\leq \beta$, even though $\beta$ is reasonably small, suggesting that $H_{0\tilde k}$ has significant posterior probability.
Since $v_{k}>\beta$ for all $k\in\{1,\ldots,K\}\backslash\{\tilde k\}$, the posterior probabilities of $H_{0k}$ for $k\in\{1,\ldots,K\}\backslash\{\tilde k\}$
are less substantial compared to that of $H_{0\tilde k}$. This indicates that $\mathcal M_{\tilde k}$ is the best model among $\mathcal M_k$; $k=1,\ldots,K$.
This key intuition is rigorously formalized in our Bayesian multiple testing procedure detailed in \ctn{Chatterjee20b}.

\subsection{Error measures for our Bayesian multiple testing procedure}
\label{subsec:Bayesian_errors}

%

To discuss appropriate measures of error for our Bayesian multiple testing procedure, first let us define
$\delta(\bd|\bar x_{T_0+1},\ldots,\bar x_T)$ to be the probability of choosing $\bd$ as the optimal decision configuration given data 
$\bar x_{T_0+1},\ldots,\bar x_T$ when a given multiple testing method is employed. 
Also, let $\mathbb D$ be the set of all $K$-dimensional binary vectors, standing for all possible decision configurations.

As suitable posterior measures of Type-I and Type-II errors, \ctn{SanatGhosh08} defined posterior false discovery rate and false non-discovery rate, respectively, 
which we denote as conditional false discovery rate (cFDR) and conditional false non-discovery rate (cFNR). 
The measures, in our current setup, are given by the following:
\begin{align*}
	cFDR
	&= E\left[\sum_{\bd\in\mathbb D}\frac{\sum_{k=1}^Kd_k(1-r_k)}{\sum_{k=1}^Kd_k \vee 1}\delta\left(\bd|\bar x_{T_0+1},\ldots,\bar x_T\right)
	\bigg |\bar x_{T_0+1},\ldots,\bar x_T \right]\\
	&= \sum_{\bd\in\mathbb{D}} 
	\frac{\sum_{k=1}^{K}d_k(1-v_{k})}{\sum_{k=1}^{K}d_k \vee 1}\delta(\bd|\bar x_{T_0+1},\ldots,\bar x_T); \notag\\ 
	cFNR
	&= E\left[\sum_{\bd\in\mathbb D} \frac{\sum_{k=1}^K(1-d_k)r_k} {\sum_{k=1}^K(1-d_k)\vee 1} 
	\delta\left(\bd|\bar x_{T_0+1},\ldots,\bar x_T\right)\bigg |\bar x_{T_0+1},\ldots,\bar x_T\right]\\
	&=\sum_{\bd\in\mathbb D}
	\frac{\sum_{k=1}^{K}(1-d_k)v_{k}}{\sum_{k=1}^{K}(1-d_k)\vee 1}\delta(\bd|\bar x_{T_0+1},\ldots,\bar x_T).\notag\\ 
\end{align*}
Note that since in our multiple testing method the decision rule is non-randomized, 
$\delta(\bd|\bar x_{T_0+1},\ldots,\bar x_T)$ is either 1 or 0 depending on data $\{\bar x_{T_0+1},\ldots,\bar x_T\}$. 

For our Bayesian purpose, following \ctn{Chatterjee20b}, we shall consider the Bayesian measures $cFDR$ and $cFNR$ as Bayesian multiple testing error rates.
These measures are also recommended by \ctn{Chandra19} and \ctn{Chandra20} 
since they are conditioned on the observed data and hence qualify as {\it bona fide} Bayesian measures.

The above error measures also point towards the best model yielded by our multiple testing procedure. Recall from the discussion toward the end of Section \ref{subsec:muller}
that the least value of $\beta\in (0,1)$ such that 
the decision configuration $\hat d_{\tilde k}=0$ and
$\hat d_k=1$ for all $k\in\{1,\ldots,K\}\backslash\{\tilde k\}$ is obtained, for some $\tilde k\in\{1,\ldots,K\}$, yields the best model $\mathcal M_{\tilde k}$.
Now, since cFDR and cFNR are step functions of $\beta$, it is clear that the first jump of the graph of either of the functions cFDR or cFNR corresponds to the same best model.

%

\section{Implementation of the Bayesian multiple testing procedure} 
\label{sec:test_implementation}

\subsection{Parallel computation of $[\bar x_1,\ldots, \bar x_{T_0}|\bar x_{T_0+1},\ldots,\bar x_T,\mathcal M_k]$ for different GCMs and climate scenarios}
\label{subsec:comp_inv_post}
Note that for conducting the Bayesian multiple hypotheses tests, we need to obtain samples from the posteriors
$[\bar x_1,\ldots, \bar x_{T_0}|\bar x_{T_0+1},\ldots,\bar x_T,\mathcal M_k]$, for all $k=1,\ldots,K$, for any given climate scenario.
These are required to evaluate the posterior probabilities of (\ref{eq:inv_test1}), associated with the inverse perspective.

The method of obtaining posterior samples from the above distributions is the same as described in Section \ref{subsec:current_given_future}, with the priors
discussed in Section \ref{sec:prior_univariate}, but we need to select the grid $\bG_n$ appropriately for creating the GP based look-up table.
Note that the input grid $\bG_n$ is a two-dimensional grid, the first component being the time component and the second being the real line. In our case,
we re-label the times $1850-2099$ as $0-249$ and further divide the re-labeled times by $250$ to have them lie in $[0,1]$. We then divide up the interval 
$[0,1]$ into $n=50$ equal sub-intervals and randomly simulate a value from each sub-interval. For the second component of $\bG_n$, gridding the interval $[0,5]$ 
instead of a large interval turned out to be more than adequate for our problem, particularly because we consider the logarithms of the time series 
rather than the actual time series. We divide up the interval
$[0,5]$ into $n=50$ equal sub-intervals and randomly simulate a value from each sub-interval. Thus, we construct $\bG_n$ using component-wise Latin hypercube sampling
with $n=50$. 

For each model $[\bar x_1,\ldots, \bar x_{T_0}|\bar x_{T_0+1},\ldots,\bar x_T,\mathcal M_k]$, $k=1,\ldots,K$, we obtain $60,000$ samples of 
$\{\bar x_1,\ldots, \bar x_{T_0}\}$ following the method described in Section \ref{subsec:current_given_future}, discarding the first $10,000$ as burn-in.
Now recall that the climate scenarios A1B, A2, B1 and Commitment consist of $21$, $17$, $21$ and $16$ GCMs, respectively. That is, in all, there are
$75$ posteriors of the form $[\bar x_1,\ldots, \bar x_{T_0}|\bar x_{T_0+1},\ldots,\bar x_T,\mathcal M_k]$, and from each of them $60,000$ realizations are to be simulated. 
This is an infeasible task if the models
are implemented separately. However, we implement our code, written in C in accordance with the MPI protocol, 
in a parallel architecture associated with a VMWare consisting of $100$ cores, running at $2.80$ GHz speed, and having $1$ TB memory. 
Specifically, we parallelize our computation by splitting $75$ model implementations into $75$ separate cores of our VMWare.
The entire exercise takes less than an hour in our parallel implementation. 

\subsection{Obtaining the posterior model probabilities using Gibbs sampling}
\label{subsec:posterior_model_probs}

Recall that our multiple testing approach also requires computation of the posterior model probabilities 
$\left[\zeta=k|\bar x_{T_0+1},\ldots,\bar x_T\right]$. We propose (see also \ctn{Chatterjee20b}) Gibbs sampling for simulation-based
computations of these probabilities, by sampling from the full conditionals $\left[\zeta|\bar x_{T_0+1},\ldots,\bar x_T,p_1,\ldots,p_K\right]$
and $\left[p_1,\ldots,p_K|\bar x_{T_0+1},\ldots,\bar x_T,\zeta\right]$ successively.

Note that given $\zeta$, the posterior distribution of $(p_1,\ldots,p_K)$ is again a Dirichlet
distribution with parameters $(\alpha_1+I(\zeta=1),\ldots,\alpha_K+I(\zeta=K))$. In other words, since $\alpha_k=1$ for $k=1,\ldots,K$, we have
\begin{equation}
	\left[p_1,\ldots,p_K|\bar x_{T_0+1},\ldots,\bar x_T,\zeta\right]\equiv Dirichlet(1+I(\zeta=1),\ldots,1+I(\zeta=K)).
	\label{eq:dir}
\end{equation}
Given $(p_1,\ldots,p_K)$, the posterior distribution of $\zeta$ is given by 
\begin{equation}
	\left[\zeta=k|\bar x_{T_0+1},\ldots,\bar x_T,p_1,\ldots,p_K\right]
	= \frac{p_k[\bar x_{T_0+1},\ldots,\bar x_T|\mathcal M_k]}{\sum_{\ell=1}^Kp_\ell [\bar x_{T_0+1},\ldots,\bar x_T|\mathcal M_\ell]};~k=1,\ldots,K,
	\label{eq:post_zeta3}
\end{equation}
where for any $k$, letting ${\bD^*_n}^{(k)}$ denote the look-up table associated with model $\mathcal M_k$,
\begin{align}
	&[\bar x_{T_0+1},\ldots,\bar x_T|\mathcal M_k]\notag\\ 
	&=\int [\bar x_{T_0+1},\ldots,\bar x_T|{\bD^*_n}^{(k)},\btheta^{(k)},\mathcal M_k]d[\btheta^{(k)}]d[{\bD^*_n}^{(k)}]\notag\\
	&=\int \prod_{t=T_0}^{T-1} [\bar x_{t+1}=f(t+1,\bar x_t)+\e_{t+1}\mid {\bD^*_n}^{(k)},\bar x_{t},\btheta^{(k)},\mathcal M_k]d[\btheta^{(k)}]d[{\bD^*_n}^{(k)}]\notag\\
	&\approx\frac{1}{N}\sum_{i=1}^N\prod_{t=T_0}^{T-1} [\bar x_{t+1}=f(t+1,\bar x_t)+\e_{t+1}\mid {\bD^*_n}^{(k)}_i,\bar x_{t},\btheta^{(k)}_i,\mathcal M_k],
	\label{eq:marginal_density}
\end{align}
where $\left\{\left({\bD^*_n}^{(k)}_i,\btheta^{(k)}_i\right):~i=1,\ldots,N\right\}$, for sufficiently large $N$, 
is a set of simulations from the prior distributions of $\btheta^{(k)}$ and the distribution of the look-up table ${\bD^*_n}^{(k)}$.

In practice, rather than simulating from the priors, we simulate $\left\{\left({\bD^*_n}^{(k)}_i,\btheta^{(k)}_i\right):~i=1,\ldots,N\right\}$ from the
posterior distributions of $\btheta^{(k)}$ and ${\bD^*_n}^{(k)}$. The reason for this is the following. Simulating from the priors would lead to many
realizations that are not well-supported by the data $\{\bar x_{T_0+1},\ldots,\bar x_T\}$, and these realizations would render the density
$[\bar x_{T_0+1},\ldots,\bar x_T|{\bD^*_n}^{(k)},\btheta^{(k)},\mathcal M_k]$ extremely small, thus significantly reducing the effective simulation size.
This issue is clearly much alleviated if the simulations correspond to the posterior distributions 
$[{\bD^*_n}^{(k)},\btheta^{(k)}|\bar x_{T_0+1},\ldots,\bar x_T,\mathcal M_k]$, since such realizations are well-supported by the data
that has been conditioned upon. 
This strategy also led to numerically stable estimates of the marginal densities in all our cases.


Using the full conditional distributions (\ref{eq:dir}) and (\ref{eq:post_zeta3}), along with the aforementioned posterior-based computation of 
(\ref{eq:marginal_density}), we obtain $100,000$ realizations from the posterior
distribution of $(\zeta,p_1,\ldots,p_K)$ using Gibbs sampling, after discarding the first $10,000$ iterations as burn-in.

\subsection{Obtaining the posterior probabilities of the alternative hypotheses $H_{1k}$}
\label{subsec:posterior_alternatives}
Note that for $k=1,\ldots,K$, the posterior probability of $H_{1k}$ is given by
\begin{align}
	v_{k}&=1-\left[\zeta=k,S^{(k)}(\bar\bx_{T_0})-S^{(k)}\left(\bx^{(0)}_{T_0}\right)
	\in [\bar\ell_{k},\bar u_{k}] \big |\bar x_{T_0+1},\ldots,\bar x_T \right]\notag\\
	&=1-\left[\zeta=k\big |\bar x_{T_0+1},\ldots,\bar x_T\right]\left[S^{(k)}(\bar\bx_{T_0})-S^{(k)}\left(\bx^{(0)}_{T_0}\right)\in 
	[\bar\ell_{k},\bar u_{k}] \big |\zeta=k,\bar x_{T_0+1},\ldots,\bar x_T\right].
	\label{eq:mc_average}
\end{align}
Hence, once we obtain realizations from the posteriors of $S^{(k)}(\bar\bx_{T_0})$ for $k=1,\ldots,K$, and $(\zeta,p_1,\ldots,p_K)$, evaluation of 
$v_{k}$; $k=1,\ldots,K$, follows simply by Monte Carlo averaging associated with the two factors
of (\ref{eq:mc_average}).

\section{GCM selection results}
\label{sec:gcm_results}
\begin{figure}
	\centering
	\subfigure [cFDR for SRES: A1B.]{ \label{fig:cfdr_a1b}
	\includegraphics[width=7.5cm,height=6.7cm]{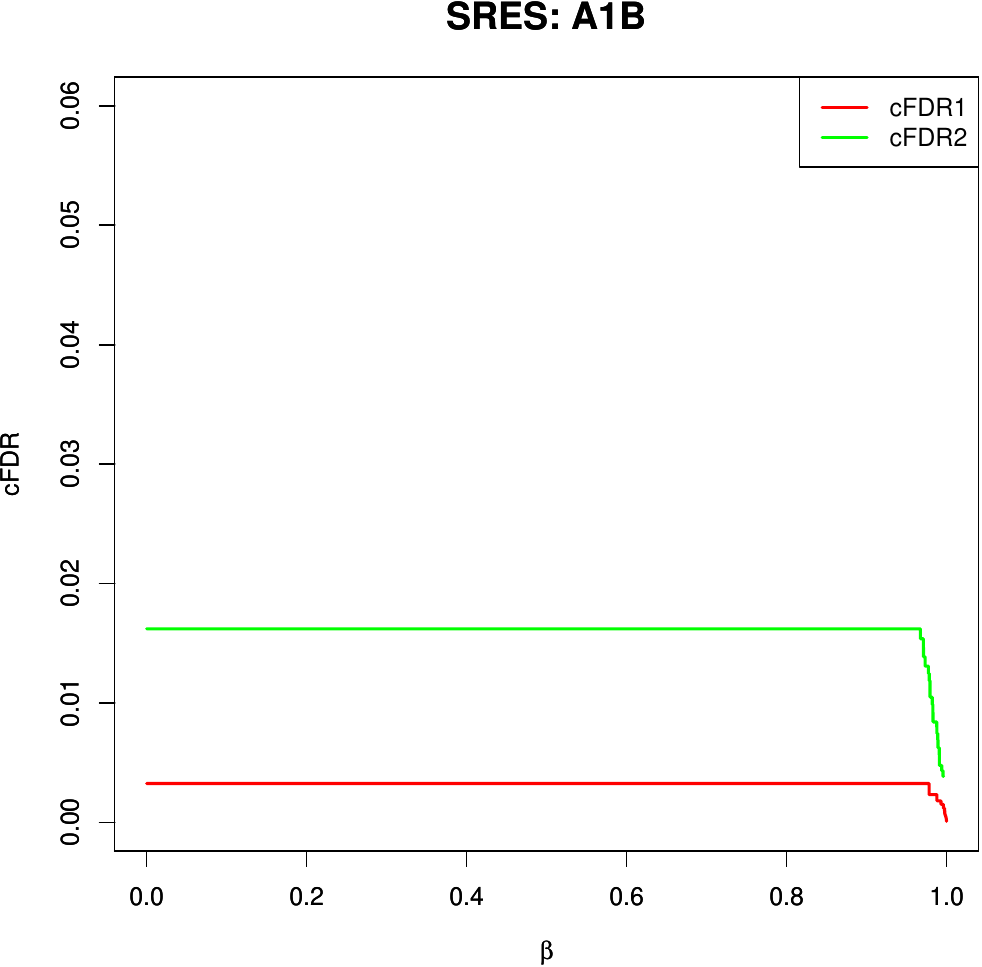}}
	\hspace{2mm}
	\subfigure [cFNR for SRES: A1B.]{ \label{fig:cfnr_a1b}
	\includegraphics[width=7.5cm,height=6.7cm]{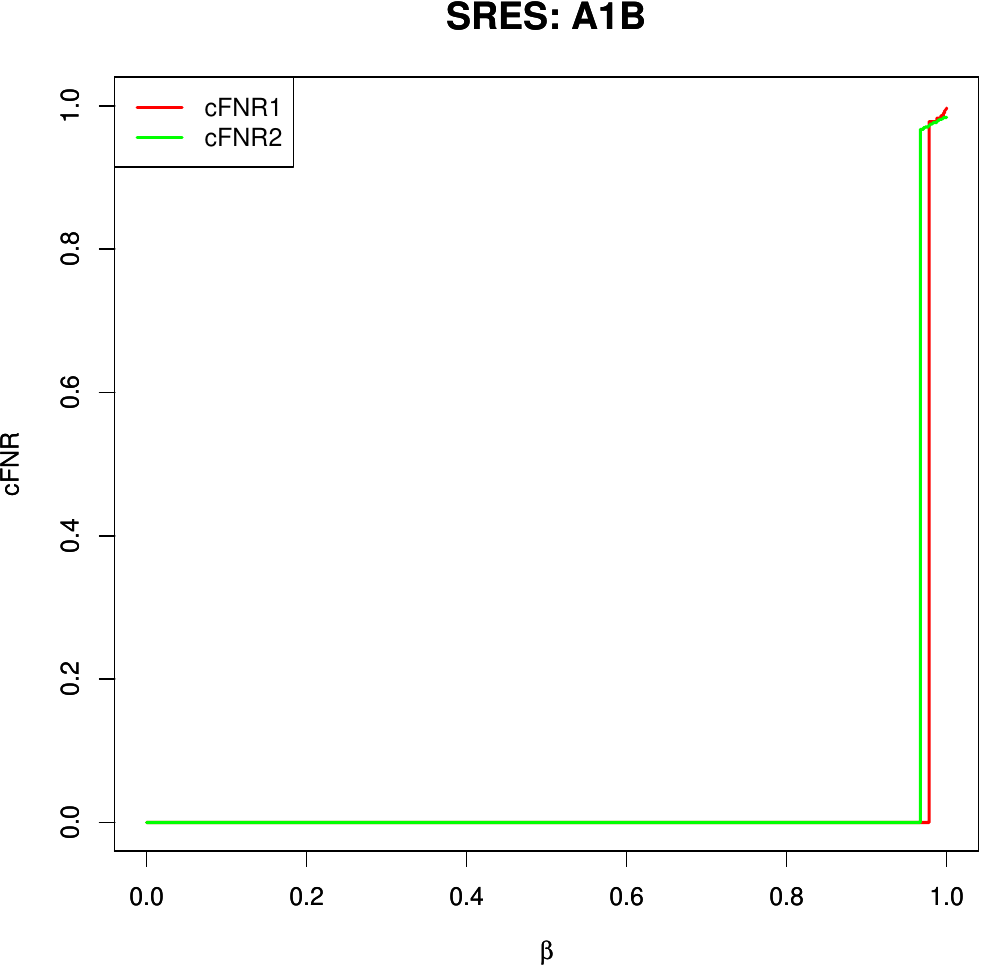}}\\
	\vspace{2mm}
	\subfigure [cFDR for SRES: A2.]{ \label{fig:cfdr_a2}
	\includegraphics[width=7.5cm,height=6.7cm]{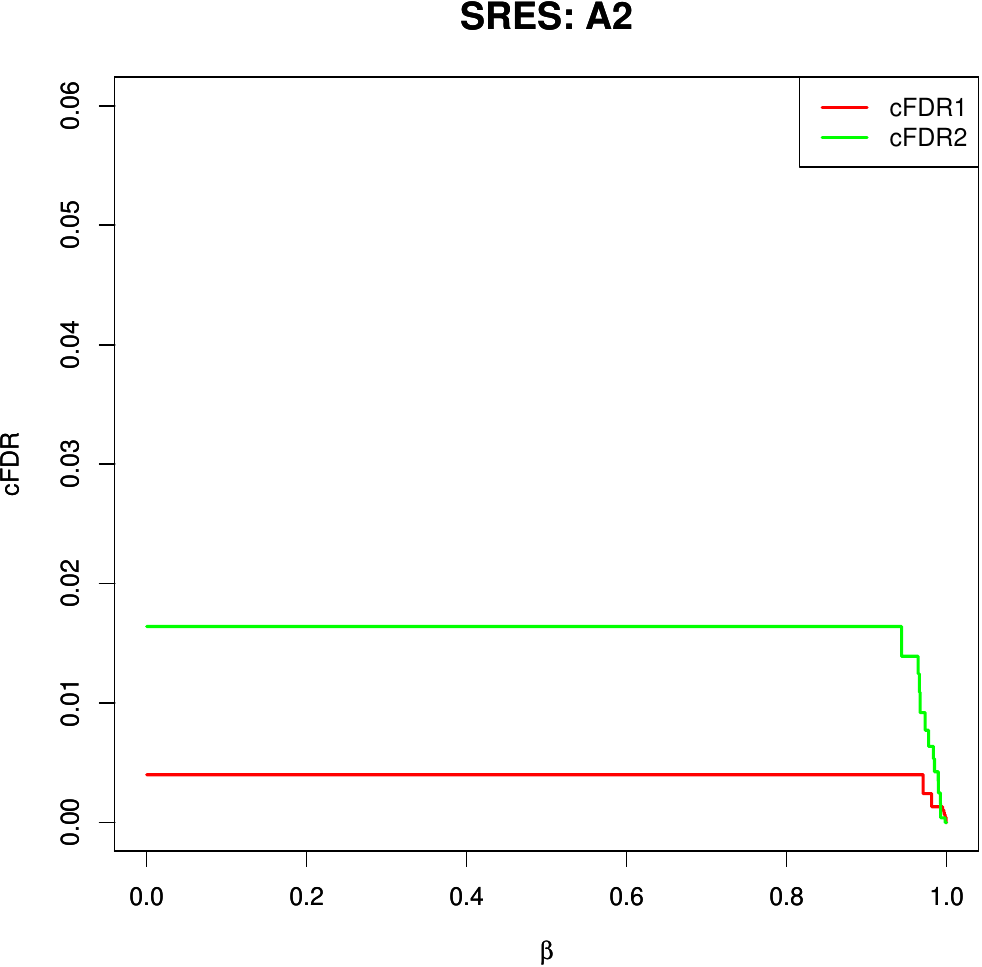}}
	\hspace{2mm}
	\subfigure [cFNR for SRES: A2.]{ \label{fig:cfnr_a2}
	\includegraphics[width=7.5cm,height=6.7cm]{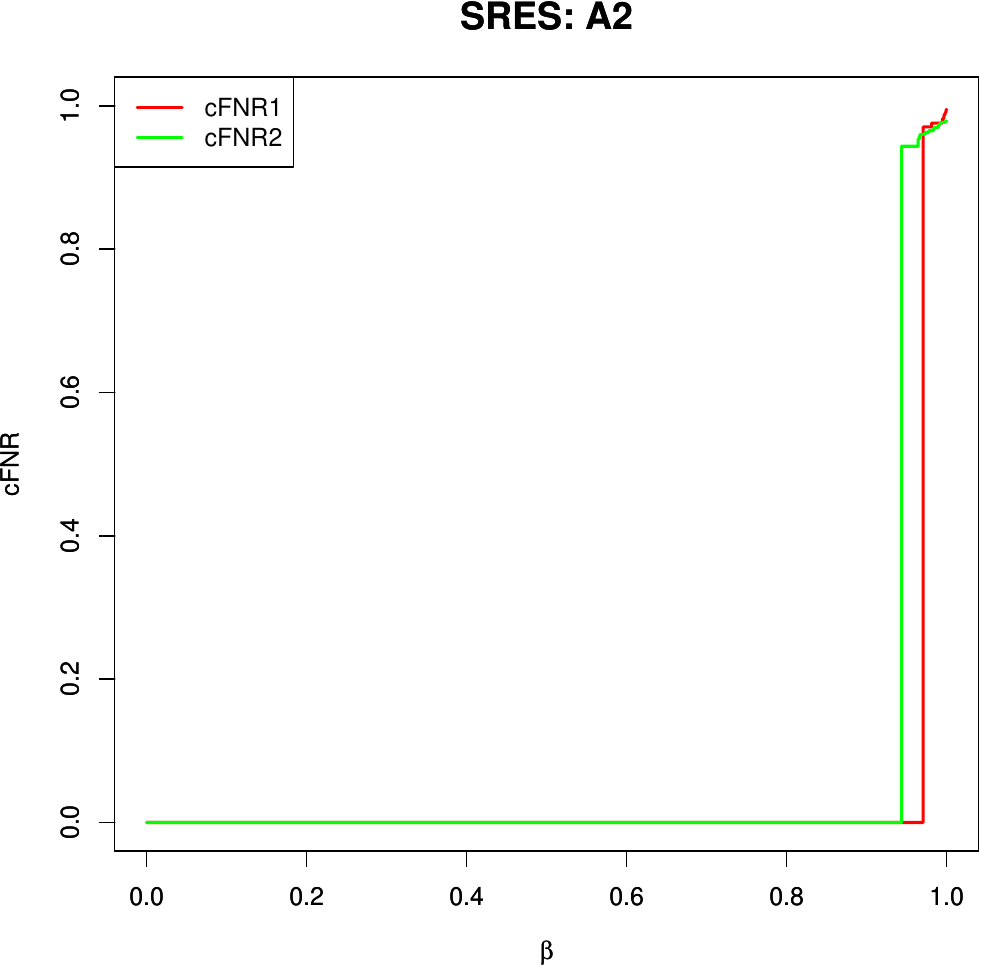}}
	\caption{cFDR and cFNR for GCM selection in the climate scenarios A1B and A2 using Bayesian multiple testing.}
	\label{fig:gcm_errors1}
\end{figure}

\begin{figure}
	\centering
	\subfigure [cFDR for SRES: B1.]{ \label{fig:cfdr_b1}
	\includegraphics[width=7.5cm,height=6.7cm]{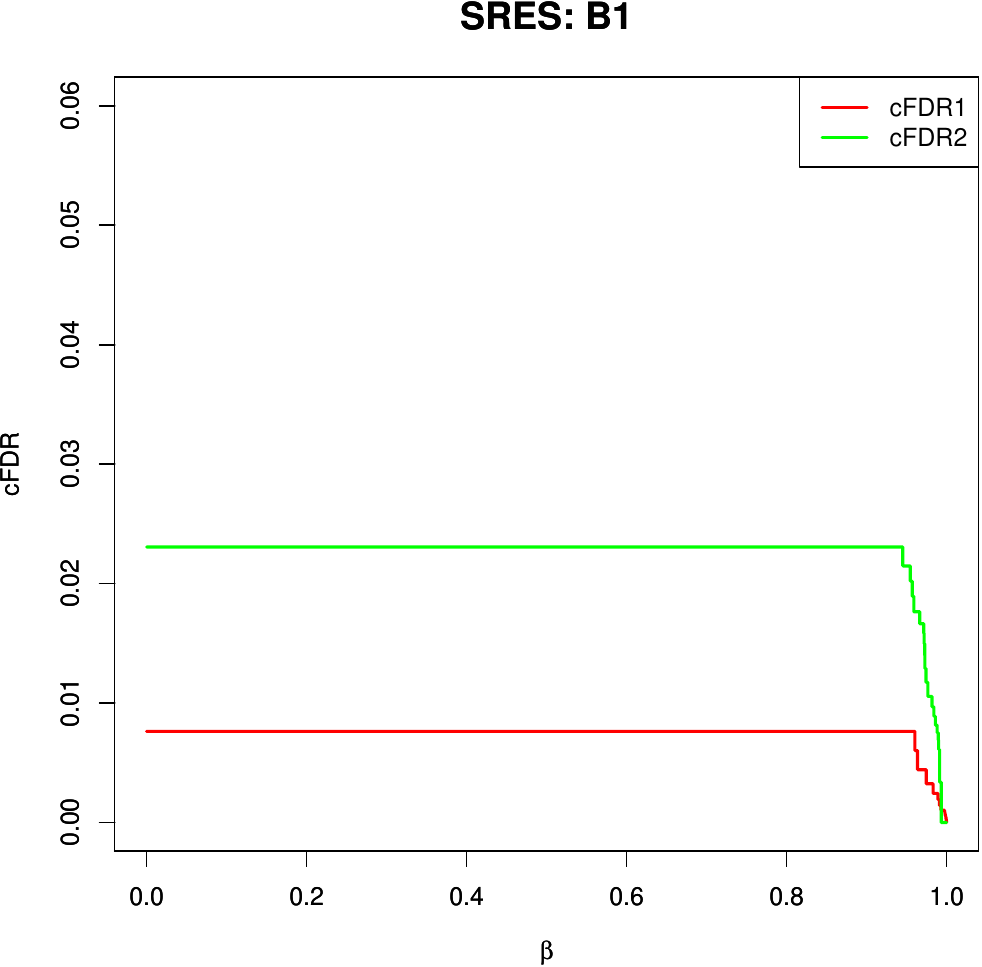}}
	\hspace{2mm}
	\subfigure [cFNR for SRES: B1.]{ \label{fig:cfnr_b1}
	\includegraphics[width=7.5cm,height=6.7cm]{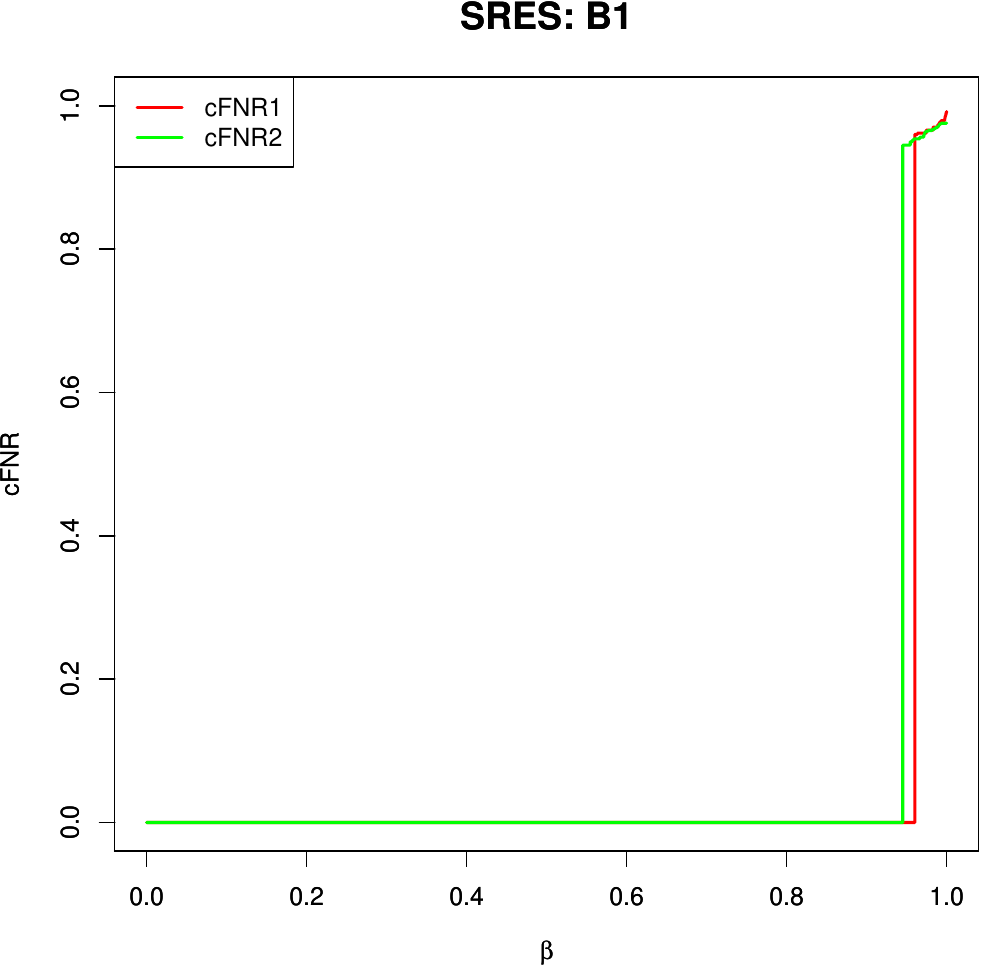}}\\
	\vspace{2mm}
	\subfigure [cFDR for Commitment.]{ \label{fig:cfdr_commit}
	\includegraphics[width=7.5cm,height=6.7cm]{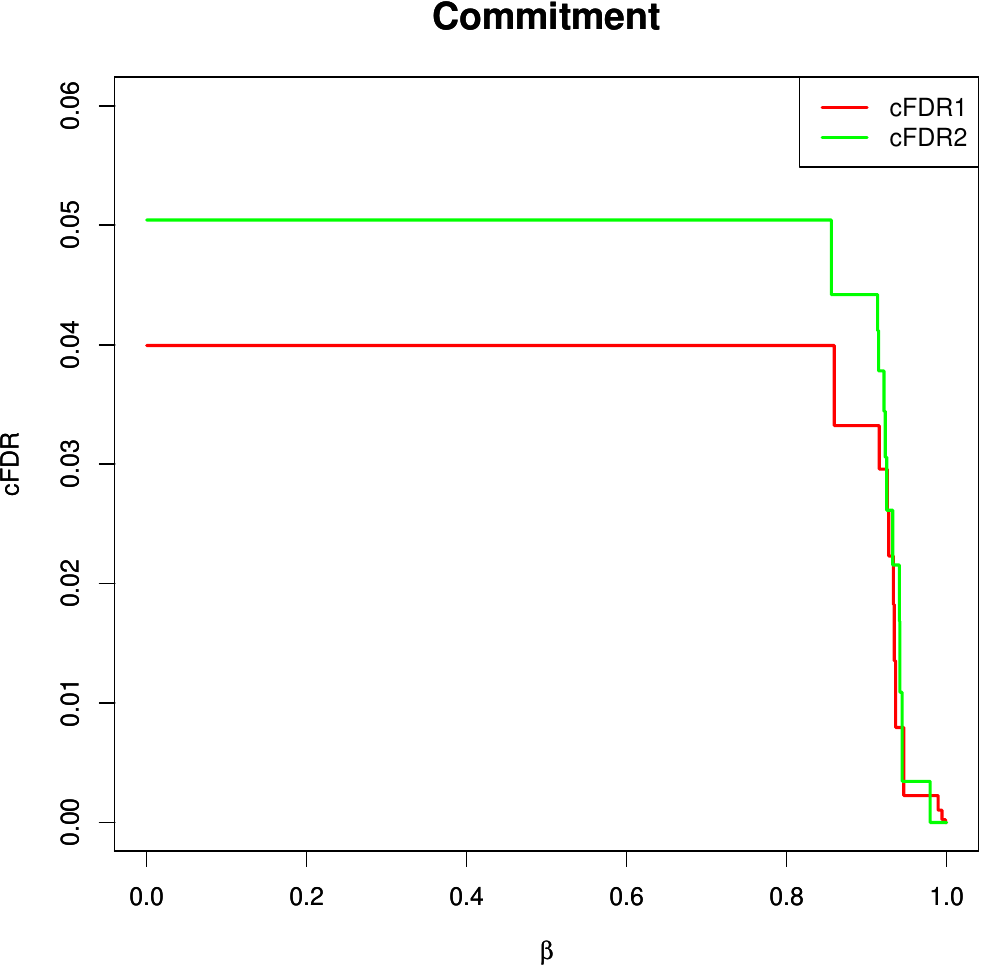}}
	\hspace{2mm}
	\subfigure [cFNR for Commitment.]{ \label{fig:cfnr_commit}
	\includegraphics[width=7.5cm,height=6.7cm]{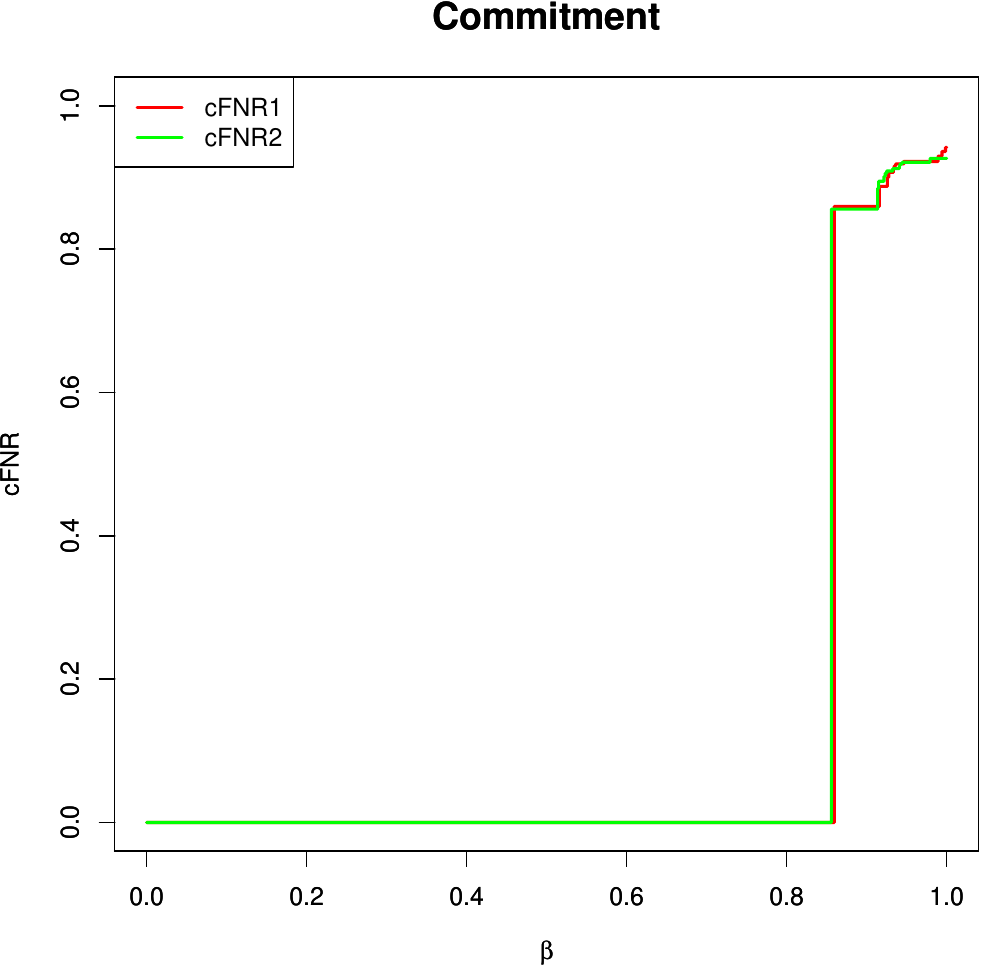}}
	\caption{cFDR and cFNR for GCM selection in the climate scenarios B1 and Commitment using Bayesian multiple testing.}
	\label{fig:gcm_errors2}
\end{figure}

We implemented our Bayesian multiple testing procedure with both the discrepancy measures $S^{(k)}_1$ and $S^{(k)}_2$ given by 
(\ref{eq:T1}) and (\ref{eq:T2}), respectively. We denote the corresponding cFDRs by cFDR1 and cFDR2 and the corresponding cFNRs by cFNR1 and cFNR2, respectively.
Figures \ref{fig:gcm_errors1} and \ref{fig:gcm_errors2} depict these Bayesian error measures as functions of the penalty $\beta$, 
for all the four climate scenarios A1B, A2, B1 and Commitment, with respect to both the discrepancy measures $S^{(k)}_1$ (red line) and 
$S^{(k)}_2$ (green line). 

The discussions toward the ends of Section \ref{subsec:muller} and \ref{subsec:Bayesian_errors} point out that the first jump occurring in either of the graphs
of cFDR or cFNR as functions of $\beta$, corresponds to the best model. In this regard, 
Figures \ref{fig:gcm_errors1} and \ref{fig:gcm_errors2} show that values of the penalty $\beta$ close to one are required to obtain the first jumps of cFDR and cFNR
for both the discrepancy measures $S^{(k)}_1$ and $S^{(k)}_2$, for all the four climate scenarios. 
Thus, none of the selected models seem to be satisfactory. Also, all the jumps occur close to each other in all the cases, indicating that the best models are not 
significantly good compared to the other competing models.

In all the cases, $S^{(k)}_2$ performs relatively better than
$S^{(k)}_1$ in the sense that the value of $\beta$ required for $S^{(k)}_2$ is somewhat less than the $S^{(k)}_1$ counterpart for selecting the best model. 
Among all the four climate
scenarios, the Commitment scenario turns out to be the best since here the best model is selected for a value of $\beta$ that is lesser than those of the
other scenarios.

In the case of A1B, $S^{(k)}_1$ and $S^{(k)}_2$ yielded two different best models, $\mbox{csiro}\_\mbox{mk3}\_0$ and $\mbox{inmcm3}\_0$, respectively. 
In the remaining climate scenarios,
both the discrepancy measures $S^{(k)}_1$ and $S^{(k)}_2$ yielded the same best models. The best GCMs selected for the scenarios A2, B1 and Commitment,
are $\mbox{ukmo}\_\mbox{hadgem1}$, $\mbox{gfdl}\_\mbox{cm2}\_0$ and $\mbox{cnrm}\_\mbox{cm3}$, respectively.

\begin{figure}
	\centering
	\subfigure [A1B: Best GCM $\mbox{csiro}\_\mbox{mk3}\_0$.]{ \label{fig:a1b_best1}
	\includegraphics[width=7.5cm,height=6.7cm]{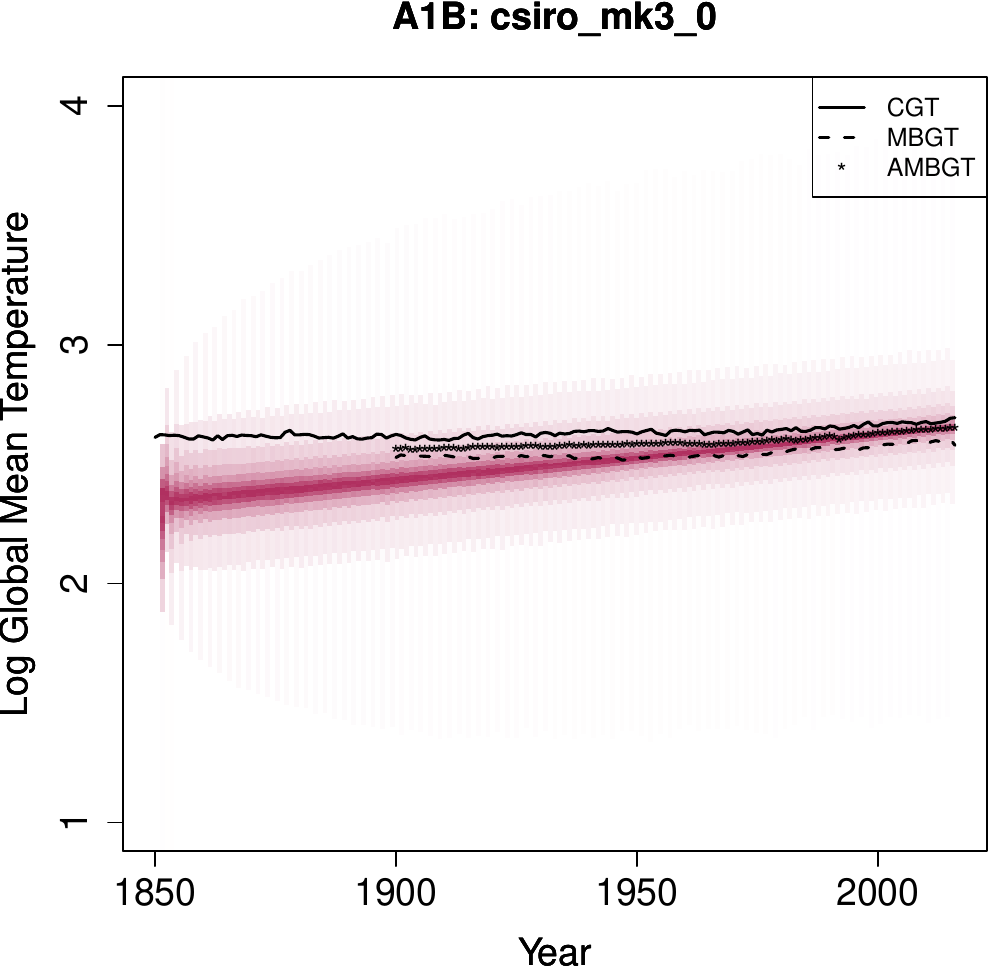}}
	\hspace{2mm}
	\subfigure [A1B: Best GCM $\mbox{inmcm3}\_0$.]{ \label{fig:a1b_best2}
	\includegraphics[width=7.5cm,height=6.7cm]{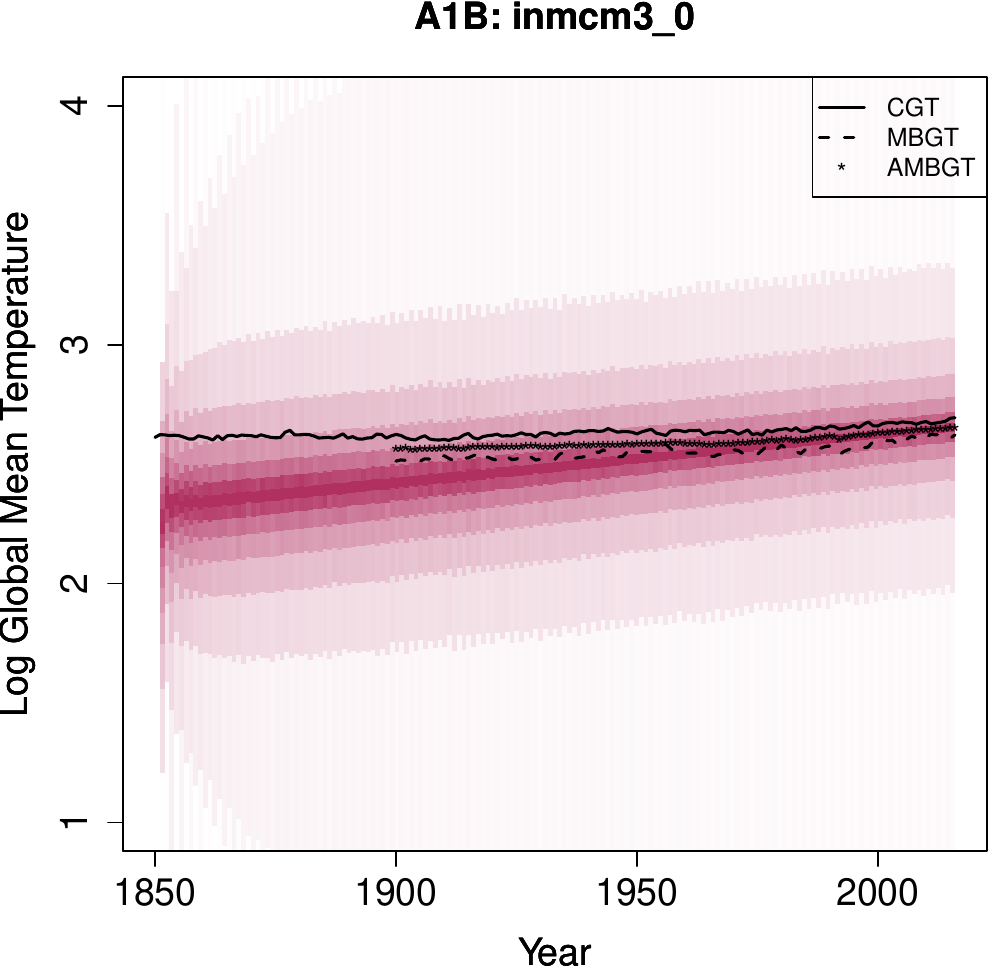}}\\
	\vspace{2mm}
	\subfigure [A2: Best GCM $\mbox{ukmo}\_\mbox{hadgem1}$.]{ \label{fig:a2_best}
	\includegraphics[width=7.5cm,height=6.7cm]{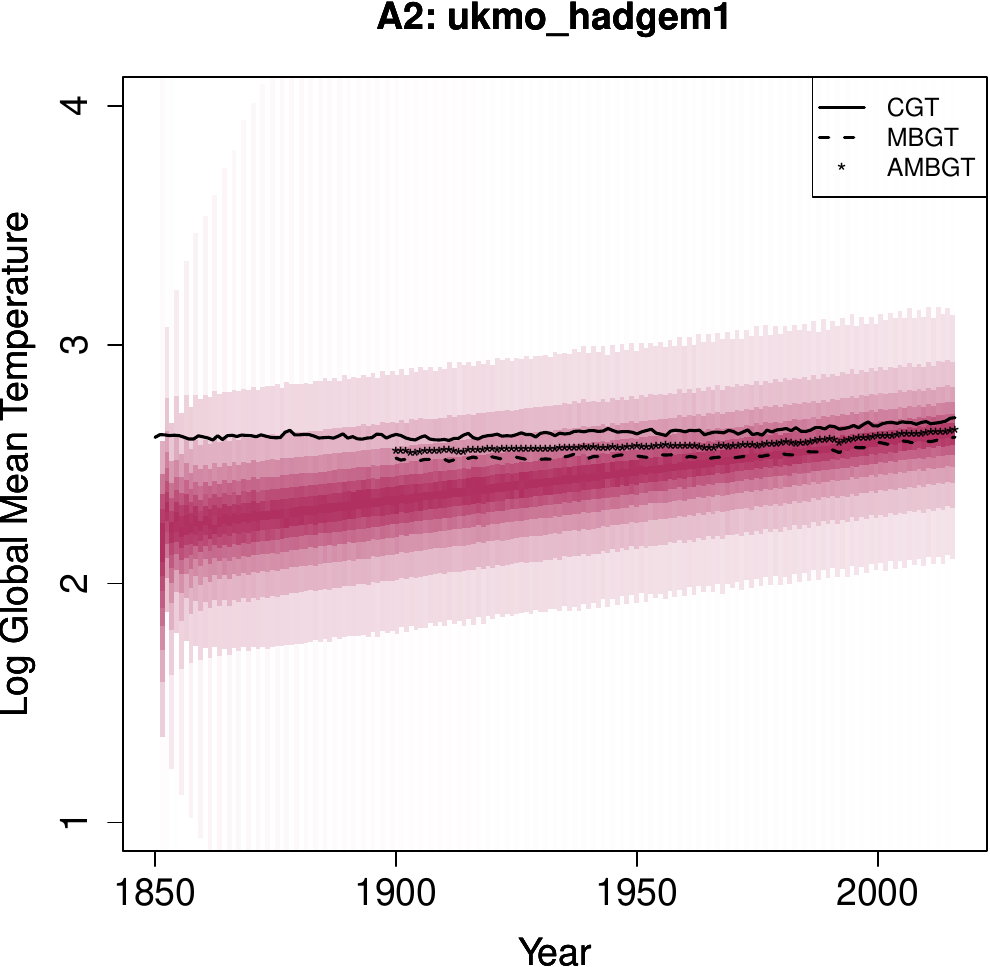}}
	\hspace{2mm}
	\subfigure [B1: Best GCM $\mbox{gfdl}\_\mbox{cm2}\_0$.]{ \label{fig:b1_best}
	\includegraphics[width=7.5cm,height=6.7cm]{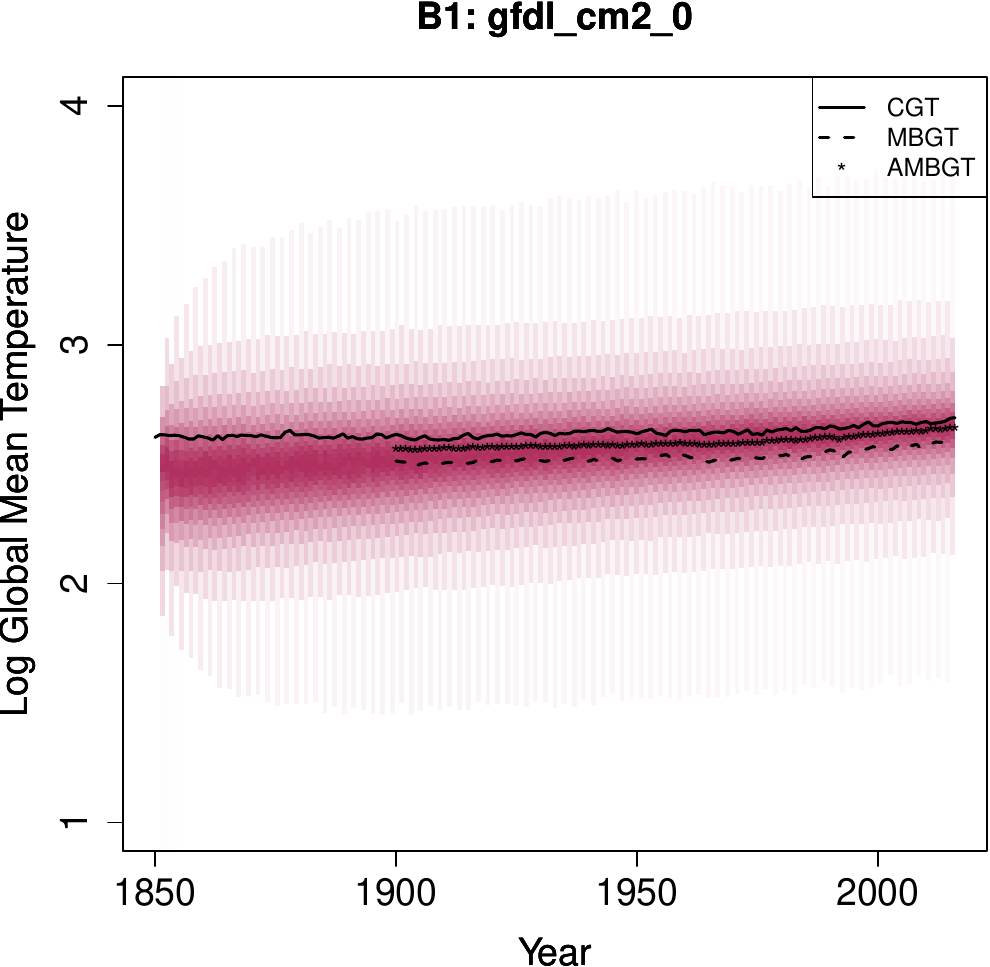}}\\
	\vspace{2mm}
	\subfigure [Commitment: Best GCM $\mbox{cnrm}\_\mbox{cm3}$.]{ \label{fig:commit_best}
	\includegraphics[width=7.5cm,height=6.7cm]{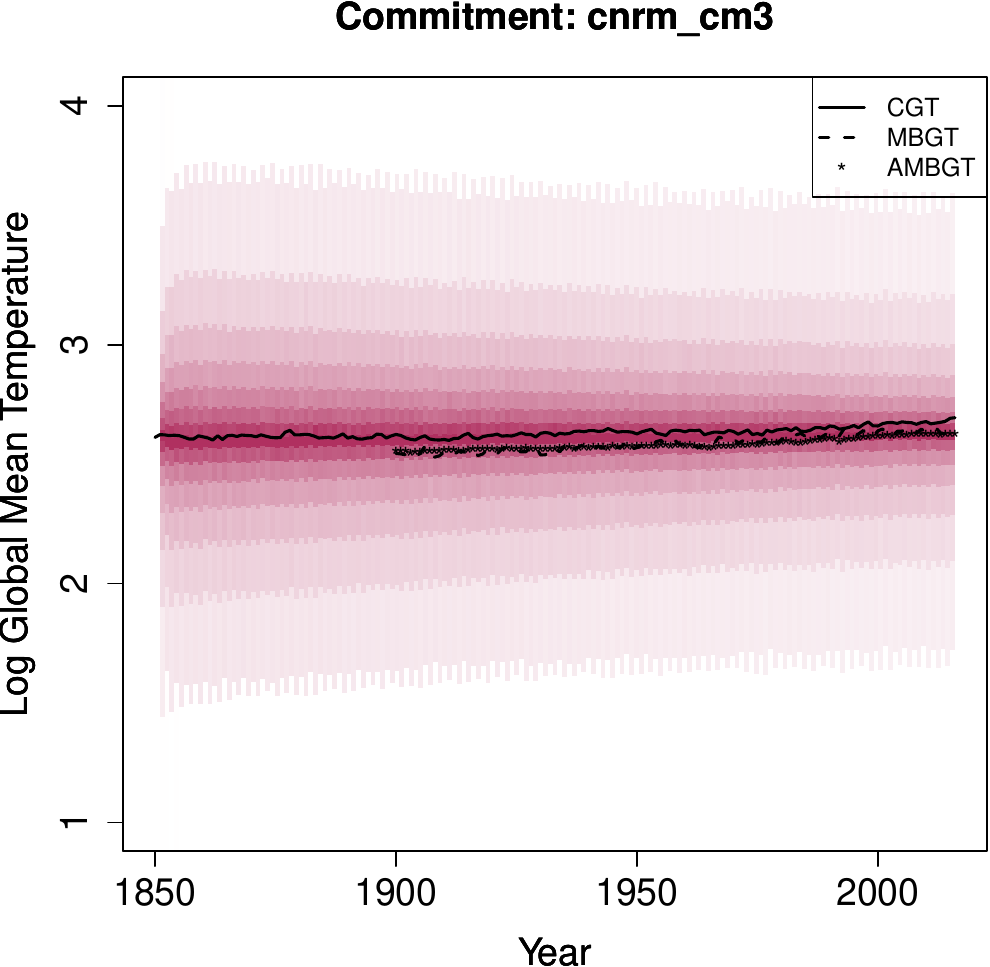}}
	\caption{The posteriors corresponding to the HadCRUT4 data or the current global temperature (CGT) 
	conditional on GCM-based average time series are shown as colour plots with progressively higher densities depicted by progressively intense colours.
	Also shown are the HadCRUT4 data (CGT), GCM based time series (MBGT) and the average of GCM based time series (AMBGT).
	The temperature is in $\degree$C and in the log-scale.}
	\label{fig:gcm_best_average}
\end{figure}

Figure \ref{fig:gcm_best_average} displays the posterior distribution of the time series 
$[\bar x_0,\bar x_1,\ldots, \bar x_{T_0}|\bar x_{T_0+1},\ldots,\bar x_T]$ (note that $\bar x_0=x_0$, since $x_0$ is assumed to be known) 
corresponding to the aforementioned best GCMs selected by our Bayesian multiple testing procedure,
as colour plots. The progressively higher densities are represented by progressively intense colours.
The thick black line is the HadCRUT4 data, which is the current global temperature (CGT) and the dashed line is the model based global temperature (MBGT), 
the simulated global temperatures
by the underlying GCM. The other starred line stands for the average model based global temperature (AMBGT), which is the average over all the GCM based 
simulated time series in the respective climate scenario. All the time series are in degree celsius and in the log scale. Recall that the 
HadCRUT4 data is associated with the years $1850-2016$ and the GCMs are associated with $1900-2099$, which is why the time scales for the HadCRUT4 data
and the GCM based simulated data are different.

Observe that except for B1 and Commitment most part of the observed HadCRUT4 data is not included in the high density regions of the corresponding posterior time series
associated with the best GCMs. In fact, except the case of Commitment, all other posteriors strongly support lower temperatures than HadCRUT4. This is not surprising 
since Figure \ref{fig:gcm_models} show that the GCM-simulated time series significantly underestimate the HadCRUT4 data during the relevant time period, and so must
the averaged GCM time series, and this is broadly consistent with the observations on Figure \ref{fig:gcm_best_average}. Also observe that MBGT and AMBGT lie closer to the high 
density regions compared to CGT, which is again not unexpected as Figure \ref{fig:gcm_models} indicates. 

\newpage
\begin{table}[h]
\centering
	\caption{Goodness-of-fit check for the best GCMs with respect to averaged time series. Here 95\% BCI stands for 95\% Bayesian credible intervals.}
\label{table:table1}
	\begin{tabular}{|c|c|c|c|c|}
\hline
		Model & $S^{(k)}_1\left(\bx^{(0)}_{T_0}\right)$ & 95\% BCI of $S^{(k)}_1\left(\bar\bx_{T_0}\right)$ 
		& $S^{(k)}_2\left(\bx^{(0)}_{T_0}\right)$ & 95\% BCI of $S^{(k)}_2\left(\bar\bx_{T_0}\right)$\\ 
\hline
		A1B ($\mbox{csiro}\_\mbox{mk3}\_0$)  &  0.126  &  [0.104,0.281]  &  0.024  &  [0.015,0.424]\\
		A1B ($\mbox{inmcm3}\_0$) &  0.001  &  [$5\times 10^{-4}$,0.023]  &  $126\times 10^{-6}$  &  [$7.324\times 10^{-6}$,0.048]\\
		A2 ($\mbox{ukmo}\_\mbox{hadgem1}$) &  0.006  &  [0.003,0.048]  &  0.001  &  [$9.047\times 10^{-5}$,0.082]\\
		B1 ($\mbox{gfdl}\_\mbox{cm2}\_0$) &  0.142  &  [0.107,1.048]  &  0.028  &  [0.017,2.420]\\   
		Commit ($\mbox{cnrm}\_\mbox{cm3}$) &  0.039  &  [0.119,1.599]  &   0.002  &  [0.021,3.848]\\ 
\hline
\end{tabular}
\end{table}

Table \ref{table:table1}, summarizing the goodness-of-fit of the posteriors to the HadCRUT4 data with respect to the discrepancy measures $S^{(k)}_1$ and $S^{(k)}_2$, 
tell a somewhat different story.
The best GCM in the Commitment scenario seems to overfit the HadCRUT4 data in the sense that the observed discrepancies are too small to be included the 95\% credible
intervals of the reference discrepancy measures. Given the large variability of the time series as shown in panel (e) of Figure \ref{fig:gcm_best_average}, which can also
be gauged by the less colour intensities compared to the other panels, this result is not unexpected in retrospect.
On the other hand, in the other cases, the observed discrepancies are included in the respective 95\% credible intervals. Although again this seems surprising at the
first glance, this is due the fact that the posterior time series
relatively closer to the year $2017$, where the GCM time series begins in our posterior formulation, well-captures the HadCRUT4 data, with relatively small posterior
variability. Hence, even though the posteriors fail to perform well for the years closer to $1850$, the overall goodness-of-fit still can not be declared as poor.

\begin{figure}
	\centering
	\subfigure [A1B: Best GCM $\mbox{csiro}\_\mbox{mk3}\_0$.]{ \label{fig:best1}
	\includegraphics[width=7.5cm,height=6.7cm]{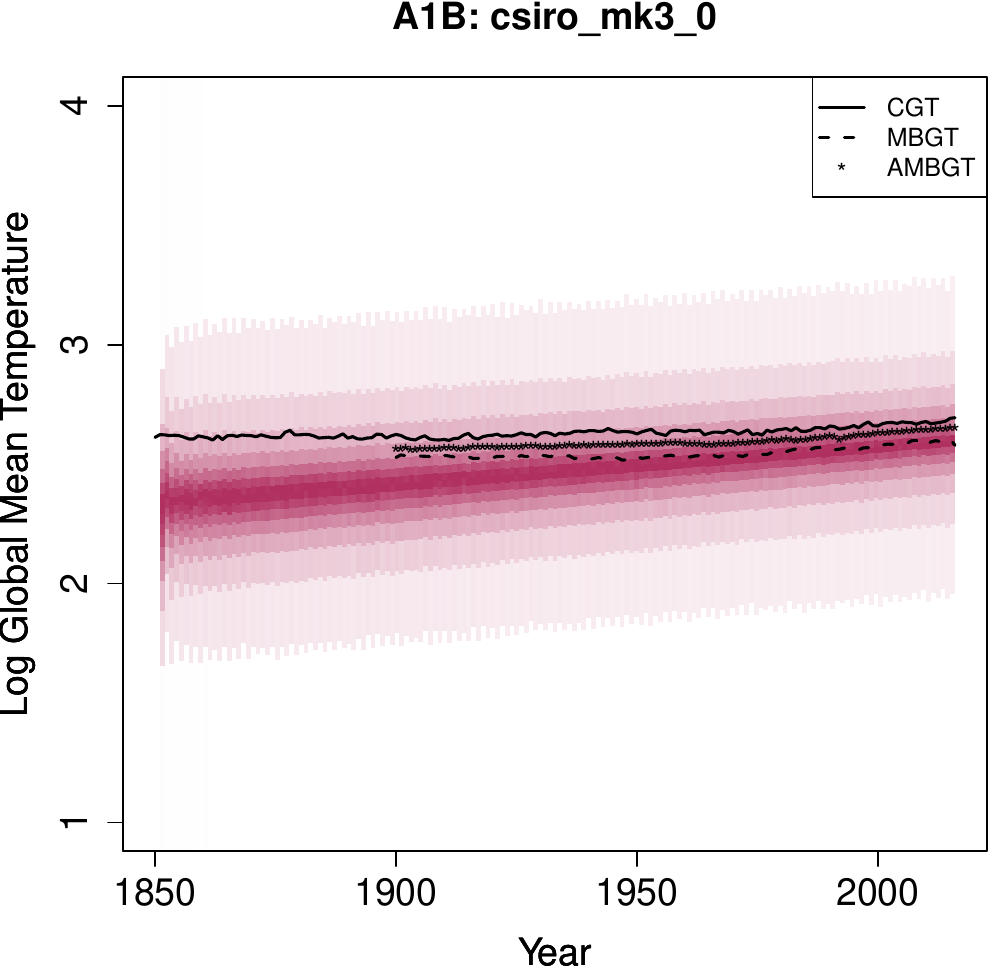}}
	\hspace{2mm}
	\subfigure [A1B: Best GCM $\mbox{inmcm3}\_0$.]{ \label{fig:best2}
	\includegraphics[width=7.5cm,height=6.7cm]{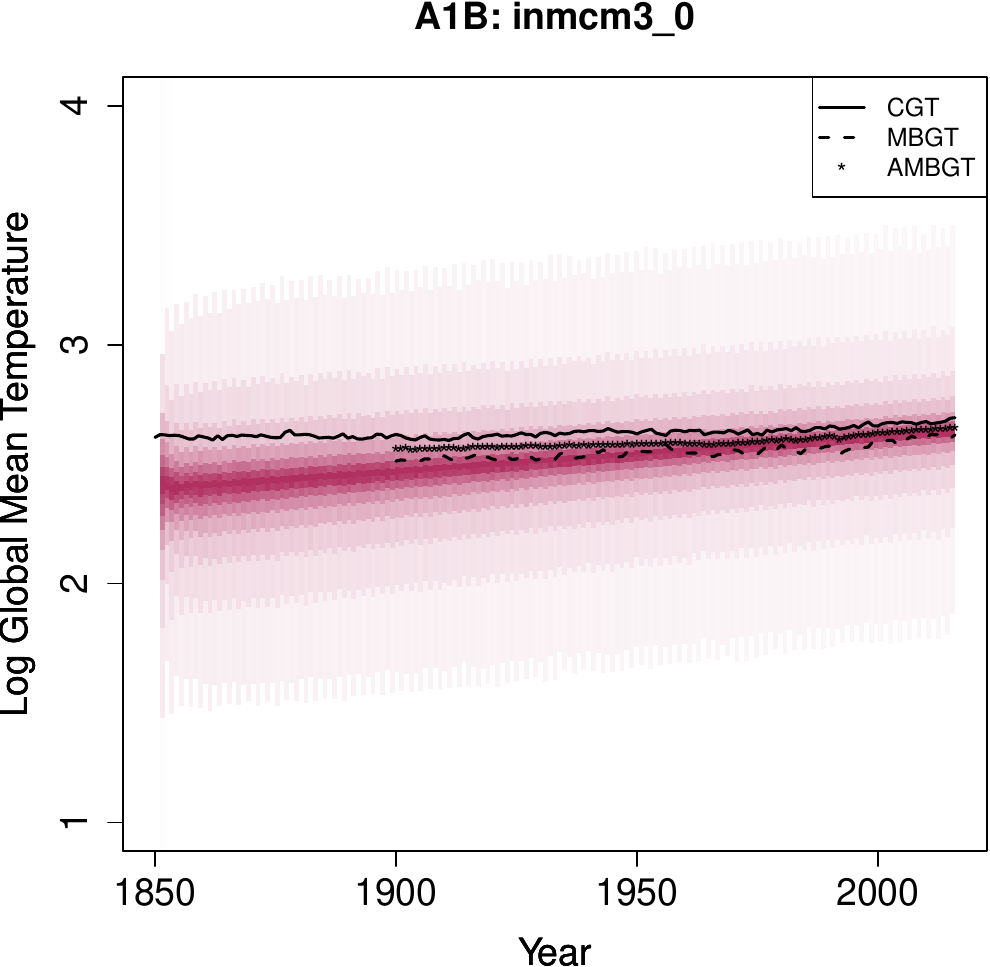}}\\
	\vspace{2mm}
	\subfigure [A2: Best GCM $\mbox{ukmo}\_\mbox{hadgem1}$.]{ \label{fig:best3}
	\includegraphics[width=7.5cm,height=6.7cm]{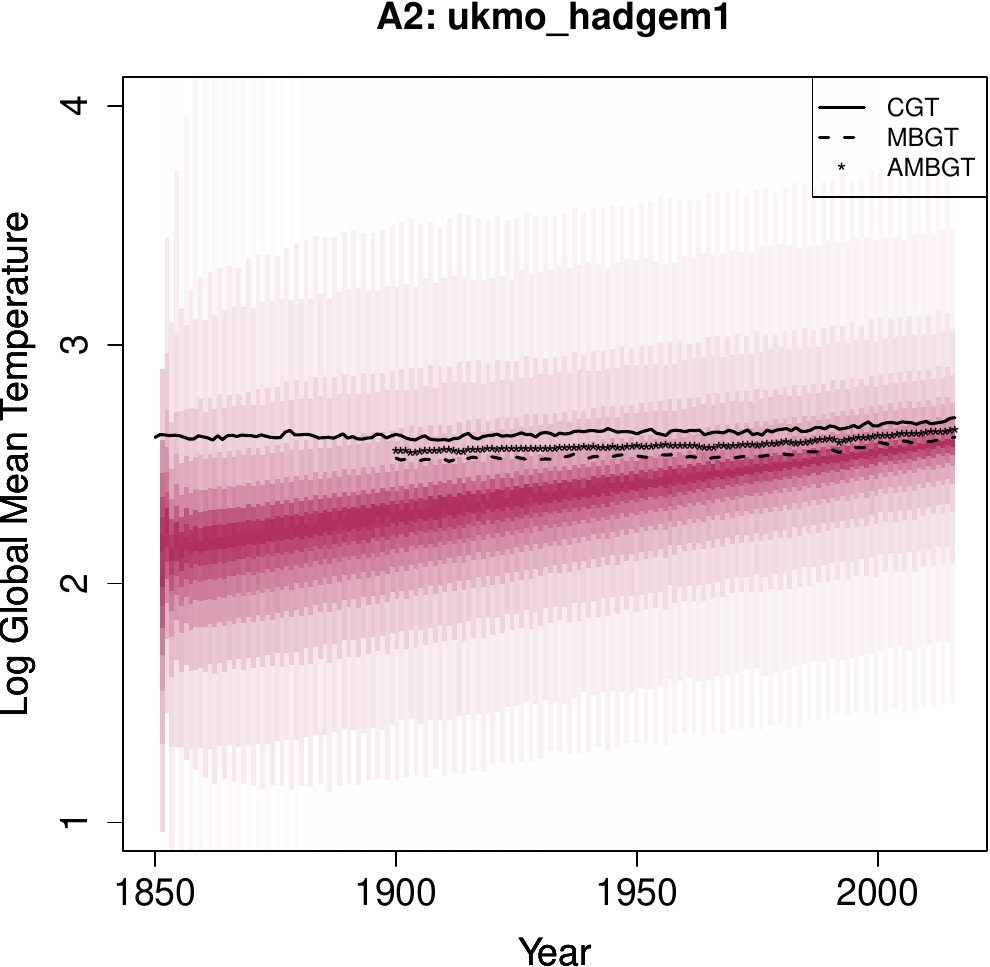}}
	\hspace{2mm}
	\subfigure [B1: Best GCM $\mbox{gfdl}\_\mbox{cm2}\_0$.]{ \label{fig:best4}
	\includegraphics[width=7.5cm,height=6.7cm]{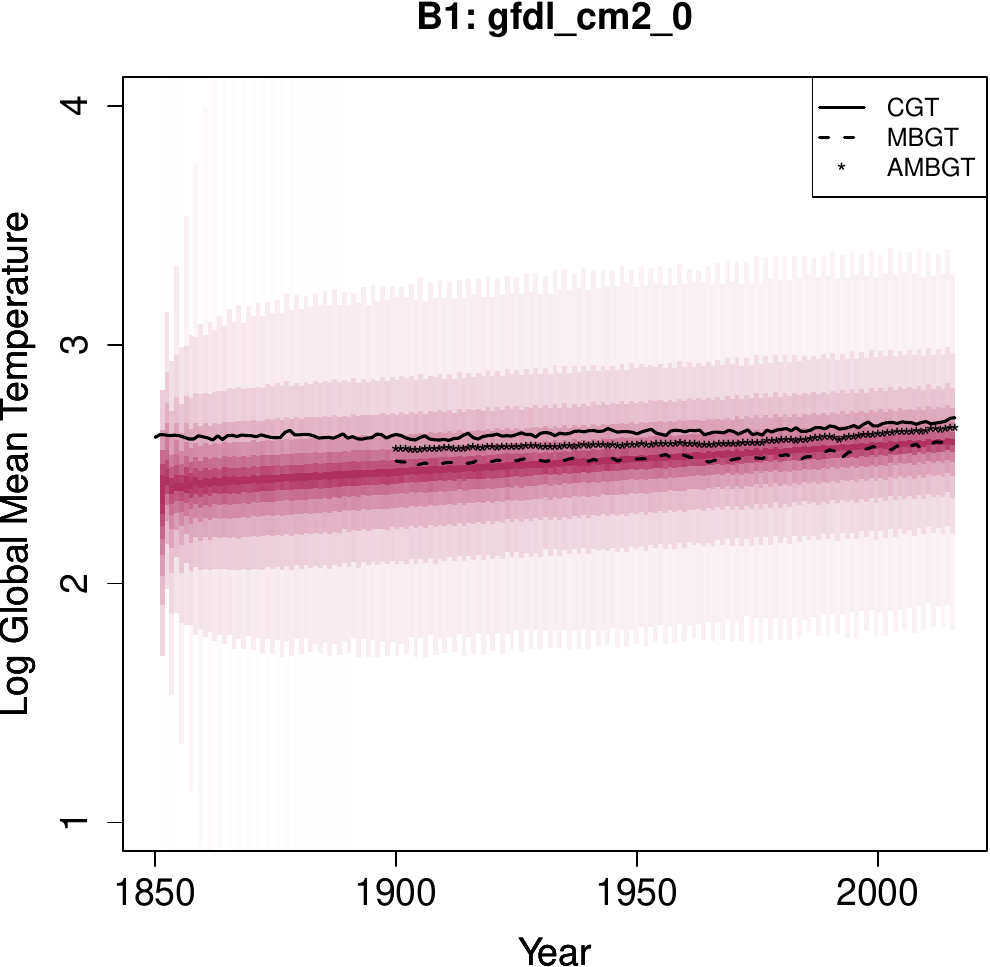}}\\
	\vspace{2mm}
	\subfigure [Commitment: Best GCM $\mbox{cnrm}\_\mbox{cm3}$.]{ \label{fig:best5}
	\includegraphics[width=7.5cm,height=6.7cm]{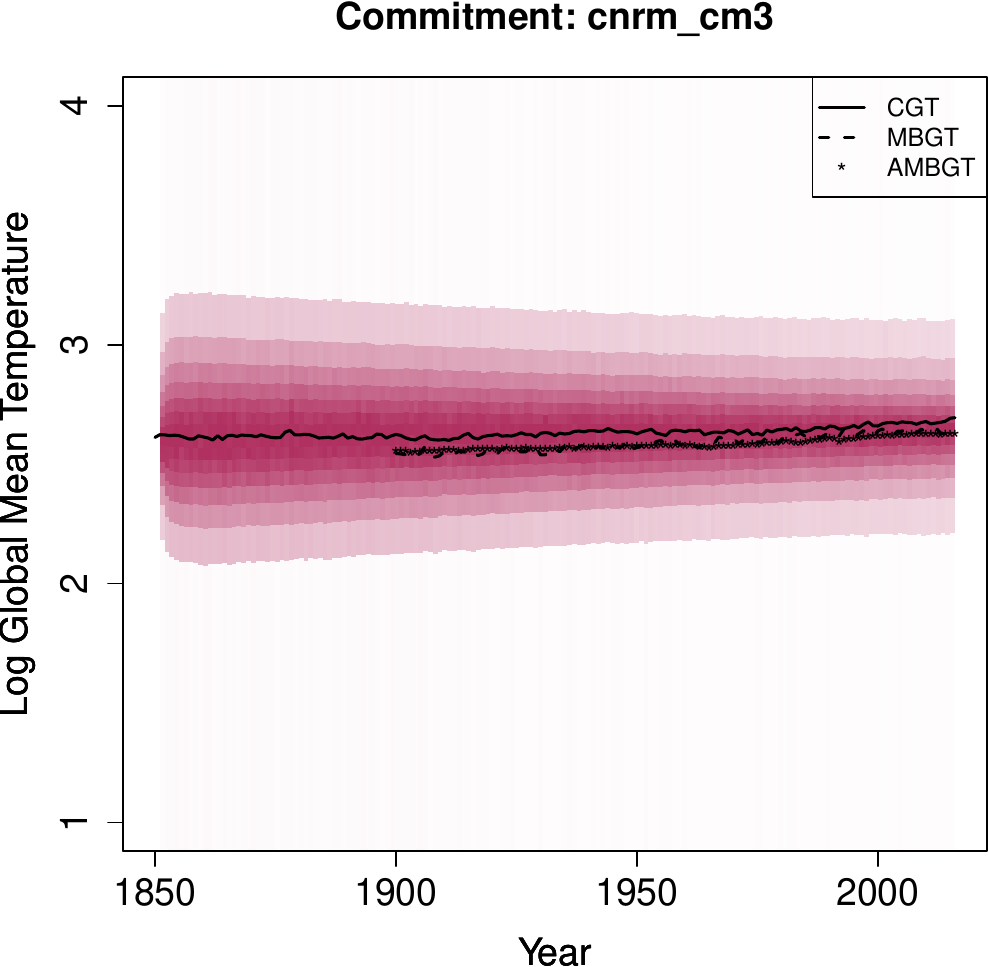}}
	\caption{The posteriors corresponding to the HadCRUT4 data or the current global temperature (CGT) 
	conditional on individual best GCM time series are shown as colour plots with progressively higher densities depicted by progressively intense colours.
	Also shown are the HadCRUT4 data (CGT), GCM based time series (MBGT) and the average of GCM based time series (AMBGT).
	The temperature is in $\degree$C and in the log-scale.}
	\label{fig:gcm_best}
\end{figure}

Figure \ref{fig:gcm_best} shows the posterior distributions of $[x_0,x_1,\ldots, x_{T_0}|x_{T_0+1},\ldots,x_T]$ associated with the individual time series 
for the best GCM models, rather than the averaged time series as shown in Figure \ref{fig:gcm_best_average}. The overall story, however, did not seem to be very
different compared to that told by Figure \ref{fig:gcm_best_average}. Table \ref{table:table2}, evaluating goodness-of-fit for these posteriors using the discrepancy measures,
also provide similar inference as Table \ref{table:table1}, where $\bx_{T_0}=(x_1,\ldots,x_{T_0})$. 

\newpage
\begin{table}[h]
\centering
	\caption{Goodness-of-fit check for the best GCMs with respect to individual time series. Here 95\% BCI stands for 95\% Bayesian credible intervals.}
\label{table:table2}
	\begin{tabular}{|c|c|c|c|c|}
\hline
		Model & $S^{(k)}_1\left(\bx^{(0)}_{T_0}\right)$ & 95\% BCI of $S^{(k)}_1\left(\bx_{T_0}\right)$ 
		& $S^{(k)}_2\left(\bx^{(0)}_{T_0}\right)$ & 95\% BCI of $S^{(k)}_2\left(\bx_{T_0}\right)$\\ 
\hline
		A1B ($\mbox{csiro}\_\mbox{mk3}\_0$)  &  0.127  &  [0.099,0.255]  &  0.028  &  [0.014,0.419]\\
		A1B ($\mbox{inmcm3}\_0$) &  0.105  &  [0.098,0.163]  &  0.016  &  [0.013,0.179]\\
		A2 ($\mbox{ukmo}\_\mbox{hadgem1}$) &  0.088  &  [0.077,0.152]  &  0.015  &  [0.009,0.186]\\
		B1 ($\mbox{gfdl}\_\mbox{cm2}\_0$) &  0.072  &  [0.062,0.202]  &  0.010  &  [0.005,0.348]\\   
		Commit ($\mbox{cnrm}\_\mbox{cm3}$) &  0.043  &  [0.179,1.496]  &  0.003  &  [0.049,3.504]\\ 
\hline
\end{tabular}
\end{table}

\section{GCM simulations as ensembles: extension of our GP emulation approach to the multivariate situation}
\label{sec:multivariate}

So far, the inference with our one-dimensional GP approach demonstrated that although even the best GCM models 
are not as adequate as desired, it is not very easy to discard them since 
Tables \ref{table:table1} and \ref{table:table2} demonstrate quantitatively that in general their overall performances in fitting the observed 
current global temperatures are not particularly poor. 
However, Figures \ref{fig:gcm_best_average} and \ref{fig:gcm_best} show that a large part of the current global temperature data, beginning from $1851$,
fails to lie in the high density regions of the relevant posterior, which is clearly very disconcerting. Even though the Commitment scenario
includes almost the entire current temperature time series in its high posterior density region, the posterior variability turns out to be too high to render
the fit satisfactory.

For further investigation we consider all the $K$ GCM-based time series in any climate scenario as an ensemble of time series, and consider modeling them
as multivariate ($K$-dimensional) time series, extending our one-dimensional GP emulation theory to multidimensional GP emulation.
In this regard, for $t=0,1,2,\ldots$, 
let $\bx_t=\left(x^{(1)}_1,\ldots,x^{(K)}_t\right)'$ be $K$-component vectors, corresponding to $K$ different GCM based log time series $x^{(k)}_t$; $k=1,\ldots,K$.  
With $\bar x_t=K^{-1}\sum_{k=1}^Kx^{(k)}_t$, we shall be interested in the posterior $[\bar x_1,\ldots, \bar x_{T_0}|\bx_{T_0+1},\ldots,\bx_T]$,
for predicting the logarithm of the observed current temperature data. Note that $[\bar x_1,\ldots, \bar x_{T_0}|\bx_{T_0+1},\ldots,\bx_T]$
is induced by $[\bx_1,\ldots, \bx_{T_0}|\bx_{T_0+1},\ldots,\bx_T]$ as the former is obtained from the latter by simply taking the averages of
the components of $\bx_t$, for each $t=1,\ldots,T_0$. It is thus sufficient to build the multivariate GP emulation theory with respect
to the $K$-dimensional vectors $\bx_t$.

Our multivariate dynamic model is of the form 
\begin{eqnarray}
\bx_t&=&\bof(\bx^*_{t,t-1})+\bepsilon_t,\hspace{2mm}\bepsilon_t\sim N_K(\bzero,\bSigma_{\e}), \label{eq:evo2_mult}
\end{eqnarray}
where $\bx_0=x_0\bone_K$ is assumed known. Here $\bone_K$ is a $K$-dimensional vector with all components $1$.

In the above, $\bof(\cdot)=(f_1(\cdot),\ldots,f_K(\cdot))'$ is a function with $K$ components. 
We assume that $\bof(\cdot)$ is a $K$-variate GP with 
mean $E[\bof(\cdot)]=\bB'_f\bh(\cdot)$ 
and covariance function $cov (\bof(\bz^*_1),\bof(\bz^*_2))=c_f(\bz^*_1,\bz^*_2)\bSigma_f$, 
for any $(K+1)$-dimensional inputs $\bz^*_1,\bz^*_2$. Here 
$\bh(\cdot)=(h_1(\cdot),\ldots,h_m(\cdot))'$ and
$\bB_f=(\bbeta_{1,f},\ldots,\bbeta_{K,f})$, where,
for $j=1,\ldots,K$, $\bbeta_{j,f}$ are $m$-dimensional column vectors. Note that $h_1(\cdot)\equiv 1$ corresponds to the intercept 
and $h_2(\cdot),\ldots,h_m(\cdot)$ correspond to the components of $(K+1)$-dimensional inputs $\bz^*$. Hence,
it is clear that $m=K+2$.
Also,
$c_f(\bz^*_1,\bz^*_2)=\exp\left\{-(\bz^*_1-\bz^*_2)'\bR_f(\bz^*_1-\bz^*_2)\right\}$,
where $\bR_f$ is a diagonal matrix consisting of $(K+1)$ smoothness parameters, denoted by 
$\{r_{1,f},\ldots,r_{(K+1),f}\}$.

\subsection{Distributions of $\bof(\bx^*_{1,0})$ and $\bD^*_n$}

Conditional on $\bx_0$, $\bof(\bx^*_{1,0})$ is $K$-variate normal with mean $\bB_f'\bh(\bx^*_{1,0})$ and covariance matrix $\bSigma_f$. 
Now, $\bD_{z^*,nK}=\left(\bof'(\bz^*_1),\bof'(\bz^*_2),\ldots,\bof'(\bz^*_n)\right)'$ has an $nK$-variate normal distribution with mean
\begin{equation}
E[\bD_{z^*,nK}\mid\bB_f,\bSigma_f,\bR_f]=\left(\begin{array}{c}\bB'_f\bh(\bz^*_1)\\\bB'_f\bh(\bz^*_2)\\\vdots\\\bB'_f\bh(\bz^*_n)\end{array}\right)
	=\bmu_{D_{z^*,nK}}\hspace{2mm}\mbox{(say)}
\label{eq:mult_Dz_mean}
\end{equation}
and covariance matrix 
\begin{equation}
V[\bD_{z^*,nK}\mid\bB_f,\bSigma_f,\bR_f]=\bA_{f,D^*_n}\otimes\bSigma_f=\bSigma_{D_{z^*,nK}}\hspace{2mm}\mbox{(say)},
\label{eq:cov_Dz_vector}
\end{equation}
where $``\otimes"$ denotes Kronecker product.
Hence, the distribution of the $n\times K$-dimensional matrix $\bD^*_n=\left(\bof(\bz^*_1),\bof(\bz^*_2),\ldots,\bof(\bz^*_n)\right)'$ is matrix normal:
\begin{equation}
	[\bD^*_n\mid\bB_f,\bSigma_f,\bR_f]\sim\mathcal N_{n,K}\left(\bH_{D^*_n}\bB_f,\bA_{f,D^*_n},\bSigma_f\right).
\label{eq:Dz_matrix_normal}
\end{equation}
Conditionally on $(\bx_0,\bof(\bx^*_{1,0}))$,
it follows that $\bD^*_n$ is $n\times K$-dimensional matrix-normal:
\begin{equation}
[\bD^*_n\mid\bof(\bx^*_{1,0}),\bx_0,\bB_f,\bSigma_f,\bR_f,\bSigma_{\e}]\sim\mathcal N_{n,K}\left(\bmu_{f,D^*_n},\bSigma_{f,D^*_n},\bSigma_f\right)
\label{eq:Dz_matrix_normal_conditional}
\end{equation}
In (\ref{eq:Dz_matrix_normal_conditional}) $\bmu_{f,D^*_n}$ is the mean matrix, given by
\begin{equation}
\bmu_{f,D^*_n}=\bH_{D^*_n}\bB_f+\bs_{f,D^*_n}(\bx^*_{1,0})(\bof(\bx^*_{1,0})'-\bh(\bx^*_{1,0})'\bB_f), \label{eq:mean1_mult}
\end{equation}
and 
\begin{equation}
\bSigma_{f,D^*_n}=\bA_{f,D^*_n}-\bs_{f,D^*_n}(\bx^*_{1,0})\bs_{f,D^*_n}(\bx^*_{1,0}).'
\label{eq:var1_mult}
\end{equation}

\subsection{Joint distribution of $\{\bx_1,\ldots,\bx_{T},\bD^*_n\}$}

Note that
\begin{align}
[\bx_1\mid \bof(\bx_0),\bx_0,\bB_f,\bSigma_f]&\sim N_K\left(\bof(\bx^*_{1,0}),\bSigma_{\e}\right),\label{eq:dist_x1_mult}
\end{align}
and for $t=1,\ldots,T$, the conditional distribution $[\bx_{t+1}=\bof(\bx^*_{t+1,t})+\bepsilon_{t+1}\mid \bD^*_n,\bx_t,\bB_f,\bSigma_f,\bR_f,\bSigma_{\e}]$
is $K$-variate normal with mean
\begin{equation}
\bmu_{x_t}=\bB_f'\bh(\bx^*_{t+1,t})+(\bD^*_n-\bH_{D^*_n}\bB_f)'\bA_{f,D^*_n}^{-1}\bs_{f,\bD^*_n}(\bx^*_{t+1,t})
\label{eq:cond_mean_xt1_mult}
\end{equation}
and variance 
\begin{equation} \bSigma_{x_t}=\left\{1-\bs_{f,\bD^*_n}(\bx^*_{t+1,t})^\prime \bA_{f,\bD^*_n}^{-1}\bs_{f,\bD^*_n}(\bx^*_{t+1,t})\right\}\bSigma_f+\bSigma_{\e}.
\label{eq:cond_var_xt1_mult}
\end{equation}
Since $\bx_0$ is assumed to be known and the distribution of $\bD^*_n$ is given by (\ref{eq:Dz_matrix_normal}),
the joint distribution is obtained by taking products of the individual distributions.

\subsection{Prior distributions}
We assume the following forms of the prior distributions:
\begin{align*}
[\bB_f\mid\bSigma_f]&\sim \mathcal N_{m,K}\left(\bB_{f,0},\bSigma_{B_f,0},\psi\bSigma_f\right);
\\
	[\bSigma_f]&\propto\left|\bSigma_{f}\right|^{-\frac{\nu_f+K+1}{2}}\exp\left[-\frac{1}{2}tr\left(\bSigma^{-1}_f\bSigma_{f,0}\right)\right],~\mbox{with}~
\nu_f>K-1;
\\
[\bSigma_{\epsilon}]&\propto\left|\bSigma_{\epsilon}\right|^{-\frac{\nu_{\epsilon}+K+1}{2}}
	\exp\left[-\frac{1}{2}tr\left(\bSigma^{-1}_{\epsilon}\bSigma_{\epsilon,0}\right)\right],~\mbox{with}~
	\nu_{\epsilon}>K-1;~\mbox{and}
	\\
\mbox{for}\ \ i=1,\ldots,(K+1),\nonumber\\
[\log(r_{i,f})]&\stackrel{iid}{\sim} N\left(\mu_{R_f},\sigma^2_{R_f}\right).
\end{align*}
%
%
%
%


For the prior of $\bB_f$ we set $\psi=1$, and except the first column of $\bB_{f,0}$, we set all other columns of $\bB_{f,0}$ to be null vectors.
We set the first column of $\bB_{f,0}$ to be the vector of means of the $K$ GCM based time series thinned by $5$ observations. Recall that these means
are also used for the corresponding prior in the one-dimensional situation for model selection.

In the priors for $\bSigma_f$ and $\bSigma_{\epsilon}$, we set $\nu_f=K$ and $\nu_\epsilon=K$.
For $\bSigma_{B_f,0}$ and $\bSigma_{\e,0}$, we first let $\hat\bSigma$ to be the empirical covariance matrix for the $K$ GCM-based time series, thinned by
$5$ observations. Then we set $\bSigma_{B_f,0}=\bSigma_{\e,0}=\hat\bSigma/2$.
Again, this choice is analogous to the previous one-dimensional setup. 


For the log-normal priors of the smoothness parameters
we set $\mu_{R_f}=-0.5$ and $\sigma^2_{R_f}=1$. 
The choices imply as in the one-dimensional situation that the prior mean and the prior variance of each of the smoothness
parameters are, respectively, 1 and 2 (approximately).

Thus, these prior choices are in keeping with the one-dimensional situation and have similar rationale as before.

\subsection{Choice of the input grid $\bG_n$}
\label{subsec:input grid}
To set up the $(K+1)$-dimensional grid $\bG_n$ for the model-fitting purpose, we considered $[-5,5]^K$ to be a grid space for the $K$-dimensional
variable $\bz$. 
We divide $[-5,5]$ into 50 equal sub-intervals
and choose a point randomly from each of the $50$ sub-intervals, in each dimension, yielding $n=50$ $K$-dimensional points corresponding to $\bz$.
For the first component of the grid, corresponding to the time
component, we follow the strategy in the one-dimensional GP situation. That is, we first re-label the times $1850-2099$ as $0-249$ and further
divide the re-labeled times by $250$ to have them lie in $[0,1]$. Then, after dividing the interval
$[0, 1]$ into $50$ equal sub-intervals we randomly simulate a value from each sub-interval, to complete construction of the input grid $\bG_n$. This grid choice
turned out to be adequate for our purpose, as the results demonstrate.

The rest of the multivariate GP emulation theory remains analogous to the corresponding univariate case, but the full conditionals of $\bB_f$
and $\bD^*_n$ are no longer available in standard form for simulating in the MCMC context, which is not analogous to the univariate context discussed in  
Section \ref{sec:current_future}; see the supplement of \ctn{Ghosh14} for details. We use additive TMCMC to update the unknowns in the multidimensional situation.
For updating the positive definite matrices $\bSigma_f$ and $\bSigma_{\epsilon}$, we represent the matrices in the Cholesky decomposition forms $\bC\bC'$, where
$\bC$ is a lower triangular matrix, and use additive TMCMC to update the
non-zero elements in a single block. We implement our codes, written in C, in our VMWare. The implementations associated with A1B, A2, B1 and
Commitment took about 30 hours 37 minutes, 18 hours 52 minutes, 29 hours 12 minutes and 16 hours 59 minutes, respectively.

\section{Results for the multivariate climate dynamics}
\label{sec:results_mult}
\begin{figure}
	\centering
	\subfigure [SRES: A1B.]{ \label{fig:best_mean1}
	\includegraphics[width=7.5cm,height=6.7cm]{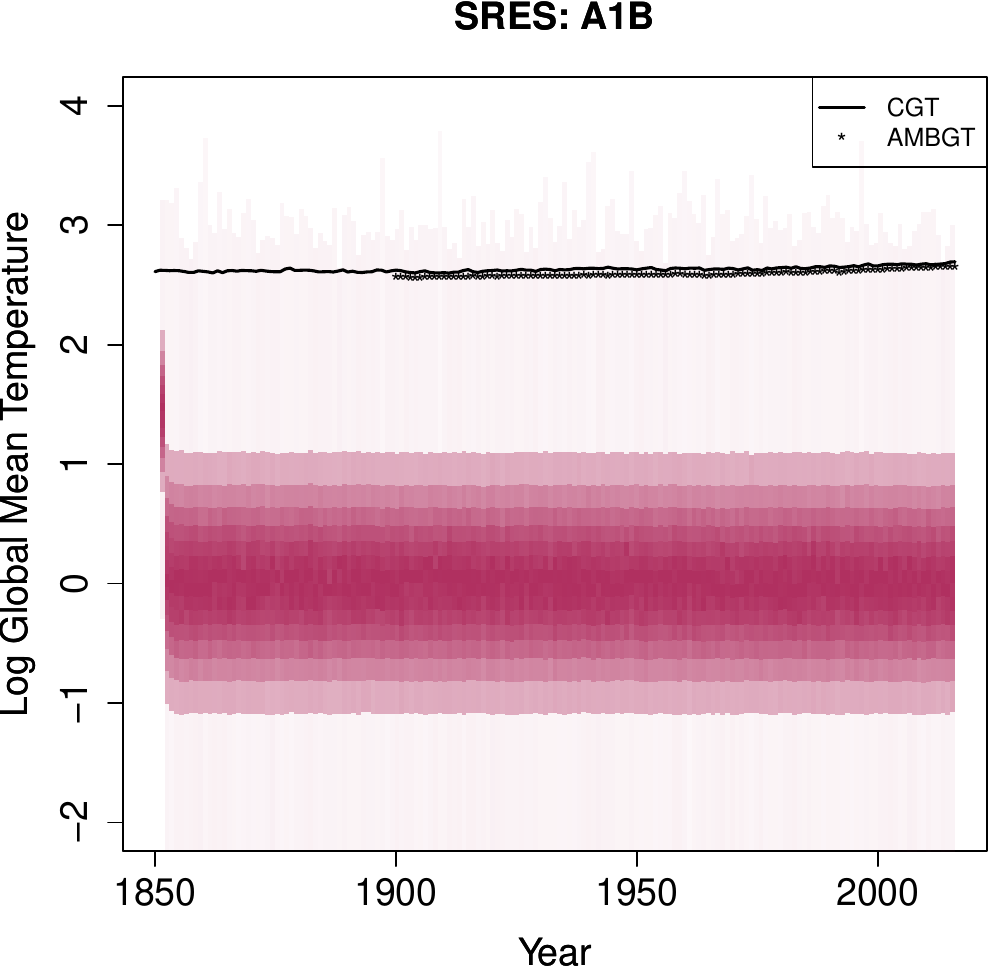}}
	\hspace{2mm}
	\subfigure [SRES: A2.]{ \label{fig:best_mean2}
	\includegraphics[width=7.5cm,height=6.7cm]{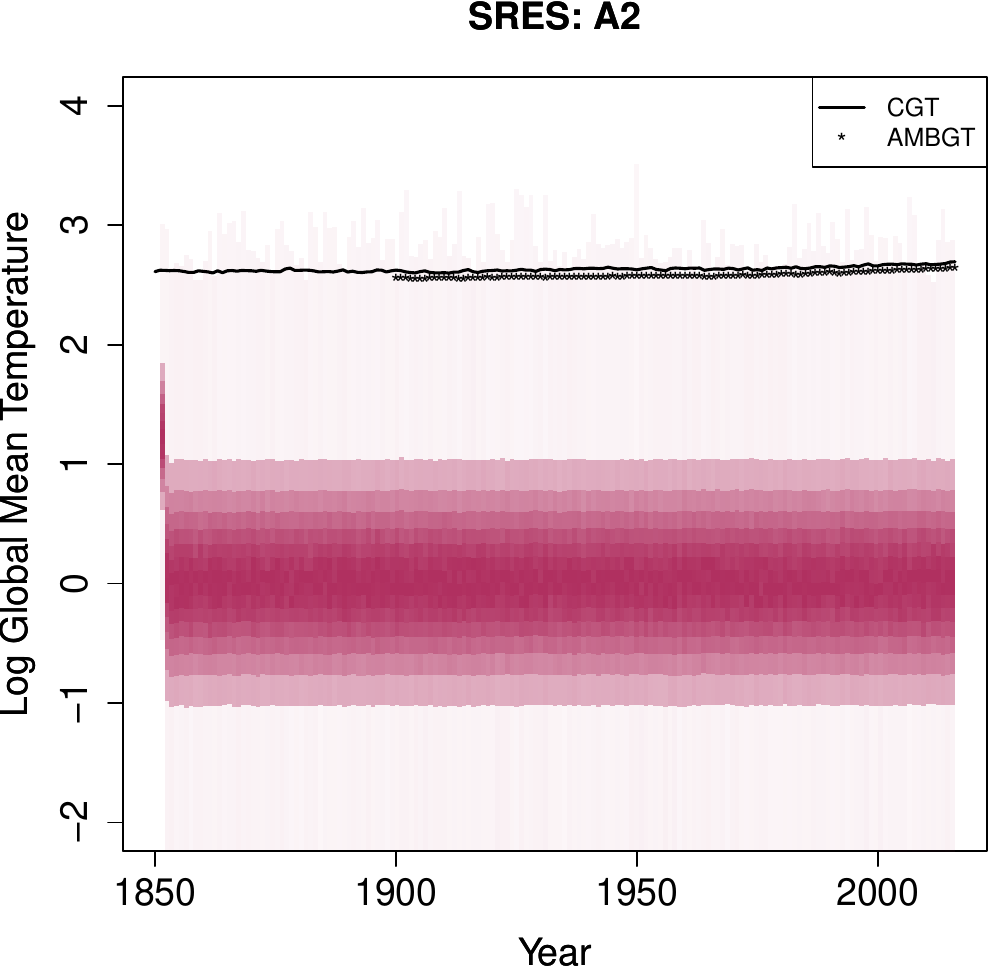}}\\
	\vspace{2mm}
	\subfigure [SRES: B1.]{ \label{fig:best_mean3}
	\includegraphics[width=7.5cm,height=6.7cm]{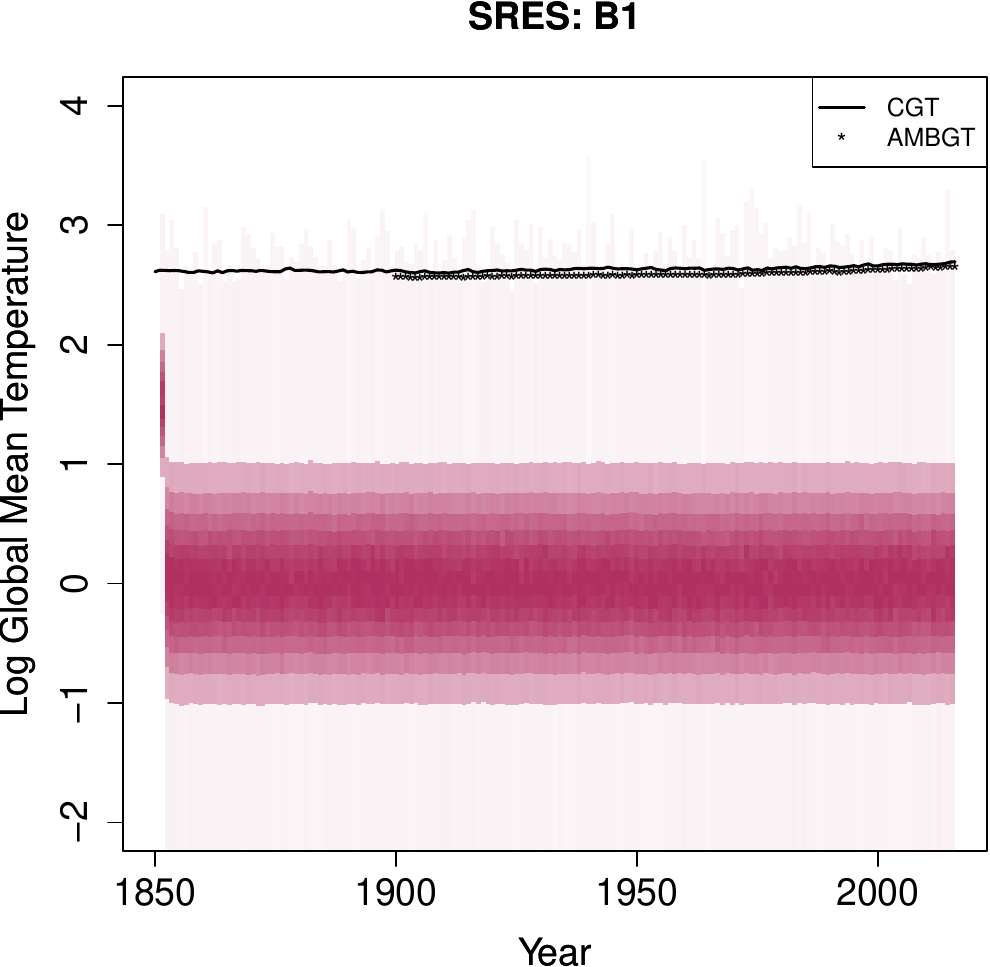}}
	\hspace{2mm}
	\subfigure [Commitment.]{ \label{fig:best_mean4}
	\includegraphics[width=7.5cm,height=6.7cm]{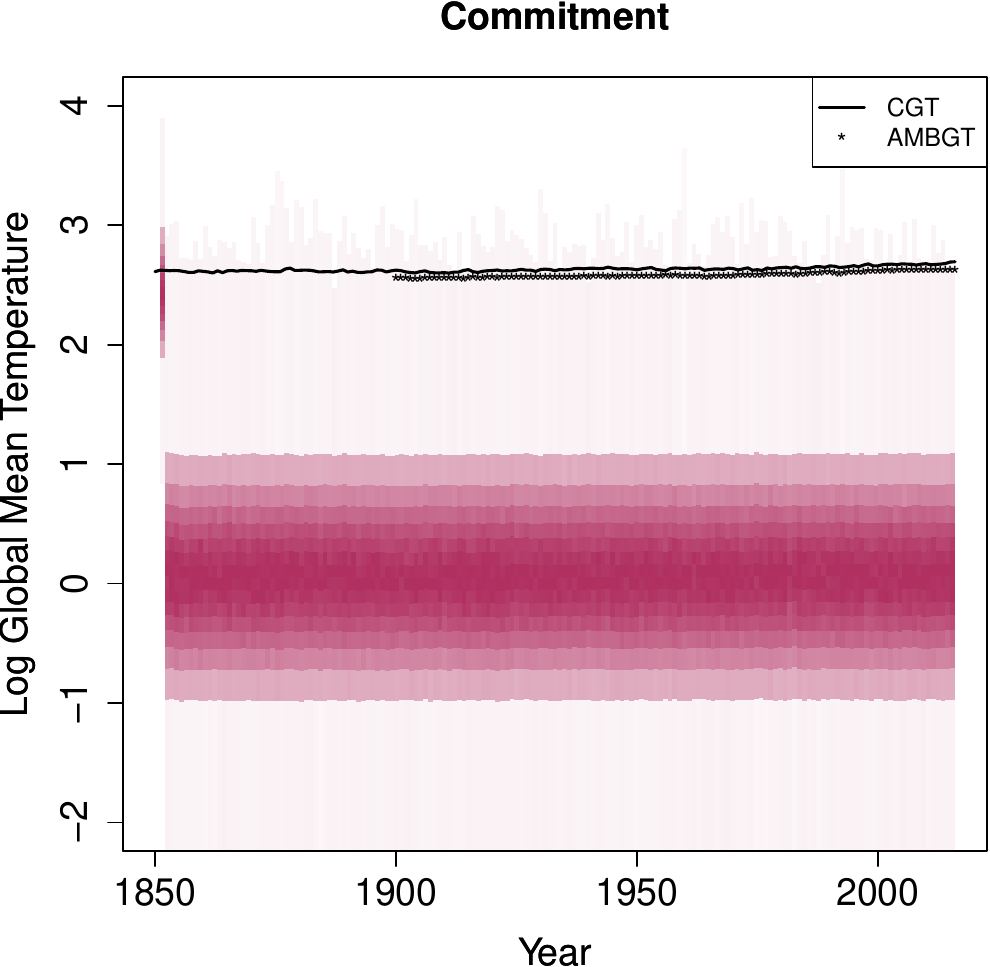}}
	\caption{The posteriors $[\bar x_0,\bar x_1,\ldots, \bar x_{T_0}|\bx_{T_0+1},\ldots,\bx_T]$ 
	are shown as colour plots with progressively higher densities depicted by progressively intense colours, along with the 
	HadCRUT4 data (CGT) and the average of GCM based time series (AMBGT).
	The temperature is in $\degree$C and in the log-scale.}
	\label{fig:mult_mean}
\end{figure}

For the four climate scenarios, the posterior distributions of 
$[\bar x_0,\bar x_1,\ldots, \bar x_{T_0}|\bx_{T_0+1},\ldots,\bx_T]$ (where $\bar x_0=x_0$, since $x_0$ is assumed to be known) are shown in Figure \ref{fig:mult_mean}.
Now, compared to the one-dimensional situations, severe under-estimation of the HadCRUT4 data by all the four climate scenarios is corroborated by this multivariate framework.
And, Table \ref{table:mult_mean} revealing severe underfits for all the four climate scenarios, confirms that 
even the discrepancy measures could not act as saviours this time.
\newpage
\begin{table}[h]
\centering
	\caption{Goodness-of-fit check for ensembles of GCM time series with respect to $[\bar x_0,\bar x_1,\ldots, \bar x_{T_0}|\bx_{T_0+1},\ldots,\bx_T]$. 
	Here 95\% BCI stands for 95\% Bayesian credible intervals.}
\label{table:mult_mean}
	\begin{tabular}{|c|c|c|c|c|}
\hline
		Model & $S^{(k)}_1\left(\bx^{(0)}_{T_0}\right)$ & 95\% BCI of $S^{(k)}_1\left(\bx_{T_0}\right)$ 
		& $S^{(k)}_2\left(\bx^{(0)}_{T_0}\right)$ & 95\% BCI of $S^{(k)}_2\left(\bx_{T_0}\right)$\\ 
\hline
		A1B   &  3.580  &  [0.690,0.872]  &  13.158  &  [0.763,1.182]\\
		A2  &  3.807  &  [0.689,0.871]  &  14.909  &  [0.759,1.179]\\
		B1  &  3.872  &  [0.688,0.870]  &  15.434  &  [0.758,1.177]\\   
		Commit  &  3.711  &  [0.690,0.870]  &  14.229  &  [0.760,1.176]\\ 
\hline
\end{tabular}
\end{table}

\begin{figure}
	\centering
	\subfigure [SRES: A1B.]{ \label{fig:best_max1}
	\includegraphics[width=7.5cm,height=6.7cm]{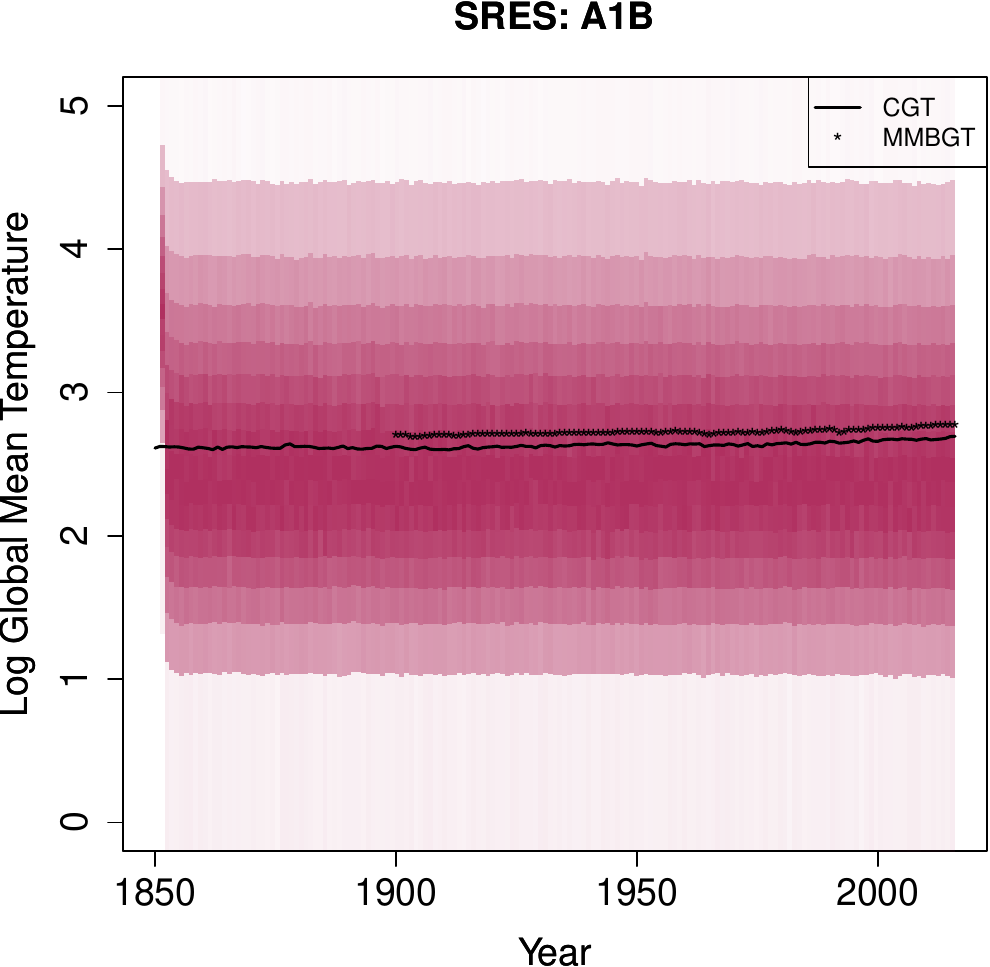}}
	\hspace{2mm}
	\subfigure [SRES: A2.]{ \label{fig:best_max2}
	\includegraphics[width=7.5cm,height=6.7cm]{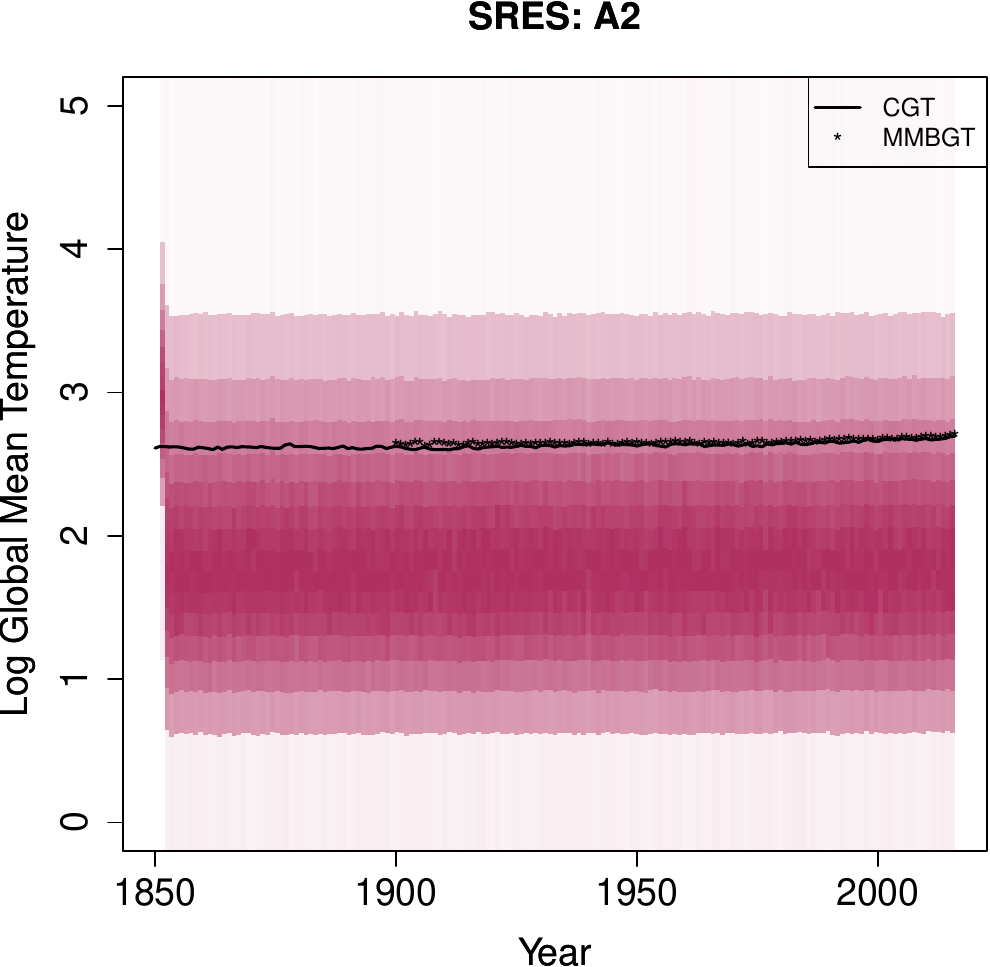}}\\
	\vspace{2mm}
	\subfigure [SRES: B1.]{ \label{fig:best_max3}
	\includegraphics[width=7.5cm,height=6.7cm]{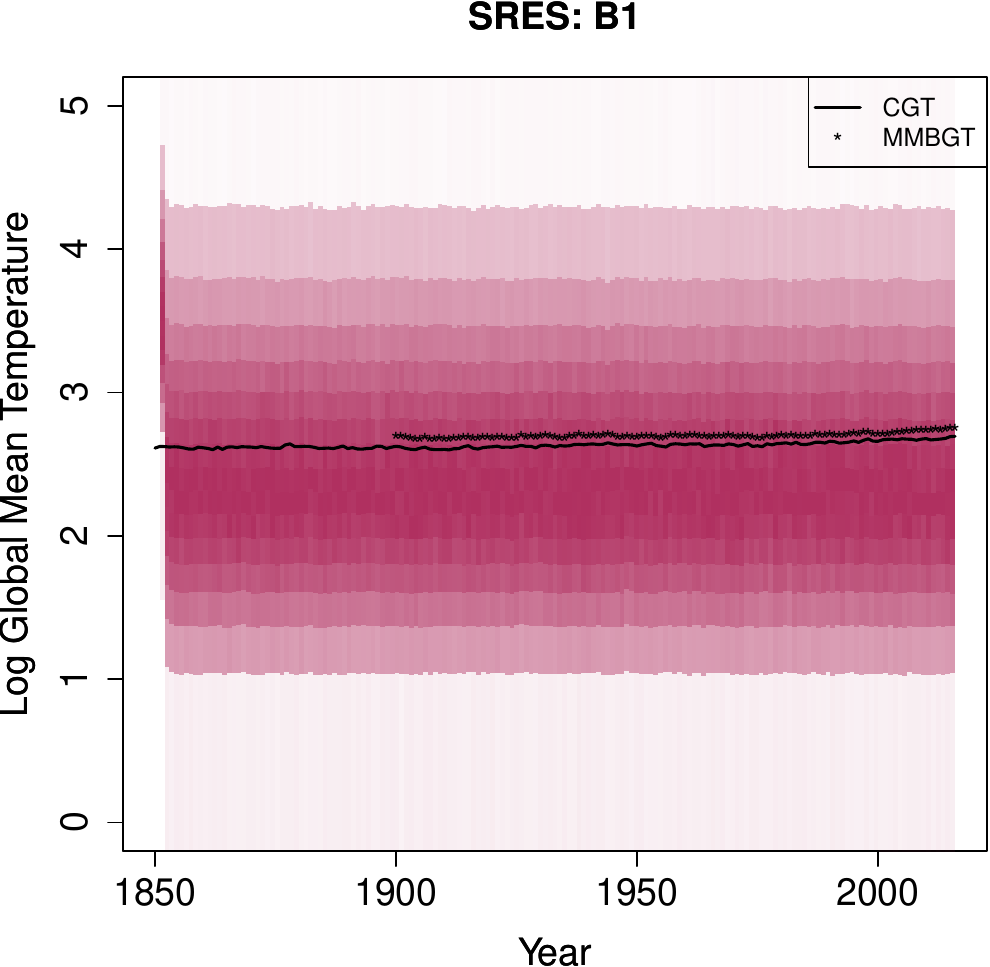}}
	\hspace{2mm}
	\subfigure [Commitment.]{ \label{fig:best_max4}
	\includegraphics[width=7.5cm,height=6.7cm]{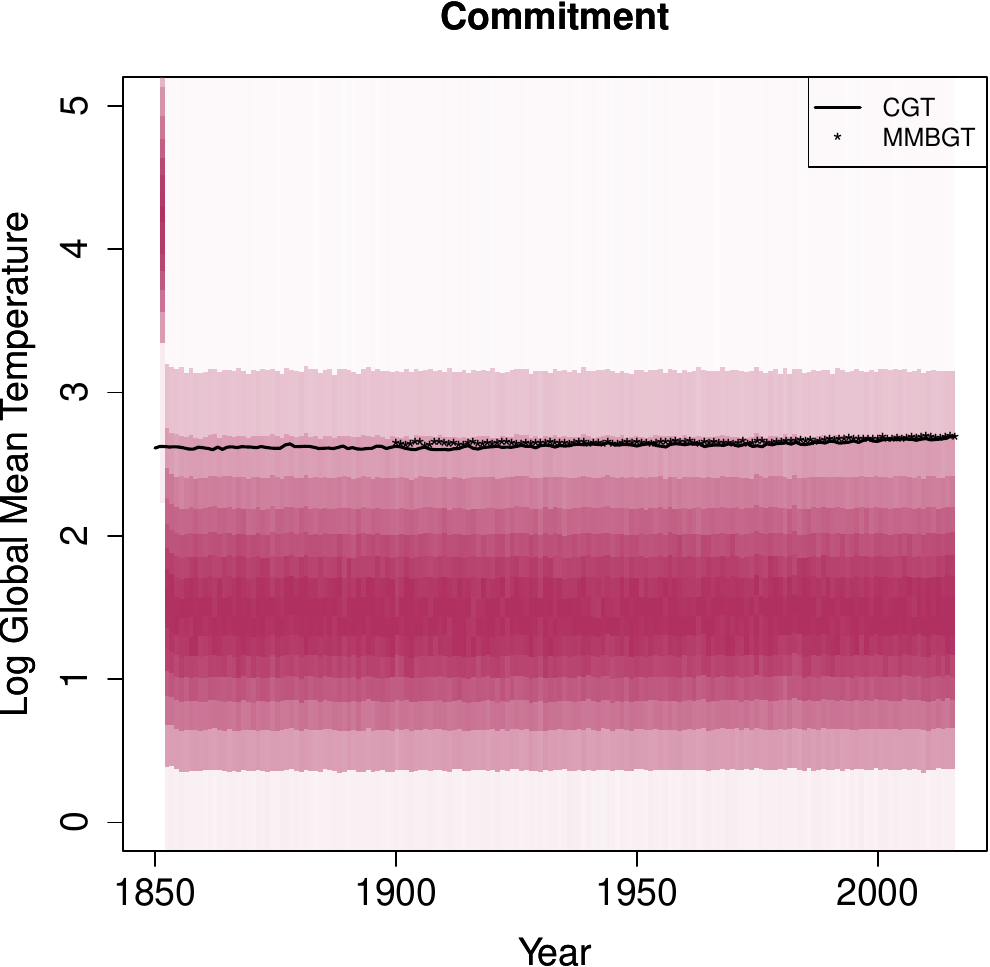}}
	\caption{The posteriors $[x^{(max)}_0,x^{(max)}_1,\ldots, x^{(max)}_{T_0}|\bx_{T_0+1},\ldots,\bx_T]$ 
	are shown as colour plots with progressively higher densities depicted by progressively intense colours, along with the 
	HadCRUT4 data (CGT) and the maximum of model based global temperature (MMBGT).
	The temperature is in $\degree$C and in the log-scale.}
	\label{fig:mult_max}
\end{figure}

Since $[\bar x_0,\bar x_1,\ldots, \bar x_{T_0}|\bx_{T_0+1},\ldots,\bx_T]$ severely under-estimates the HadCRUT4 data, we now investigate how well
the posterior $[x^{(max)}_0,x^{(max)}_1,\ldots, x^{(max)}_{T_0}|\bx_{T_0+1},\ldots,\bx_T]$ can capture the observed current temperature data, where
for $t=0,1,2,\ldots$, $x^{(max)}_t$ is the maximum of the components of $\bx_t$. 
Figure \ref{fig:mult_max} displays the relevant posterior time series as colour plots,
along with the HadCRUT4 data (CGT) and the maximum of model based global temperature (MMGT) associated with the GCM simulations, in the log scales.
Observe that CGT and MMGT are included in the supports, but it is doubtful how good the fits are, since the posterior variances are high and moreover for A2 and Commitment
CGT and MMGT fall in low density regions. Table \ref{table:mult_max} shows that the fits are indeed not encouraging.
Observe that A1B overfits with respect to both $S^{(k)}_1$ and $S^{(k)}_2$. With respect to $S^{(k)}_1$, A2 slightly underfits, while the fit is adequate
with respect to $S^{(k)}_2$. Since $S^{(k)}_2$ is generally a better performer than $S^{(k)}_1$, one can consider the fit of A2 to be adequate. B1 seriously
overfits with respect to both the discrepancy measures, while Commitment seriously underfits with respect to both $S^{(k)}_1$ and $S^{(k)}_2$.
\newpage
\begin{table}[h]
\centering
	\caption{Goodness-of-fit check for ensembles of GCM time series with respect to $[x^{(max)}_0,x^{(max)}_1,\ldots, x^{(max)}_{T_0}|\bx_{T_0+1},\ldots,\bx_T]$. 
	Here 95\% BCI stands for 95\% Bayesian credible intervals.}
\label{table:mult_max}
	\begin{tabular}{|c|c|c|c|c|}
\hline
		Model & $S^{(k)}_1\left(\bx^{(0)}_{T_0}\right)$ & 95\% BCI of $S^{(k)}_1\left(\bx_{T_0}\right)$ 
		& $S^{(k)}_2\left(\bx^{(0)}_{T_0}\right)$ & 95\% BCI of $S^{(k)}_2\left(\bx_{T_0}\right)$\\ 
\hline
		A1B   &  0.216  &  [0.693,0.891]  &  0.061  &  [0.787,1.313]\\
		A2  &  0.893  &  [0.692,0.888]  &  0.816  &  [0.786,1.318]\\
		B1  &  0.303  &  [0.690,0.891]  &  0.104  &  [0.785,1.332]\\   
		Commit  &  1.256  &  [0.671,0.879]  &  1.617  &  [0.755,1.376]\\ 
\hline
\end{tabular}
\end{table}

\section{Future climate forecast with our Bayesian GP dynamics model}
\label{sec:forecast}
Our detailed analyses of the GCM forecasts so far failed to justify their credibilities. This failure, however, seems to hold a great deal of positivity
since the rapid future global warming foreboding that might eventually threaten life on earth, need not become the reality. 
However, it is not clear yet then what kind of climate change we can expect in the future. We attempt to answer this question, again with our Bayesian GP
emulation theory, now forecasting the log global average temperature in the years $2017-2099$ given the log HadCRUT4 dataset for the years $1850-2016$, using the
theory and strategies proposed in Section \ref{subsec:future_given_current}. Here we let the prior distributions remain the same as detailed in 
Section \ref{sec:prior_univariate}, except that the first component of $\bbeta_{f,0}$ and $\hat\sigma^2$ are now based upon thinning the log HadCRUT4 data by 
$5$ observations. The input grid $\bG_n$ remains the same as in the one-dimensional setup detailed in Section \ref{subsec:comp_inv_post}.

\begin{figure}
	\centering
	\subfigure [A1B: Best GCM $\mbox{csiro}\_\mbox{mk3}\_0$.]{ \label{fig:future1}
	\includegraphics[width=7.5cm,height=6.7cm]{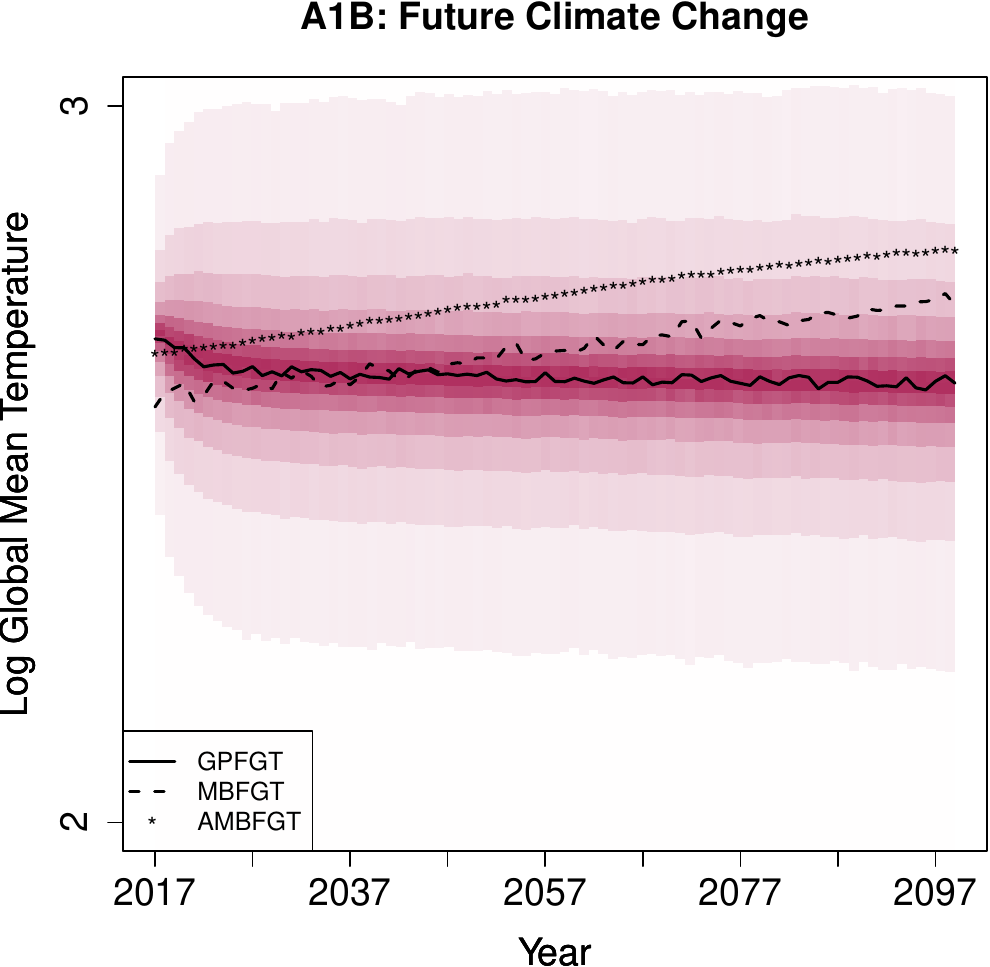}}
	\hspace{2mm}
	\subfigure [A1B: Best GCM $\mbox{inmcm3}\_0$.]{ \label{fig:future2}
	\includegraphics[width=7.5cm,height=6.7cm]{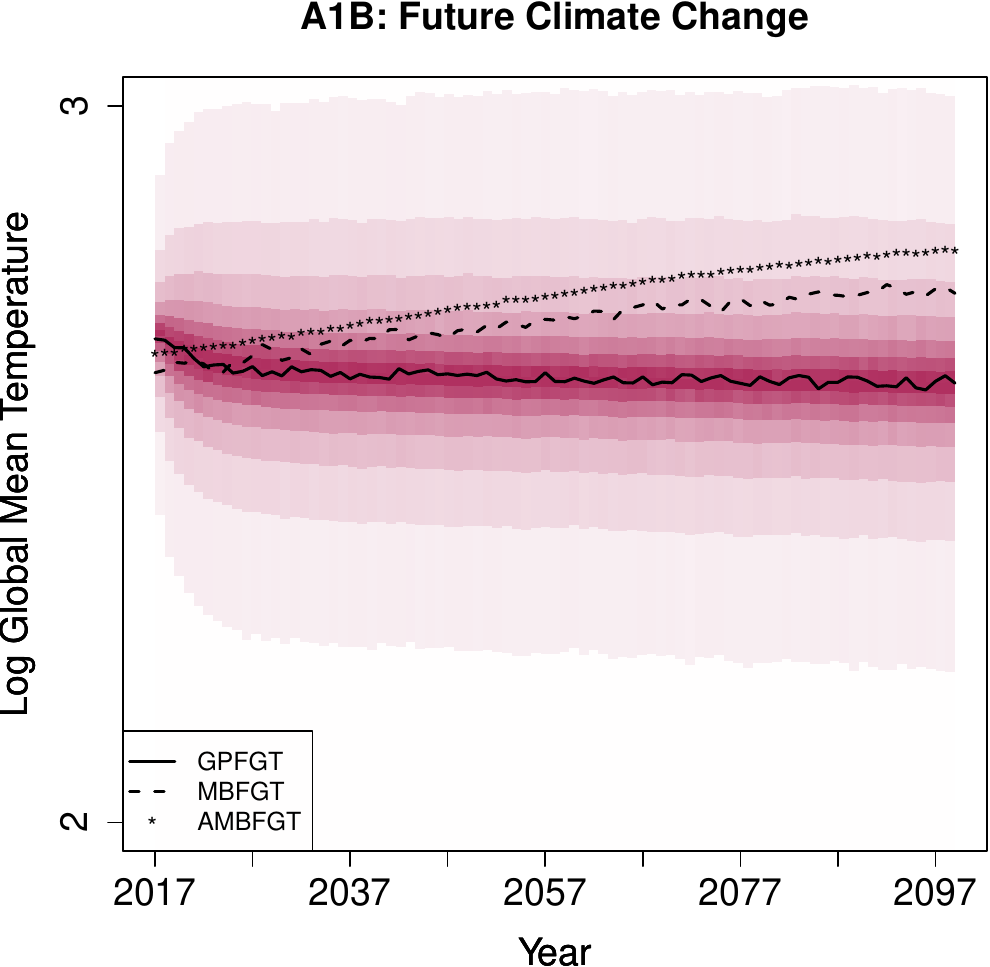}}\\
	\vspace{2mm}
	\subfigure [A2: Best GCM $\mbox{ukmo}\_\mbox{hadgem1}$.]{ \label{fig:future3}
	\includegraphics[width=7.5cm,height=6.7cm]{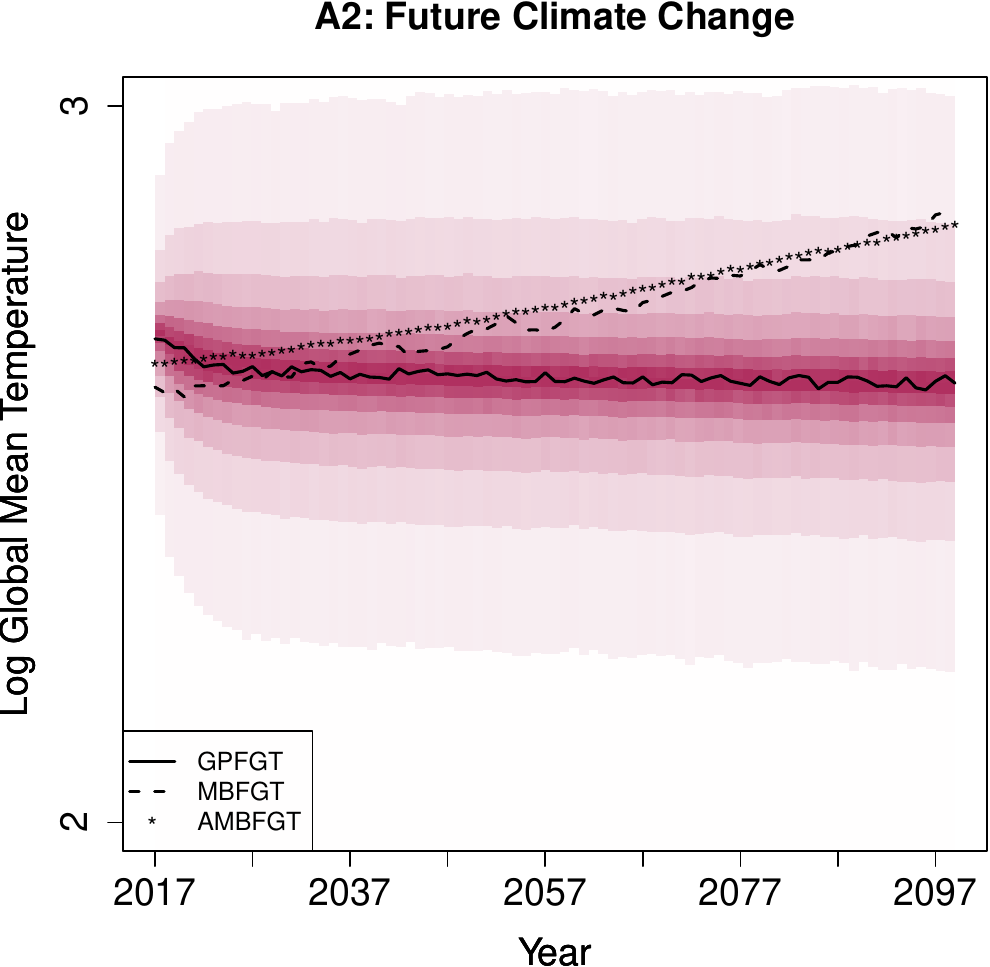}}
	\hspace{2mm}
	\subfigure [B1: Best GCM $\mbox{gfdl}\_\mbox{cm2}\_0$.]{ \label{fig:future4}
	\includegraphics[width=7.5cm,height=6.7cm]{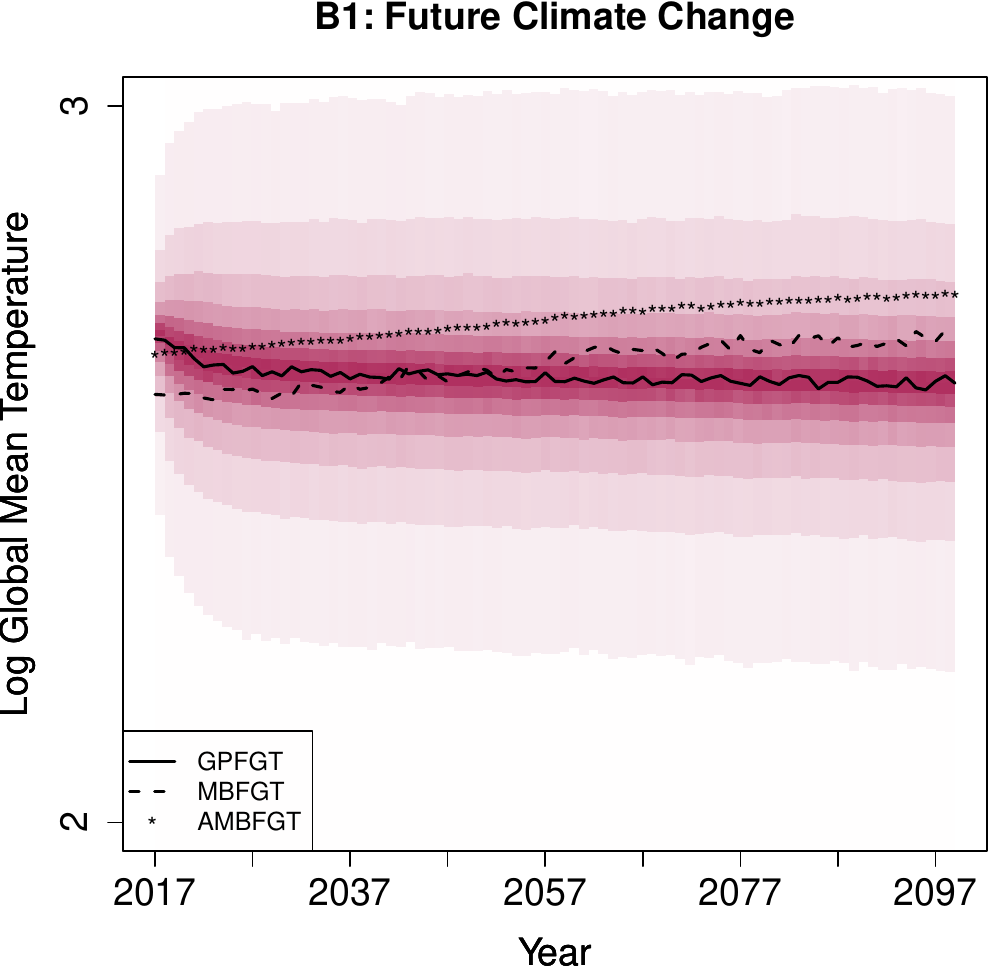}}\\
	\vspace{2mm}
	\subfigure [Commitment: Best GCM $\mbox{cnrm}\_\mbox{cm3}$.]{ \label{fig:future5}
	\includegraphics[width=7.5cm,height=6.7cm]{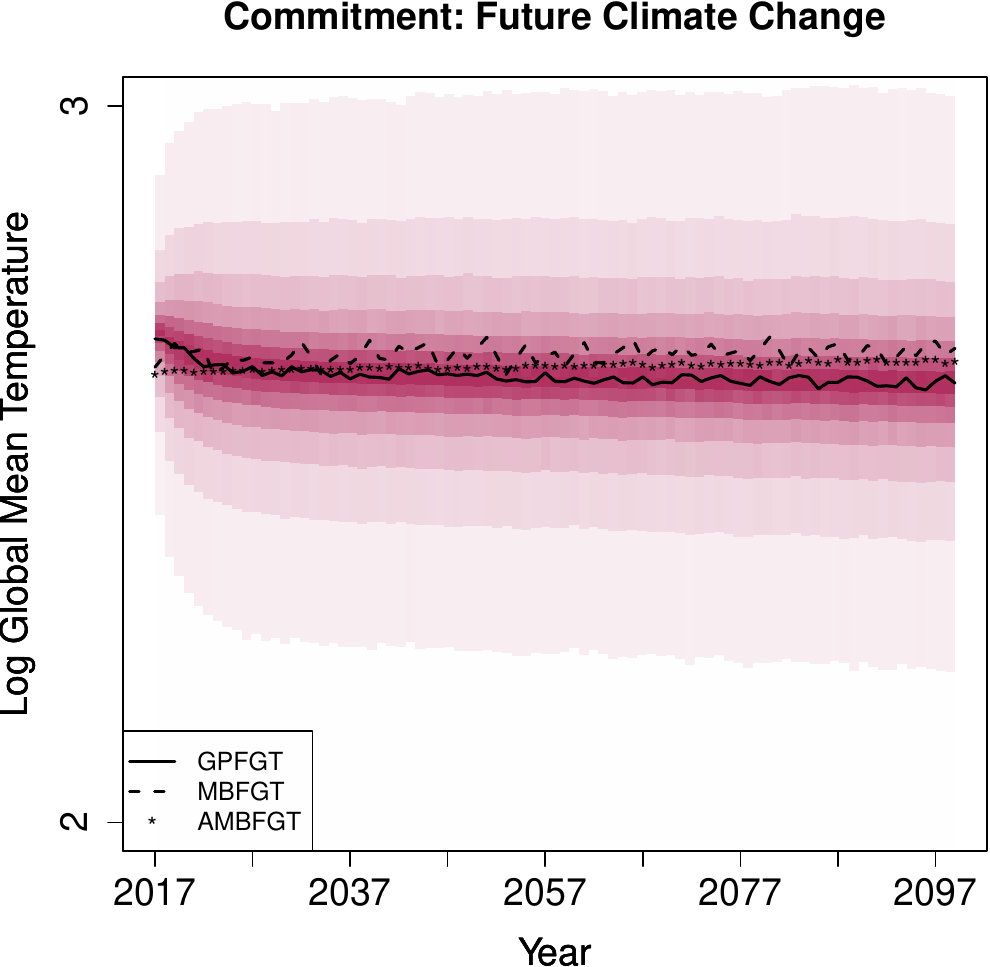}}
	\caption{The posteriors $[x_{T_0+1},\ldots, x_{T}|x_{1},\ldots,x_{T_0}]$ for future climate prediction 
	are shown as colour plots, 
	along with the posterior modes of the 
	GP forecasted global temperature (GPFGT), best GCM-specific 
	model based forecasted global temperature (MBFGT) and average model based forecasted global temperature (AMBFGT).
	The temperature is in $\degree$C and in the log-scale.}
	\label{fig:future}
\end{figure}

Our future climate prediction results are presented in Figure \ref{fig:future}, along with the posterior modes associated with the 
GP forecasted global temperature (GPFGT), the best GCM-specific model based forecasted global temperature (MBFGT) and 
average model based forecasted global temperature (AMBFGT). In stark contrast with MBFGT and AMBFGT which show steep increase in the temperature in panels (a)-(d),
the high posterior density regions of our Bayesian forecasts 
do not support increasing future global temperature. Only in the case of Commitment (panel (e)) MBFGT and AMBFGT tend to fall within the
high posterior density regions of our Bayesian forecasts.

According to \ctn{Green09}: ``{\it The benchmark forecast is that the global mean temperature for each year for the rest of this century
will be within $0.5\degree$C of the 2008 figure.}" Thus, according to their prediction, the future global temperature should lie in the interval $[13.895,14.895]\degree$C.
This interval is included even within all the $50\%$ credible intervals of our year-wise Bayesian posterior forecast distributions for $2017-2099$. 
Thus, our results are broadly in agreement with the forecast of \ctn{Green09}, and clearly do not support drastic global warming as projected by the GCMs.


\section{Summary and discussion}
\label{sec:conclusion}

As stated in \ctn{Lupo13} (see also the references therein), ``{\it When physicists, biologists, and other scientists
who are unaware of the rules of forecasting attempt to make climate predictions, their forecasts are at risk of
being no more reliable than those made by non-experts, even when they are communicated through
complex computer models}". The GCMs are indeed complex computer models built by  physicists, biologists, and other scientists. The future global warming
forecasts yielded by such models have great bearing on the current world and particularly on the IPCC policymakers. But as discussed by \ctn{Lupo13} in great detail,
major scientists of the world do not find much reason to pin faith on the global warming foreboding, and most of them, based on their experiments and
experiences, are strongly critical of the abilities of GCMs to adequately model so complex a system as world climate. 

However, we are unaware of any significant and rigorous statistical research that evaluates the GCM-based global warming projections. Such a task, which
is of global importance, must be seriously undertaken, and no wonder statistics is the only discipline that can promise to make justice to such an issue where
quantification of uncertainties (in the predictions by the GCMs) plays the most important role. It is also very well-established that the Bayesian
statistical paradigm is the most well-equipped to coherently deal with uncertainty quantifications.

In this study, we have developed and applied a novel Bayesian framework for evaluating climate model projections, with a specific focus on global warming. 
Our approach combines both inverse regression -- in which we assess the plausibility of the observed past given assumed futures -- with forward forecasting, 
where we make data-driven predictions of future climate trends based solely on historical data.

Such assessments are previously contemplated upon in the climate context by other researchers: for example, \ctn{Lupo13}, quoting \ctn{Reifen09}, write
``Expounding on this principle, \ctn{Reifen09} note, ``{\it with the ever increasing number of models, the question arises of how to make a best
estimate prediction of future temperature change.}" That is to say, which model should one use? With
respect to this question, they note, ``{\it one key assumption, on which the principle of performance-based selection rests, is that a model which performs
better in one time period will continue to perform better in the future.}" In other words, if a model predicts past climate fairly well, it should predict
future climate fairly well. The principle sounds reasonable enough, but does it hold true?" 

At the heart of our methodology lies a nonparametric, compositional GP-based model of the global temperature time series. 
This flexible, black-box model avoids restrictive parametric assumptions and allows us to emulate complex climate dynamics over time. 
Our inverse regression formulation, an unexplored paradigm in time-series analysis, enables rigorous testing of general circulation models (GCMs) 
by evaluating how well their projected futures explain the known past.

Our empirical results reveal substantial inconsistencies between the forecasts generated by most IPCC-endorsed GCMs and the historical temperature records. 
Specifically, under the inverse Bayesian model testing framework, we find that the majority of GCMs—regardless of scenario -- assign low posterior probability 
to the actual global warming pattern observed from 1850 to 2016. In other words, if their future projections were correct, the present as we know it would be highly unlikely.

These findings are strongly supported by our forward modeling approach. Using compositional GP regression trained solely on historical data, we forecast global mean 
temperatures through the end of the 21st century. Our predictions suggest more moderate warming trajectories than those forecasted by the GCMs. 
Strikingly, only the Commitment scenario shows partial alignment with our forecasts; most others lie well outside the high-probability regions of our posterior distributions.

Taken together, these results cast doubt on the fidelity of GCM projections and underscore the importance of independent, statistically grounded evaluation frameworks. 
While our analysis does not dispute the reality of current global warming, it raises important concerns about the extent to which current GCMs capture the 
true structure of future climate evolution.

We emphasize that our methodology is not meant to replace GCMs, but to complement them -- by offering a rigorous statistical lens through which their forecasts can be 
tested and refined. This work demonstrates the value of Bayesian inverse thinking in climate science and highlights the need for closer integration between 
physical modeling and data-driven statistical inference.

Given our Bayesian analysis and future climate projections, what should be the right climate policy?
In this regard, recall that (see Section \ref{sec:forecast}) our Bayesian forecast results are broadly in agreement with those 
of \ctn{Green09}, the only other statistical research on global warming, as per our knowledge. The latter's forecast results clearly
do not support future global warming. Hence, we are in agreement with their recommendation that the best policy would be to {\it do nothing} about global warming!
At least until stronger Bayesian statistical evidences of future global warming emerge, in other significant climate data analyses. 

Looking ahead, our framework could be extended to incorporate spatio-temporal models, dynamic covariates, or hybrid approaches that integrate physical constraints 
into the statistical emulation process. We also anticipate that future work may apply this framework to other environmental processes, such as sea-level rise 
or precipitation extremes, where model uncertainty remains high and decision-making stakes are critical.
In this regard, some key sources of climate data are National Aeronautics and Space Administration (NASA), which provides data on temperature, ice sheet and glacial melt,
and greenhouse gas concentrations; National Oceanic and Atmospheric Administration (NOAA), which provides data on global surface temperatures, including land and 
ocean temperature records, as well as historical weather and climate data; World Bank, that provides access to global, regional and country-level climate data;
National Snow and Ice Data Center, providing data on sea ice, ice sheet, glaciers and permafrost, etc. Combining various data sources coherently would itself be a very
challenging undertaking, which would subsequently require a very complex nonparametric Bayesian multivariate spatio-temporal model for future forecasts. Practical implementation
of such a model would likely require supercomputing facilities, with very sophisticated parallel MCMC strategy.

The concluding remarks would not be complete without the mention of the work by \ctn{Roy20}, who consider a completely novel approach to analyzing past and future 
climate dynamics using their novel theory on Bayesian assessment of random series convergence. Strikingly, their results indicate, in their words 
``{\it there does not seem to have been instances of prolonged global warming or cooling in the past, and nor such adverse climatic conditions are likely to 
prevail in the future.  Indeed, global climate dynamics is subject to temporary variations only, and the current global warming
phenomenon is just an instance of such variation."}

\section*{Acknowledgments}
We are sincerely grateful to the reviewer whose comments have led to improved presentation of our manuscript. We also thank ChatGPT for help with proofreading.

\bibliography{irmcmc}

\end{document}